\documentclass[fleqn,usenatbib]{mnras}

\usepackage{newtxtext,newtxmath}
\usepackage[T1]{fontenc}
\usepackage{graphicx}
\usepackage{amsmath}
\usepackage{ulem}
\usepackage{soul}
\usepackage[usenames,dvipsnames]{xcolor}
\usepackage{hyperref}

\usepackage{macros} % Frank's personal macros + paper-specific additions

\graphicspath{{./}{Figures/}}

\title[Satellite orbits in triaxial haloes]
      {Scatter, bias, and chaos of satellite orbits in triaxial dark matter haloes}

\author[B. T. Chiang, F. C. van den Bosch and M. A. Keim]
{Barry T. Chiang$^{1}$\thanks{E-mail: barry.chiang@yale.edu},
Frank C. van den Bosch$^{1}$, and Michael A. Keim$^{1}$
\vspace*{8pt}
\\
$^{1}$Department of Astronomy, Yale University, New Haven, CT 06511, USA.}

\date{Accepted 2026 October 6. Received 2026 October 2; in original form 2026 August 28}

\pubyear{2026}

\begin{document}

\label{firstpage}
\pagerange{\pageref{firstpage}--\pageref{lastpage}}
\maketitle

%%%%%%%%%%%%%%%%%%%%%%%%%%%%%%%%%%%%%%%%%%%%%%%%%%%%%%%%%%%%%%%%%%%%%%%%%%%%%

\begin{abstract}
The pericentric distances of satellite galaxies govern their tidal stripping, quenching, and survival, yet their orbits are almost universally computed in spherical host haloes, whereas dark matter haloes are generically triaxial. We quantify, orbit by orbit, the error this simplification incurs. We integrate $10^6$ satellites, drawn from a cosmological infall distribution, in static NFW hosts of systematically varying triaxiality at fixed mass profile. Triaxiality leaves the population medians of pericentre, apocentre, and orbital period essentially unchanged. Instead, successive pericentres of an individual orbit scatter by $5$--$23\%$ depending on host shape, irrespective of orbital energy; the minimum pericentre distance reached within a Hubble time shrinks systematically by up to $21\%$, enhancing the peak tidal mass loss. This triaxiality-driven orbital dephasing is overwhelmingly regular rather than chaotic; chaotic satellite orbits are common and dominate the near-centre pericentric passages in strongly flattened hosts, but diverge on time-scales far exceeding the Hubble time. Satellite orbits spherically reconstructed via direct backward integration in a truly triaxial host diverge by a tenth of the virial radius within $2$--$4\Gyr$, with comparable uncertainties sourced separately by the unconstrained shape of the host and by its unknown absolute orientation. For the Milky Way dwarf Triangulum~II, the unknown halo shape and orientation alone spread the inferred pericentre by $\sim\!50\%$, five times its reported uncertainty and more than ten times the LMC-induced shift, and bias it by $6$--$70\%$. This host-shape uncertainty can dominate satellite orbital error budgets and should be incorporated in future inferences.
\end{abstract}


%%%%%%%%%%%%%%%%%%%%%%%%%%%%%%%%%%%%%%%%%%

\section{Introduction}
\label{sec:intro}

In hierarchical structure formation, dark matter haloes and the galaxies they host are continuously accreted onto larger systems. As such, these ubiquitous subhaloes and satellite galaxies encode information on both the underlying cosmology and the nature of dark matter. Population-level satellite kinematics have yielded stringent constraints on the galaxy--halo connection \citep[e.g.,][]{vandenBosch2004, More2011, Lange2019} and on cosmological parameters \citep{Mitra2024, Mitra2025}. The observed satellite abundance, from the Milky Way satellites to cluster-scale substructures \citep[e.g.,][]{Nadler2019, Nadler2021, Dekker2022, Liu2026, Natarajan2026}, together with the internal kinematics and tidal structure of individual satellites \citep[e.g.,][]{Zavala2013, Schive2014, Calabrese2016, Chen2017, Marsh2019, Chiang2021, Chiang2026b, Correa2021, Dalal2022}, provides unique probes of dark matter microphysics (see also \citealp{Nadler2026} for a recent review). The fidelity of every such inference rests on accurately modelling the post-infall dynamical evolution of satellites.

After infall, satellites are transformed by the host environment. Tidal stripping unbinds dark matter and stars \citep[e.g.,][]{Mo2010}; ram-pressure stripping removes the cold gas \citep[e.g.,][]{Gunn1972, Boselli2022, Souchereau2025} and strangulation starves subsequent star formation \citep{Balogh2000}, together quenching the satellite \citep[e.g.,][]{Fillingham2019, Samuel2022, Geha2024}. The pace at which these processes unfold is fundamentally linked to the satellite's orbit about the host, and particularly to the minimum pericentric distance that sets the relevant tidal radius \citep[e.g.,][]{Binney1987, vandenBosch2018}.

Satellite orbits are almost universally computed in spherical host halo potentials, except in full cosmological simulations. This holds for the analytic and semi-analytic treatments used to characterise infall orbits \citep[e.g.,][]{Tormen1997, vandenBosch1999, Zentner2005, Benson2005, Khochfar2006, Wetzel2011, Jiang2015, vandenBosch2017}, for semi-analytic models of subhalo and satellite evolution such as \textsc{SatGen} \citep{Jiang2021, Green2022, Monzon2026} and \textsc{Galacticus} \citep{Benson2012, Du2024}, and for the backward integration of observed satellite galaxy orbits from \textit{Gaia} proper motions, in Galactic potentials with axisymmetric baryonic components and a spherical dark matter halo \citep[e.g.,][]{GaiaCollaboration2016, Fritz2018, Battaglia2022, Pace2022}. The assumption is one of convenience; in a spherical potential, energy and angular momentum fully determine the orbit, and the two classical turning points (i.e., pericentre and apocentre) are obtained by computationally trivial 1D root finding. In an aspherical host, however, the orbit additionally depends on the instantaneous position and velocity vectors relative to the host, and must be obtained by direct numerical 3D integration. 

Generically, dark matter haloes are aspherical. $\Lambda$-Cold Dark Matter cosmological simulations consistently find triaxial density profiles with typical minor-to-major axis ratios $c/a \sim 2/3$ and a broad distribution of shapes \citep[e.g.,][]{Franx1991, Bailin2005, Allgood2006, Bett2007, VeraCiro2011, Chua2019}. Halo triaxiality also persists in alternative dark matter models, as reported in warm \citep[e.g.,][]{Bose2016, Giocoli2026} and self-interacting \citep[e.g.,][]{Peter2013, Vargya2022, Giocoli2026} dark matter simulations. Observationally, the first direct characterisations of the Milky Way's halo shape are now emerging \citep[e.g.,][]{Woudenberg2024, Nibauer2025}. Halo shapes are, moreover, dynamic; the infalling LMC alone distorts and tilts the Galactic halo \citep[e.g.,][]{GaravitoCamargo2019, Dillamore2026}.

The discrepancy between spherically modelled orbits and their counterparts in a triaxial host, however, has never been systematically quantified simultaneously at the orbit-by-orbit and the population level for cosmologically infalling satellites. Pioneering studies in static triaxial hosts addressed specific sub-populations, including the tidal evolution of globular clusters within dwarf galaxies \citep{Penarrubia2009} and the stability of an idealised thin plane of satellite galaxies \citep{Bowden2013}. Comparisons with cosmological simulations, whether object by object \citep{Lux2010} or at the population level \citep{Smith2022}, remain inconclusive, given the difficulty of cleanly isolating the role of the host shape from triaxiality-independent effects (evolving potentials, dynamical friction, substructure; e.g., \citealp{Santistevan2023}); \citet{Smith2022}, for example, found satellite pericentres consistent with, or even slightly larger than, spherical expectations. A faithful comparison requires integrating identical satellite populations through host potentials that differ only in shape.

In this work, we construct nine static hosts of varying shapes at fixed virial mass, with matched spherically averaged mass profiles following the NFW density profile \citep{Navarro1997}. Through each host we integrate the same $10^6$ satellites, drawn from the universal infall distribution measured in cosmological simulations by \citet{Li2020}. As we will demonstrate, halo shape leaves the population-level orbital elements essentially untouched while dephasing individual orbits, scattering their successive pericentres and loosening the tight period--energy linkage of the spherical host. 

The paper is organised as follows. In \S\ref{sec:methods}, we describe the satellite infall population, the triaxial host potentials, and the orbit integration. \S\ref{sec:radii} presents the impact of halo shape on the distributions of pericentres and apocentres, on the passage-to-passage scatter of pericentres along individual orbits, and on the radial-period distribution. \S\ref{sec:chaos} quantifies the fraction of chaotic satellite orbits in triaxial hosts and their Lyapunov time-scales, together with the position error incurred by spherical orbit reconstruction. We summarise our findings in \S\ref{sec:conclusions}, together with their implications for the numerical convergence of simulated subhaloes, tidal stripping in semi-analytic models, and the backward integration of observed satellite orbits. Appendix~\ref{app:lyapunov} details the drift-corrected Lyapunov estimator employed in \S\ref{sec:chaos}. Throughout this paper, we follow the setup of \citet{Chiang2025} and adopt $H_0 = 70\kmsmpc$, giving a Hubble time of $\tH \equiv 1/H_0 = 13.97\Gyr$.

%%%%%%%%%%%%%%%%%%%%%%%%%%%%%%%%%%%%%%%%%%
\section{Methodology}
\label{sec:methods}

%%%%%%%%%%%%%%%%%%%%%
\subsection{Host halo potentials}
\label{ssec:haloes}

We explicitly integrate individual subhalo orbits in a set of static background potentials (i.e., the `host' haloes) that correspond to the NFW density profile generalised to ellipsoidal isodensity surfaces,
\begin{equation}
\rho(\Rell) = \frac{\rho_0}{(\Rell/R_{\rm s})\left(1 + \Rell/R_{\rm s}\right)^2}\,, \qquad
\Rell \eee \sqrt{x^2 + \frac{y^2}{q^2} + \frac{z^2}{s^2}}\,,
\label{eq:nfw}
\end{equation}
where $q = b/a$ and $s = c/a$ are the intermediate and minor axis ratios, with $a \ge b \ge c$ the major, intermediate, and minor axes of the triaxial system. The corresponding triaxiality is defined as \citep[e.g.,][]{Franx1991}
\begin{align}\label{eqn:Triaxiality}
	T \eee \frac{1-(b/a)^2}{1-(c/a)^2},
\end{align}
with $0 < T < 1/3$ corresponding to oblate-like, $1/3 < T < 2/3$ to triaxial, and $2/3 < T < 1$ to prolate-like morphologies. 

Despite the scale-free nature of purely gravitational dynamics, the kinematics of infalling satellites carry a weak dependence on host and subhalo mass, due to the competition between the host's self-gravity and the tidal field of large-scale structure \citep{Li2020}. As a fiducial choice, we consider a host-to-subhalo mass ratio of $1000:1$, as in \citet{Chiang2026a}, for which dynamical friction and self-friction are negligible over a Hubble time \citep[e.g.,][]{Mo2010, Miller2020}. We fix the host virial mass to $\Mvir = 10^{12}\Msun$ and the concentration to $c_{\rm vir} = \Rvir/R_{\rm s} = 10$, and adopt the virial convention $\Delta_{\rm vir} = 97$ \citep{Bryan1998}, giving $\Rvir \simeq 262\kpc$.

\begin{figure}
    \includegraphics[width=0.98\linewidth]{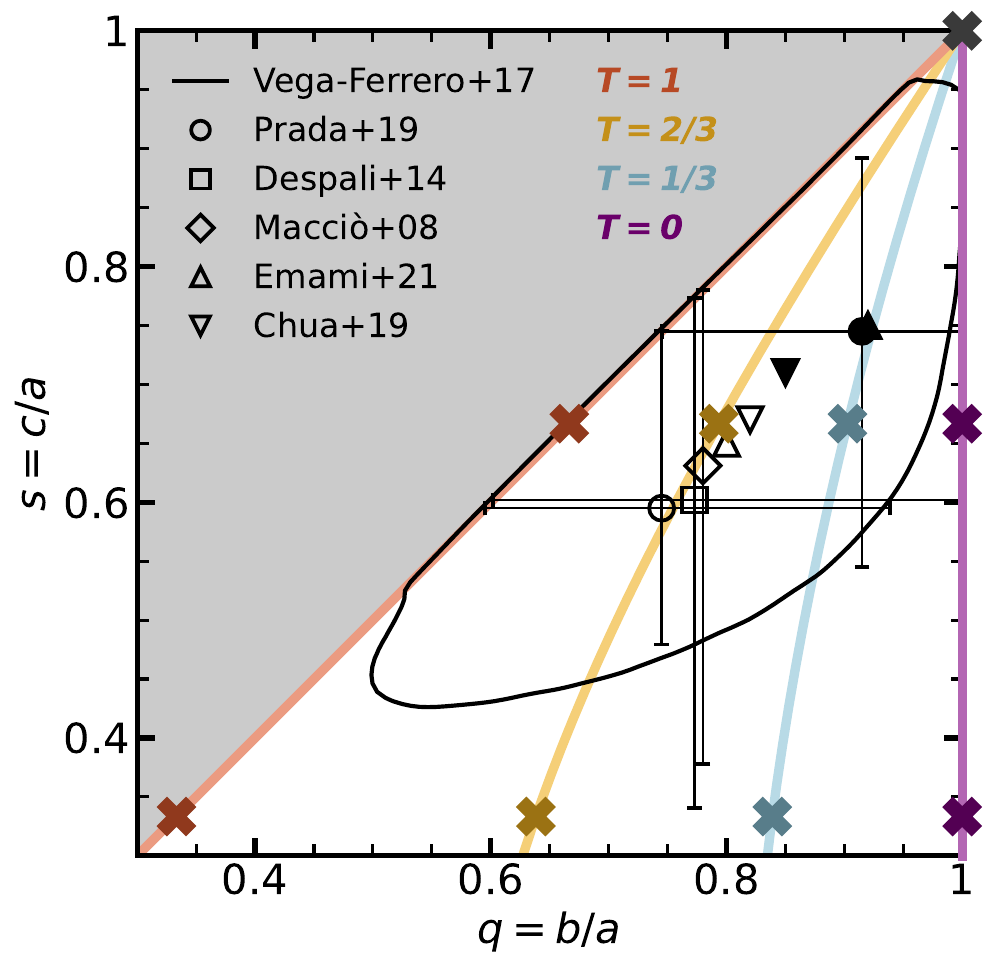}
    \caption{Spherical (grey cross) and triaxial (coloured crosses) host halo shapes considered in this work, specified by the intermediate $q$ and minor $s$ axis ratios of the density. Thick curves of constant triaxiality $T$ (colour-coded as indicated) demarcate the oblate-like ($0 < T < 1/3$), triaxial ($1/3 < T < 2/3$), and prolate-like ($2/3 < T < 1$) regimes. The unphysical regime ($s > q$) is grey-shaded. We overlay previous literature halo shape measurements for $\Mvir = 10^{12}\Msun$ hosts at $z = 0$ from dark-matter-only (open black symbols) and hydrodynamical (filled symbols) cosmological simulations, with their associated $95\%$ halo-to-halo scatter shown as error bars where available. The solid black contour shows the same $95\%$ scatter for \citet{VegaFerrero2017} in place of a median symbol; see text for details.}
    \label{fig:shapes}
\end{figure}

\fref{fig:shapes} shows the nine host halo shape parameters explored in this work, compared against halo shapes measured in cosmological simulations at the same halo mass, drawn from \citet{VegaFerrero2017} (MultiDark; \citealp{Klypin2016}), \citet{Despali2014} (GIF2, Baby, and Flora; \citealp{Gao2004}), \citet{Maccio2008} (WMAP1/3/5 $N$-body suite), \citet{Prada2019} (Auriga; \citealp{Grand2017}), \citet{Chua2019} (Illustris; \citealp{Vogelsberger2014}), and \citet{Emami2021} (IllustrisTNG50; \citealp{Pillepich2019})\footnote{We adopt the full halo sample where studies report relaxed and unrelaxed populations separately; \citet{VegaFerrero2017} is published for relaxed haloes only. Scatter quoted as $16$th--$84$th percentile or interquartile ranges is rescaled to the central $95\%$ assuming Gaussian scatter, and truncated at the physical boundaries $q = 1$ and $s = q$.}. Where a hydrodynamical counterpart to the dark-matter-only run is also available, we plot both measurements. All of these works characterise the halo out to the virial radius, except for \citet{Emami2021}, whose shape profiles terminate at $150\kpc$. These studies, albeit differing slightly in measurement conventions, quote single whole-halo shape values, directly comparable to our radius-independent parameterisation $(T, s)$. Empirically, the typical Milky Way-mass halo sits close to our $T = 2/3$ curve at a minor axis ratio slightly below $s = 2/3$. Recent measurements by \citet{Chemaly2026}, the first population-level constraints on halo flattening from extragalactic streams, yield a mean flattening of $\sim 0.72$; as an effective flattening of the total potential, this implies an even flatter density shape\footnote{\label{fn:shapeconv}Observational inferences from streams and other dynamical tracers constrain only the acceleration field, i.e., the shape of the \textit{potential}, whereas cosmological simulations, and the $(T, s)$ parameters of this work, specify the shape of the \textit{density}. Individual observational studies report either the axis ratios of a fitted density model \citep[e.g.,][]{Vasiliev2021, Woudenberg2024}, an effective flattening of the potential \citep[e.g.,][]{Chemaly2026}, or both \citep{Nibauer2025}. The potential is an integrated quantity and substantially rounder than the underlying density distribution; in our most flattened hosts with density axis ratio $s_\rho = 1/3$, we measure equipotential axis ratios of $s_\Phi = 0.59$--$0.80$ within $\Rvir$. A reported potential flattening can therefore imply a substantially more aspherical density.}, plausibly between our two flattenings. The $s = 1/3$ hosts lie beyond the $95\%$ contour and serve as a conservative lower limit on $s$.

To cleanly quantify the effect of host halo triaxiality at a fixed mass budget, we enforce shell-averaged mass profiles identical to the spherical benchmark. Specifically, we first apply the substitution
\begin{equation}
\Rell \;\to\; (qs)^{1/3}\,\Rell
\label{eq:rescale}
\end{equation}
in \eref{eq:nfw}, equivalent to a rescaling of the scale radius, $R_{\rm s} \to (qs)^{-1/3}\,R_{\rm s}$, at fixed axis ratios. Each isodensity surface of \eref{eq:nfw}, at constant $\Rell$, is an ellipsoid with semi-axes $(\Rell,\, q\Rell,\, s\Rell)$ and hence encloses the volume $\tfrac{4}{3}\pi\,qs\,\Rell^3$ of a sphere of radius $(qs)^{1/3}\,\Rell$; the volume-preserving rescaling of \eref{eq:rescale} therefore expands every isodensity surface to enclose the same volume as its equal-density counterpart in the spherical host, absorbing the leading-order change of the enclosed-mass profile by construction. Second, we apply a constant renormalisation $\rho_0 \to C \rho_0$, where $C$ is determined by numerically solving the required condition $M(<\Rvir) = \Mvir$. Empirically, the halo shapes explored in this work yield $C = 1.01$--$1.09$; the resulting spherically averaged mass profiles of the triaxial hosts typically agree with the spherical benchmark at the $0.3$--$2\%$ level for $r > 0.02\,\Rvir$.

%%%%%%%%%%%%%%%%%%%%%
\subsection{Initial conditions of subhalo infall orbits}
\label{ssec:population}

The $10^6$ satellite orbits are initialised at first infall, i.e., as they cross the virial radius $\Rvir$ of the host. Infall positions are drawn uniformly over the virial sphere; the orientation of the tangential velocity within the tangent plane is also drawn uniformly. This identical set of $10^6$ phase-space initial conditions is integrated in every host, so that any difference between the resulting orbit populations is caused by halo shape alone; the shared initial conditions also enable the orbit-by-orbit pairing in \S\ref{ssec:divergence}. In sampling positions uniformly and independently of velocities, we neglect anisotropic accretion and assume the infall kinematics to be uncorrelated with the launch position on the virial sphere, given the invariance in the spherical benchmark case; additionally, we are not aware of a published quantification of such correlations for satellite infall relative to the host principal axes at virial crossing\footnote{Existing studies of anisotropic satellite accretion either measure only infall directions relative to the large-scale cosmic web \citep{Libeskind2011, Libeskind2014}, or the present-day alignment of satellite positions relative to the host major axis \citep{Libeskind2005, Zentner2005b, Mezini2025}. However, one cannot cleanly attribute this latter alignment effect to either anisotropic accretion or the orbital evolution of initially isotropically accreted satellites in a triaxial host \citep[e.g.,][]{Zentner2005b}.}. Velocities are sampled from the distribution quantified by \citet{Li2020}, which gives the joint probability of the normalised infall velocity $u \eee v/\Vvir$ and the infall angle $\cos^2\theta = v_\rmr^2/v^2$ measured for satellite haloes in cosmological simulations, conditioned on the host peak height $\nu_{\rm peak} \eee \delta_{\rm c}/\sigma(\Mvir)$, where $\delta_{\rm c} \simeq 1.686$ is the critical linear overdensity for spherical collapse and $\sigma(\Mvir)$ the rms linear density fluctuation on the mass scale $\Mvir$ ($\nu_{\rm peak} \simeq 0.80$ at $z = 0$) and the sub-to-host mass ratio (\S\ref{ssec:haloes})\footnote{Infall orbits have previously been characterised by the circularity distribution \citep{Zentner2005, Wetzel2011}, a derived quantity presuming a spherical NFW or even point-mass host potential. We instead adopt \citet{Li2020}, who measure the infall velocity vector directly at virial crossing over a far larger halo sample. Our hosts share a common $\Mvir$ and spherically averaged mass profile, so one infall population applies to all shapes by construction. Here, $\nu_{\rm peak}$ is evaluated under the cosmology adopted in this work; the calibration cosmology of \citet{Li2020} instead gives $\nu_{\rm peak} \simeq 0.74$, a difference to which the sampled distribution is insensitive (the mean of $\cos^2\theta$ shifts by $< 0.005$).}. Here $\Vvir \eee \sqrt{G\Mvir/\Rvir}$ denotes the circular velocity at $\Rvir$, where $G$ is the gravitational constant, and $v_\rmr$ the radial component of the satellite orbital velocity with respect to the host centre.

\begin{figure}
    \includegraphics[width=0.98\linewidth]{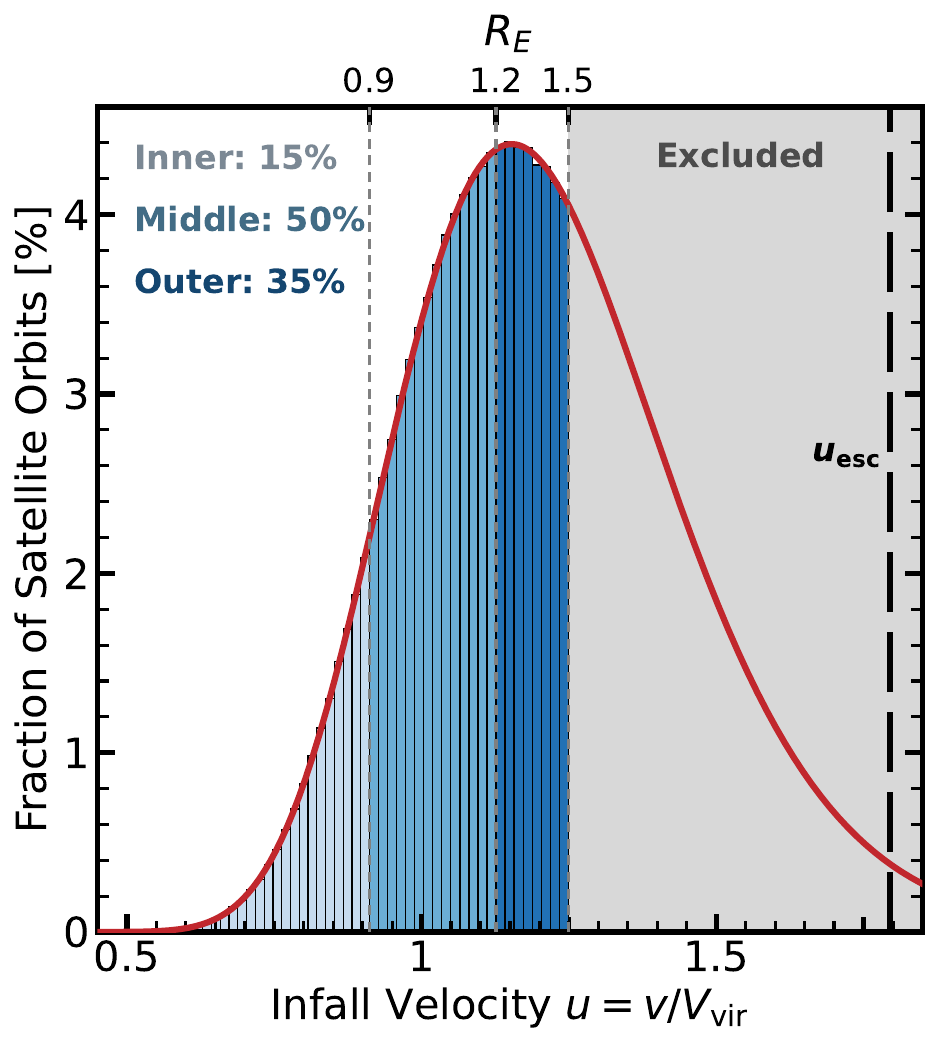}
    \caption{Histogram of the normalised infall velocity $u$ of the $10^6$ sampled satellite orbits, colour-coded by energy bin as annotated, overlaid with the \citet{Li2020} model distribution (red curve). The top axis gives the dimensionless orbital radius $\RE$, onto which $u$ maps one-to-one; grey dashed lines mark the bin boundaries at $\RE = 0.9$, $1.2$, and $1.5$, and the annotated fractions are those of the retained sample. The grey band marks the excluded regime $\RE > 1.5$, comprising $42\%$ of the raw infall population; see text for details. The model curve is drawn across this regime to show the distribution of the excluded orbits, and the black long-dashed line marks the normalised escape velocity, $u_{\rm esc} \eee v_{\rm esc}/\Vvir = 1.79$.}
    \label{fig:infall}
\end{figure}

\begin{figure*}
    \includegraphics[width=\linewidth]{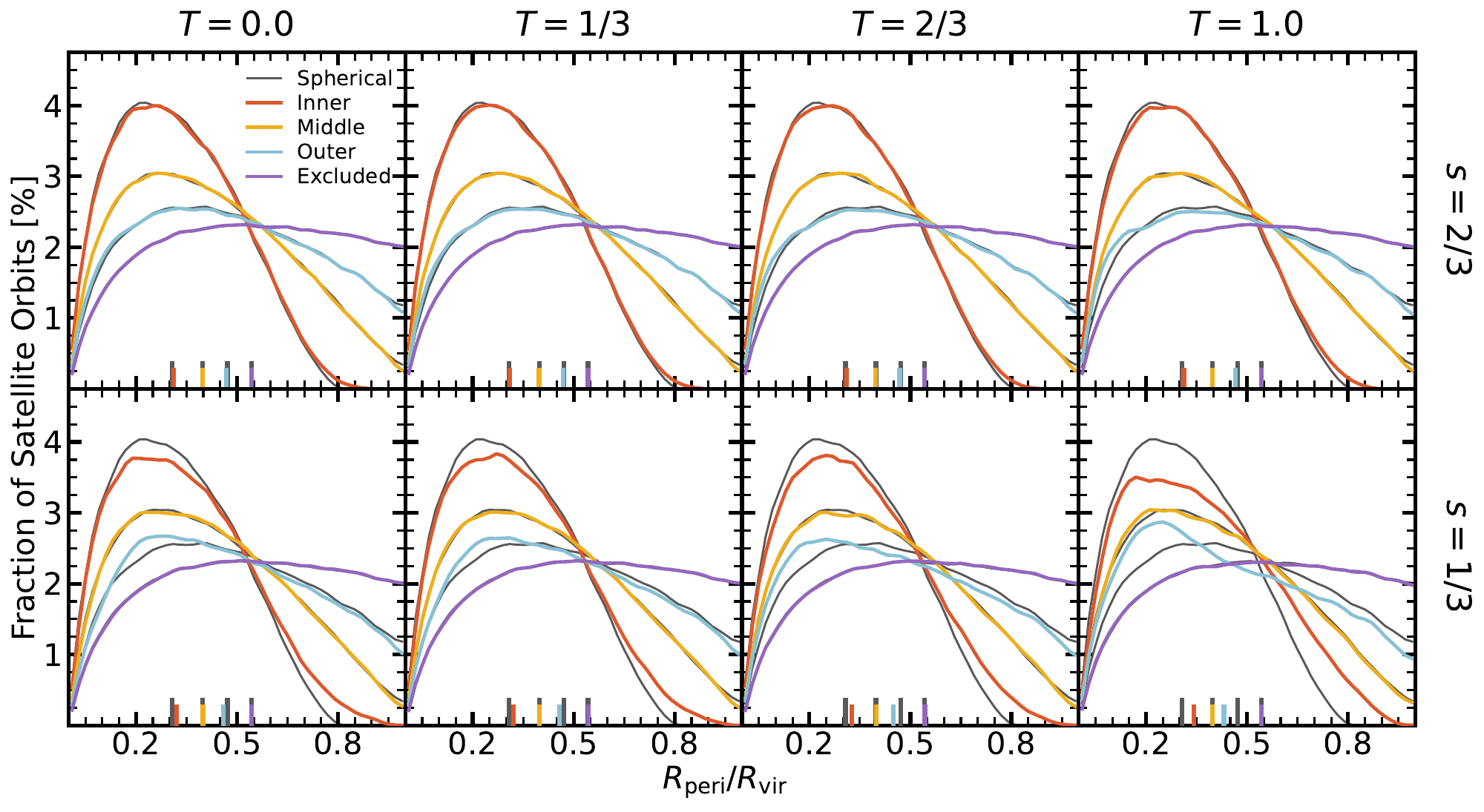}
    \caption{Distributions of the last satellite pericentre completed before $\tH$, for the eight triaxial hosts, arranged by triaxiality $T$ (columns) and sphericity $s = c/a$ (rows). In each panel, the red, yellow, and cyan curves show the inner, middle, and outer orbital-energy bins (\fref{fig:infall}) evolved in that host, and the grey curves show the benchmark spherical counterparts, identical in every panel. The purple curves show the excluded high-$u$ tail ($\RE > 1.5$; \fref{fig:infall}), which is essentially shape-insensitive and identical to its spherical counterpart. Each curve is normalised to its own orbit population, and the short vertical ticks along the bottom edge mark the corresponding medians (grey: spherical; coloured: triaxial).}
    \label{fig:peridist}
\end{figure*}

\fref{fig:infall} shows the sampled infall-velocity distribution together with the \citet{Li2020} model. To aid the interpretation of our results, each satellite is labelled by the dimensionless orbital radius $\RE \eee \Rcir(E)/\Rvir$, where $\Rcir(E)$ denotes the radius of a circular orbit of energy $E$ in the benchmark spherical host potential; for satellites initialised at $r = \Rvir$, $\RE$ is monotonically mapped to the infall speed $u$ alone. First, orbits with $\RE > 1.5$, which account for $42\%$ of the raw infall population, including the $2.2\%$ that is formally unbound, are excluded from the sample (grey-shaded in \fref{fig:infall}). These have apocentres $\ge 1.5\,\Rvir$, comparable to the splashback radius, which for haloes of this mass spans $1.0$--$1.9\,\Rvir$ depending on their mass accretion rate \citep[e.g.,][]{DiemerKravtsov2014, Adhikari2014, More2015, Diemer2017}, and orbital periods exceeding $\tH$. As we seek to quantify the effect of halo shapes on subhalo orbital evolution, this excluded population, with at most one pericentric passage over $\tH$, is still on first infall rather than repeatedly orbiting within the host. It is moreover insensitive to the host shape altogether; integration through all nine hosts yields essentially identical pericentre distributions (purple curves in \fref{fig:peridist}), with medians shifting by $< 0.2\%$. For these reasons, we focus on the remaining populations detailed below throughout this work.

Next, the remaining sampled population is partitioned into three energy bins, $\RE \le 0.9$ (`inner', $15\%$ of the retained sample), $0.9$--$1.2$ (`middle', $50\%$), and $1.2$--$1.5$ (`outer', $35\%$), marked in \fref{fig:infall}. Each satellite's bin membership is assigned once through the benchmark mapping of $\RE$ and carried over to its counterparts in the triaxial hosts. The inner bin spans the present-day orbital energies of much of the well-observed Milky Way dwarf satellite population (see Fig.~14 of \citealp{Chiang2025}). The inner satellites are also the most susceptible to resolution-limited numerical artefacts in cosmological simulations, owing to their small pericentres \citep{vandenBosch2018b, Martin2024, Chiang2026a}.

%

%%%%%%%%%%%%%%%%%%%%%
\subsection{Orbit integration}
\label{ssec:integration}

All satellites are treated as massless test particles evolving in the static host potentials, and they exert no back reaction on the host. Orbits are integrated with the triaxial NFW implementation in \textsc{galpy} \citep{Bovy2015}, using an adaptive, error-controlled eighth-order Dormand--Prince scheme, and sampled at $1\Myr$ resolution. Each satellite is followed for $50\Gyr$; all diagnostics quoted in this work are evaluated within $\tH$, while the extended baseline is used only where at least three complete radial cycles per orbit are required. Pericentres and apocentres are extracted as sign changes of the radial velocity along the trajectory\footnote{In the spherical benchmark host, the first pericentres agree with the analytic turning points \citep[e.g.,][]{Binney1987} to a median of $8 \times 10^{-5}\,\Rvir$. We have additionally verified that integrations with a fourth-order symplectic integrator agree orbit by orbit to better than $10^{-7}\,\Rvir$.}. The finite $1\Myr$ output sampling contributes a pericentre error of $\sim\!2\times10^{-4}\kpc$, negligible compared to every effect discussed below. Unless specified otherwise, we characterise each orbit by its \textit{last} turning points, i.e., the final pericentre and apocentre completed before $\tH$.

To isolate the effect of the host shape, our set-up adopts the following simplifications: (i) the hosts are static and smooth, neither growing in mass nor evolving in shape, and contain no substructure, so that we also do not account for subhalo--subhalo interactions \citep[e.g.,][]{Penarrubia2025}; (ii) the main runs are dark matter only; baryons are included only in the Milky Way case study of \S\ref{ssec:divergence}; (iii) satellites are massless tracer particles, without dynamical friction (negligible for our $1000:1$ mass ratio; \S\ref{ssec:haloes}), tidal mass loss, and the back reaction on the host; (iv) satellites are initialised at first infall, with launch positions drawn isotropically and independently of their velocities (\S\ref{ssec:population}); and (v) massive satellites such as the LMC are not modelled (see \S\ref{ssec:divergence}). Time-evolving potentials extracted from cosmological simulations include several of these effects jointly \citep[e.g.,][]{Arora2026}, but do not by themselves isolate the role of the host shape. We emphasise that the present set-up is not intended to reproduce a best-fitting Milky Way model; our aim is to quantify statistically the host-shape-driven effects on infalling satellite orbits in generic aspherical hosts, and to separately compare them in a case study against the increasingly well-characterised orbital kinematics of the observed Milky Way satellites (\S\ref{ssec:divergence}).

%%%%%%%%%%%%%%%%%%%%%%%%%%%%%%%%%%%%%%%%%%
\section{Triaxiality-driven orbital dephasing}
\label{sec:radii}

Satellites at infall carry a natal kinematic distribution in orbital energy $E$ and angular momentum $\mathbf{L}$, seeded by the cosmological accretion of large-scale structure (\S\ref{ssec:population}); how this distribution subsequently evolves depends sensitively on the exact shape of the host halo. In a (static) spherical host, $E$ and $\mathbf{L}$ fully determine the orbit \citep[e.g.,][]{Binney2008}. The conserved $\mathbf{L}/|\mathbf{L}|$ fixes the orbital plane, while $E$ and $|\mathbf{L}|$ together pin the two turning points (the pericentre $\rperi$ and apocentre $\rapo$) and the radial orbital period $\Trad$. These isolating integrals render the natal orbital properties time-invariant; orbits of common $E$ and $|\mathbf{L}|$ but different orientations are physically identical copies of one another, sharing the same constant $\rperi$, $\rapo$, and $\Trad$.

In a generic triaxial host, by contrast, the exact launch location on the virial sphere becomes a dynamical parameter. Orbits entering with identical $E$ and $|\mathbf{L}|$ but at different locations traverse physically distinct orbits, and may even belong to different orbit families \citep{deZeeuw1985, Statler1987}. Tube orbits loop around the major or minor principal axis in a fixed direction and stay away from the centre; box orbits instead oscillate back and forth along all three principal axes and can pass arbitrarily close to the centre. In triaxial potentials with a central density cusp, many box orbits are replaced by resonant orbits (`boxlets'), for which the fundamental orbital frequencies are commensurable, and by irregular chaotic orbits that have energy as the only isolating integral \citep[e.g.,][]{MiraldaEscude1989, Merritt1999, MerrittValluri1999}. This carries two consequences operating on two distinct time-scales. First, $\Trad$ is now orientation-dependent. The period is a time integral over the radial cycle, so orbits plunging along different principal axes acquire different periods on their very first cycle, broadening the period distribution at fixed energy (\S\ref{sec:periods}; \fref{fig:periods}). In contrast, the $\Trad$ median remains largely unchanged, as the orientation-induced offsets average out across the population to leading order. Second, $|\mathbf{L}|$ becomes time-varying. Each orbit fills a volume of the host allowed by its integrals of motion \citep[cf.][]{deZeeuw1985}, and the aspherical mass distribution torques the orbit passage by passage, so successive closest approaches sample a range of pericentric radii within that volume, rather than repeating a single value. A pericentre \textit{distribution} per orbit is therefore the natural outcome in any aspherical host. Some orbits in a triaxial potential moreover have no third integral and are chaotic, as quantified in detail in \S\ref{sec:chaos}.

We adopt the following terminology throughout this work. Two orbits are said to \textit{dephase} when they enter the host with identical orbital elements $(E, |\mathbf{L}|)$ but at different locations on the virial sphere, and subsequently differ in their instantaneous orbital properties: turning radii, radial periods, and positions along the radial cycle. Such differences are absent in a spherical host by symmetry. Dephasing is thus a divergence between orbits launched from different phase-space coordinates, conceptually distinct from chaos, which instead denotes the exponential divergence of initially infinitesimally close orbits in the same host (\S\ref{sec:chaos}). Triaxiality-driven dephasing is the central process quantified in this section; we measure its imprints on the last pericentres (\S\ref{ssec:phase_mixing}), the minimum pericentre (\S\ref{ssec:minperi}), and the radial periods (\S\ref{sec:periods}).

%%%%%%%%%%%%%%%%%%%%%
\subsection{Halo-shape-driven pericentre scatter}
\label{ssec:phase_mixing}

\fref{fig:peridist} shows the distributions of the last pericentre completed within $\tH$ for the eight triaxial hosts with $T = 0, 1/3, 2/3, 1$ (left to right columns) and $s = 2/3, 1/3$ (top and bottom rows, respectively). In each panel, we compare the inner (red), middle (yellow), and outer (cyan) orbit distributions against their reference spherical counterparts, together with the excluded high-$u$ population (purple; see \S\ref{ssec:population}). The corresponding medians are marked by the ticks along the bottom edge of each panel. 

By and large, the pericentre distributions are only weakly affected by the host halo shapes. Physically, the orbital energy is always conserved in a static host; the aspherical host shape can thus only redistribute the angular momentum of individual orbits, scattering individual pericentres about their spherical values rather than driving a coherent net shift. For example, in the $T = 2/3$ and $s = 2/3$ host, the last pericentre of the inner-bin orbits lies within $0.79$--$1.29$ times its spherical value ($16$--$84$th percentiles) around a nearly unbiased median of $1.02$. Centrophilic orbits, i.e., those passing arbitrarily close to the centre, play little role here, as box-like orbits make up $\lesssim\!9\%$ of the orbits in the $s = 2/3$ hosts and are completely absent in the axisymmetric host haloes (see \S\ref{ssec:census} and footnote~\ref{fn:boxlike}). The medians shift by at most $1.7\%$ relative to the spherical case for the cosmologically typical $s = 2/3$ hosts across all three orbital populations. Even for the more flattened $s = 1/3$ hosts, the shifts remain below $6\%$ except in the inner energy bin of the most triaxial hosts, where the median pericentre rises by up to $11\%$. 

At a population level, the pericentre distribution widths similarly appear insensitive to halo shapes. Specifically, the $16$--$84$ distribution half-width negligibly increases from $0.187\,\Rvir$ in the spherical host to $0.189$--$0.192\,\Rvir$ across the $s = 2/3$ hosts ($+1$--$3\%$), reaching $0.222\,\Rvir$ ($+19\%$) only at $T = 1$ and $s = 1/3$. The last apocentres (not shown) are also essentially unaffected, with medians of $1.286$--$1.290\,\Rvir$ across all nine hosts. Halo triaxiality therefore does not alter the ensemble-averaged satellite orbits. At fixed virial mass and fixed radial mass profile, the satellite orbit distribution of a spherical model is \textit{unbiased}, consistent with \citet{Smith2022}, who showed that satellite pericentres measured in the triaxial hosts of cosmological simulations are compatible with spherical expectations.

However, the detailed shapes of the pericentre distributions are systematically deformed in the flattened $s = 1/3$ row of \fref{fig:peridist}. In particular, the last-pericentre distributions of the inner and outer populations both converge towards that of the middle population, more strongly with increasing $T$, and only marginally in the $s = 2/3$ row. This apparent convergence is a blurring effect. Each spherical distribution is effectively convolved with the shape-driven per-orbit pericentre scatter. Because the pericentre depends much more strongly on $|\mathbf{L}|$ than on $E$, the $|\mathbf{L}|$-driven spread within a single bin ($16$--$84$ range $0.14$--$0.51\,\Rvir$ for the inner orbits) dwarfs the energy-driven offsets between the bin medians ($0.31$, $0.40$, $0.47\,\Rvir$), washing out the bin-to-bin distinction. Indeed, the first-passage pericentre distributions are indistinguishable from their spherical counterparts, with medians agreeing to better than $0.5\%$ in every host and energy bin; the blurring accumulates only over subsequent passages.

In contrast, host triaxiality does markedly alter the orbit-by-orbit behaviour. The measured passage-to-passage scatter of both $\rperi$ and $\rapo$ in the spherical host is, as expected, zero to our numerical precision, whereas in a triaxial host successive turning points of the same orbit can differ. We next quantify this with the per-orbit relative scatter $\sigma_{\rm peri}/\langle\rperi\rangle$, the standard deviation over each orbit's first three pericentric passages normalised by their mean, and take the median over the population.

\begin{figure}
    \includegraphics[width=\linewidth]{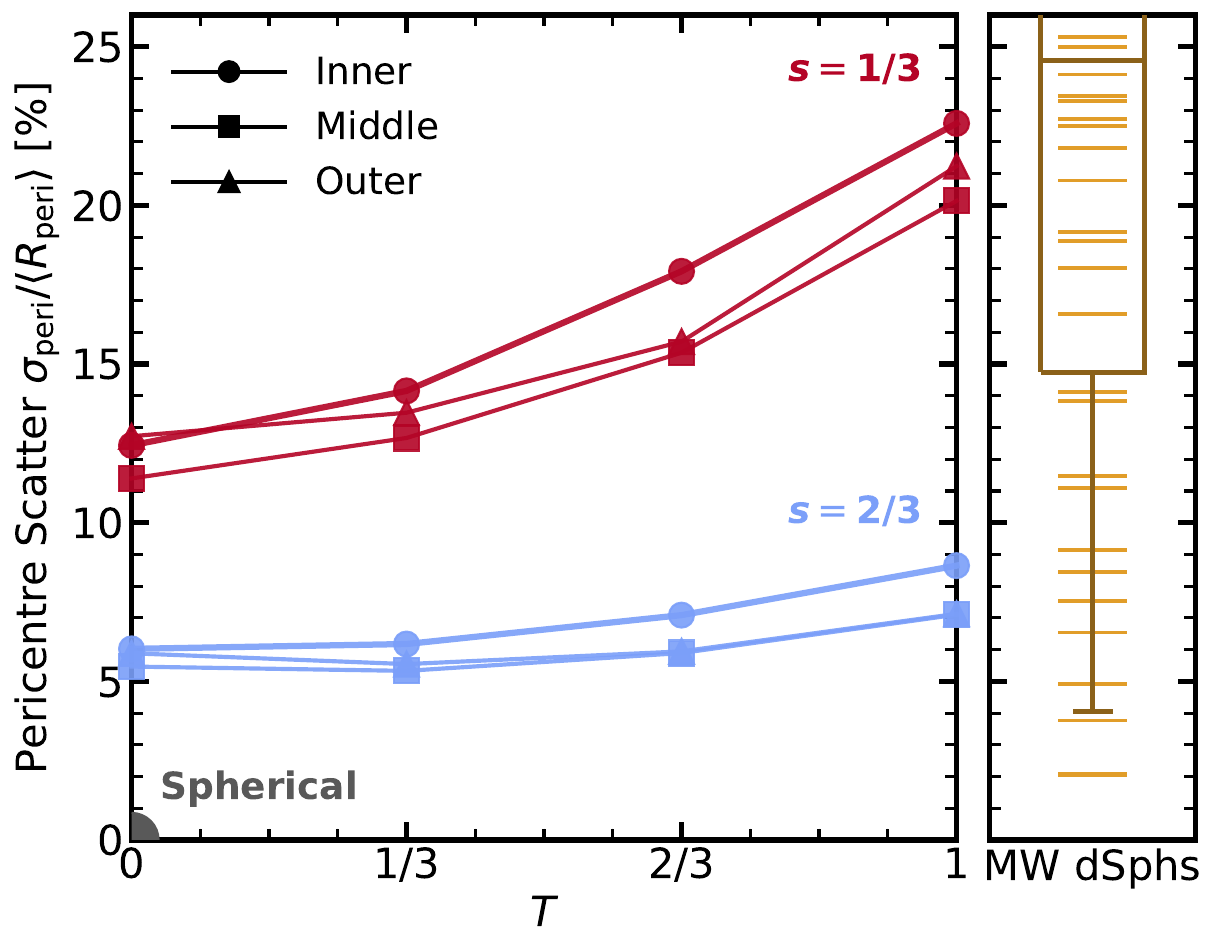}
    \caption{\textit{Left:} Median per-orbit pericentre scatter $\sigma_{\rm peri}/\langle\rperi\rangle$, computed over each orbit's first three pericentric passages, as a function of host triaxiality $T$ for the $s = 1/3$ (red) and $s = 2/3$ (blue) hosts. Symbols distinguish the three orbital-energy bins. The grey point is the measured spherical value, zero to better than $10^{-4}$, and nominally placed at $T = 0$. \textit{Right:} Fractional symmetrised $1\sigma$ inference uncertainties in the \textit{most recent} pericentre distances (horizontal dashes) of the $46$ Milky Way dwarf spheroidals reported by \citet{Pace2022}, from their fiducial with-LMC models; $21$ dwarfs with uncertainties beyond the plotted range are not shown. The box marks the quartiles of the full $46$-dwarf distribution, the horizontal line its median, and the whiskers its $5$th and $95$th percentiles.}
    \label{fig:cv}
\end{figure}

\fref{fig:cv} shows this scatter as a function of host triaxiality $T$, for the two flattenings $s = 1/3$ (red) and $s = 2/3$ (blue). Overall, the pericentre-by-pericentre variation of an orbit increases monotonically with triaxiality $T$ and, much more strongly, with flattening $s$. This scatter is $5.3$--$8.6\%$ for the typical $s = 2/3$ shapes and $11.4$--$22.6\%$ in the more flattened $s = 1/3$ hosts. Furthermore, the response across different orbital energy bins is remarkably consistent in all cases. \textit{Fractional pericentre scatter is thus a function of host halo shape alone}; the host shape $(T, s)$ suffices to robustly assign a per-passage uncertainty to any spherically computed pericentre. Note that this per-orbit scatter is nearly invisible in \fref{fig:peridist}, because the population's intrinsic $(E, \mathbf{L})$ spread is far broader and the two independent widths combine in quadrature. Quantitatively, this quadrature suppression predicts that the per-orbit scatter increases the distribution widths in \fref{fig:peridist} by only $\simeq\!+1\%$ for the $s = 2/3$ hosts relative to the spherical benchmark, which is consistent with the measured $+1$--$3\%$. Population statistics are thus intrinsically blind to this per-passage scatter, which also explains why population-level comparisons in cosmological simulations \citep[e.g.,][]{Smith2022} could not have reliably detected the effect.

\begin{figure*}
    \includegraphics[width=\linewidth]{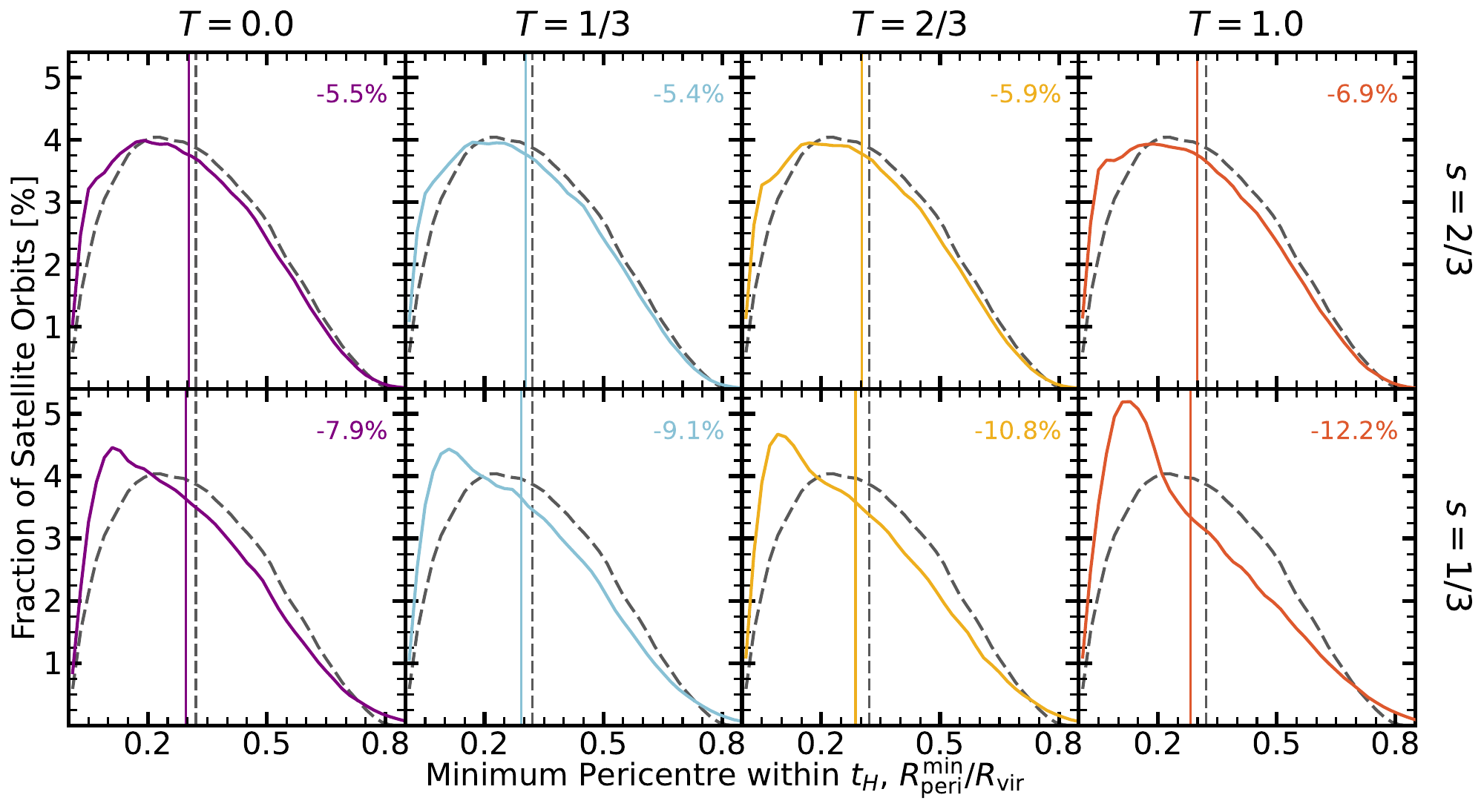}
    \caption{Distribution of the minimum pericentre reached within a Hubble time, $R^{\rm min}_{\rm peri}/\Rvir$, for the inner orbital-energy bin (\fref{fig:infall}), shown for the eight triaxial hosts (solid coloured curves, colour-coded by triaxiality; panels ordered as in \fref{fig:peridist}) against the spherical control (grey dashed, identical in every panel). Vertical lines mark the means of the two distributions, and each panel is annotated with the relative change of the mean.}
    \label{fig:minperi}
\end{figure*}

The right-hand panel of \fref{fig:cv} places these numbers in observational context by compiling the fractional inference uncertainties in the distances of the \textit{most recent} pericentre passages for the $46$ Milky Way dwarf spheroidals with \textit{Gaia}-EDR3-based orbit models from \citet{Pace2022}, whose error budget explicitly samples the observational uncertainties, the LMC mass, and the posterior of the \citet{McMillan2017} Galactic potential. Crucially, the Galactic dark matter halo is assumed spherical throughout, so the quoted uncertainties omit the host-shape-induced scatter altogether, effectively placing every dwarf at the zero-scatter spherical point of \fref{fig:cv}. At the current precision, $7$ of the $46$ dwarfs ($\simeq\!15\%$) already have $\rperi$ characterised more tightly than the expected physical scatter of the cosmologically typical $s = 2/3$ hosts; for flattened hosts with $s = 1/3$, the fraction of dwarfs whose quoted uncertainties fall below the intrinsic shape-driven scatter, and are hence overly optimistic, rises to $\simeq\!20$--$41\%$, depending on $T$. Every term of this error budget is, moreover, actively shrinking; proper-motion uncertainties, unchanged between EDR3 and DR3 \citep{GaiaDR3}, will fall by factors of a few with the forthcoming DR4 and DR5 \citep{McKinnon2026}, and the mass and trajectory of the LMC are increasingly better constrained by stellar streams and other kinematic probes \citep[e.g.,][]{Erkal2019, Petersen2021, Vasiliev2021, JimenezArranz2023}. As such, ever more Milky Way satellites will have their inferred $\rperi$ uncertainties drop below this currently neglected scatter imprinted by the unmapped halo triaxiality (however, see \S\ref{ssec:divergence} for discussion on the Milky Way--LMC uncertainty coupling). Accurate shape characterisation of the Milky Way's dark matter halo is essential for reconstructing the most recent satellite pericentres to better than the $5$--$23\%$ level, as demonstrated further in \S\ref{ssec:divergence}.

%%%%%%%%%%%%%%%%%%%%%
\subsection{Minimum pericentre}
\label{ssec:minperi}

The same host-shape-driven scatter additionally lowers each orbit's \textit{minimum} pericentric distance $R^{\rm min}_{\rm peri}$ over time. As a single orbit can now traverse the volume allowed by its integrals of motion and progressively sample the full range of pericentric radii permitted in a triaxial host, $R^{\rm min}_{\rm peri}$ necessarily decreases with time towards its asymptotic lower bound (which can approach zero for centrophilic orbits) and falls below the single $\rperi$ value in the spherical counterpart. Importantly, this minimum sets the smallest tidal radius experienced, a key quantity for the tidal mass-loss rate \citep[e.g.,][]{Penarrubia2008, Jiang2016, Jiang2021, Stucker2023, Errani2024} and the numerical force-convergence requirement \citep{vandenBosch2018b, Chiang2026a}.

\fref{fig:minperi} shows the distribution of the minimum pericentre attained within $\tH$ for the eight triaxial hosts with $T = 0, 1/3, 2/3, 1$ (left to right columns) and $s = 2/3, 1/3$ (top and bottom rows, respectively), against the spherical benchmark (grey dashed). We show the inner orbital-energy bin; the same trend holds for the middle and outer bins at reduced amplitude. In each panel, we mark the distribution means (vertical lines) and annotate the deviation of the mean from the spherical case. Overall, the $R^{\rm min}_{\rm peri}$ distributions shift systematically towards smaller radii with increasing triaxiality and flattening; the mean minimum pericentre decreases by $5$--$7\%$ across the $s = 2/3$ hosts and by $8$--$12\%$ for $s = 1/3$, with the median reduction reaching $21\%$ in the most aspherical host. This systematic reduction follows directly from the per-orbit pericentre scatter of \fref{fig:cv}. The amplitude of this reduction is set by the number of passages completed within $\tH$. Inner orbits complete two to three pericentric passages while outer orbits complete barely one (\S\ref{sec:periods}), so the reduction diminishes from the inner to the outer populations, despite the per-passage scatter itself being essentially energy-independent (\fref{fig:cv}). Indeed, we have explicitly verified that, at a fixed number of pericentric passages, the reduction of $R^{\rm min}_{\rm peri}$ is comparable across the three energy bins in each host.

\begin{figure*}
    \includegraphics[width=\linewidth]{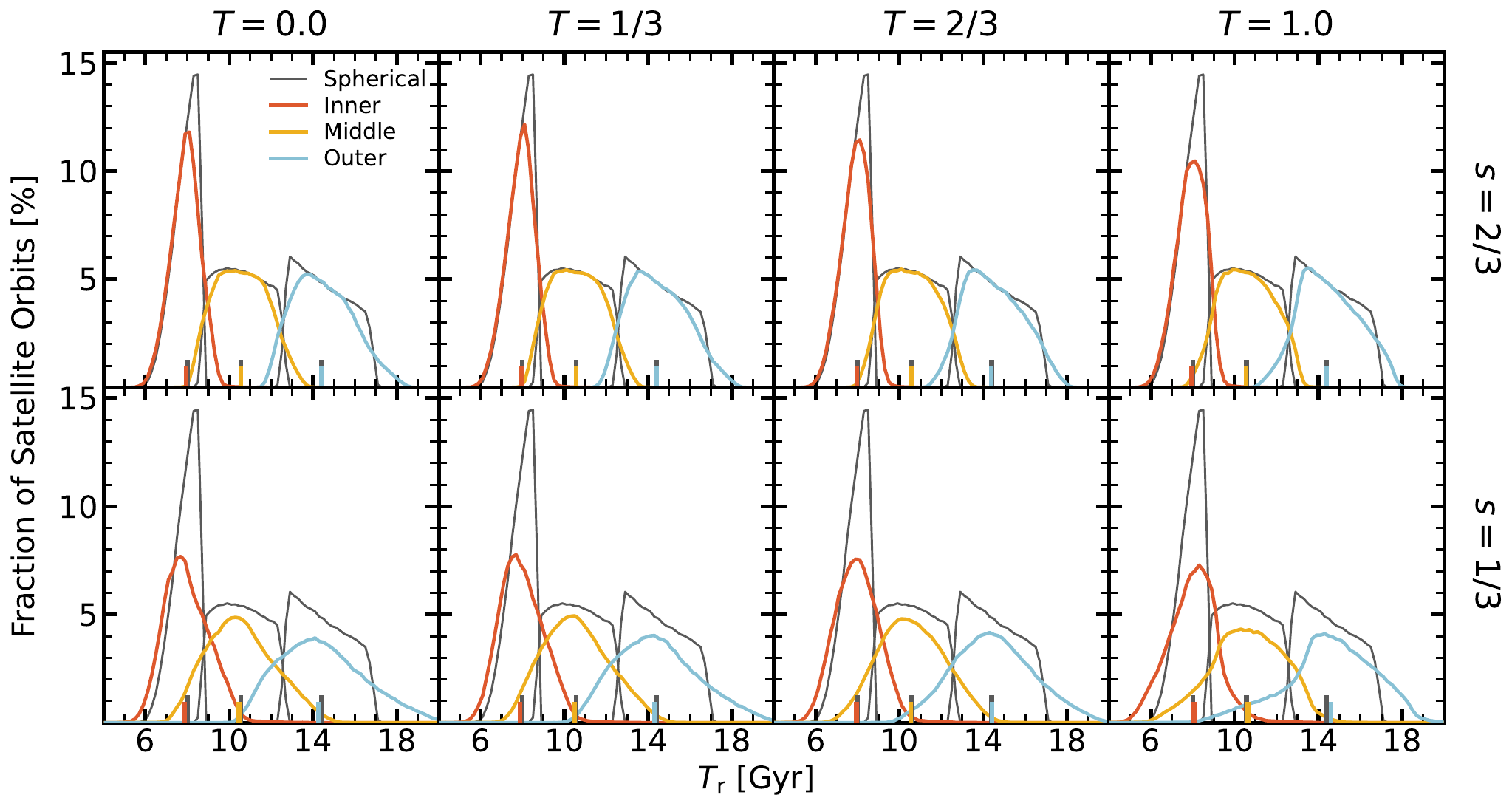}
    \caption{Distributions of the first radial period $\Trad$ for the eight triaxial hosts, arranged by triaxiality $T$ (columns) and sphericity $s = c/a$ (rows). In each panel, the red, yellow, and cyan curves show the inner, middle, and outer orbital-energy bins (\fref{fig:infall}) evolved in that host, and the grey curves show the benchmark spherical counterparts, identical in every panel. Each curve is normalised to its own orbit population, and the short vertical ticks along the bottom edge mark the corresponding medians, as in \fref{fig:peridist}.}
    \label{fig:periods}
\end{figure*}

Strikingly, in the flattened $s = 1/3$ hosts the $R^{\rm min}_{\rm peri}$ distributions develop a pronounced `pile-up' at $\simeq\!0.1\,\Rvir$, up to $\sim\!30\%$ higher in amplitude than the spherical counterpart, whereas in the $s = 2/3$ hosts the peak merely shifts inward, from $0.21\,\Rvir$ to $0.15$--$0.17\,\Rvir$. The pile-up is a direct consequence of the per-orbit pericentre scatter of \S\ref{ssec:phase_mixing}; convolving the spherical $R^{\rm min}_{\rm peri}$ distribution with the measured scatter of each triaxial host reproduces the location and height of the peak in every panel of \fref{fig:minperi}, and the feature persists across halo concentrations $c_{\rm vir} = 4$--$40$. At first sight it seems plausible that this can be attributed to centrophilic box orbits \citep[e.g.,][]{Valluri2010}. However, this explanation is ruled out by the pile-up being strongest at $T = 1$ and persisting at $T = 0$, the two axisymmetric limits whose potentials permit no box orbits at all \citep{deZeeuw1985}. Instead, the pile-up is often dominated by chaotic orbits that plunge deepest into the central cusp, comprising $82$--$96\%$ of all orbits within $R^{\rm min}_{\rm peri} < 0.1\,\Rvir$ in the $s = 1/3$ hosts (\S\ref{ssec:census}), with per-orbit pericentre scatter stronger by a factor of three to four than that of regular orbits at fixed orbital energy. Quantitatively, chaotic orbits account for $75$--$81\%$ of the systematic $R^{\rm min}_{\rm peri}$ deepening of the inner population in the triaxial $s = 1/3$ hosts, but for $\lesssim\!30\%$ in the $s = 2/3$ hosts, where the deepening is carried by regular orbits. Furthermore, the exact peak locations are an instantaneous snapshot of an inward-diffusing process recorded at $\tH$, en route towards the asymptotic lower limits. Following these infall orbits over $1000\Gyr$, $R^{\rm min}_{\rm peri}$ continues to decrease, by a median of $18$--$68\%$ across hosts for chaotic orbits but only by $3$--$12\%$ for regular orbits. For example, in the $T = 2/3$, $s = 1/3$ host, the peak of the distribution migrates inward from $0.07$ to $0.01\,\Rvir$ between $14$ and $1000\Gyr$. In short, triaxiality does not shift but only disperses satellite pericentres, deepening each orbit's closest approach ever experienced.

In contrast, the apocentre is remarkably insensitive to the varying host shapes. The measured per-orbit scatter is $\lesssim\!1\%$ in every host and energy bin, consistent with the shape-invariant apocentre medians of $1.286$--$1.290\,\Rvir$ across all nine hosts (\S\ref{ssec:phase_mixing}). This robustness comes from two combined physical effects, one geometric and one dynamical. Geometrically, although the density shape of each host is constant at all radii by construction, the potential is rounder than the density (see footnote~\ref{fn:shapeconv}) and increasingly so at large radii. Quantitatively, in the flattened $s = 1/3$ hosts for example, the equipotential radius varies across the principal axes by $\pm(15$--$21)\%$ at typical pericentric radii ($\simeq\!0.35\,\Rvir$) but only by $\pm(10$--$16)\%$ at apocentric radii ($\simeq\!1.3\,\Rvir$). Dynamically, $\rperi$ and $\rapo$ carry nearly opposite dependences on the orbital integrals, with the pericentric barrier controlled directly by the non-conserved $|\mathbf{L}|$, whereas $\rapo$ is primarily set by the conserved $E$ through the potential. Satellite apocentre inferences under the spherical host assumption are therefore comparatively robust against the unknown true host halo shape.

%%%%%%%%%%%%%%%%%%%%%
\subsection{Satellite orbital periods}
\label{sec:periods}

Next, we quantify the radial period $\Trad$, whose broadening represents another physical aspect of the orbital dephasing. \fref{fig:periods} shows the $\Trad$ distributions, measured for each orbit as twice the time between its first pericentre and first apocentre\footnote{The choice is justified by the fact that the $\Trad$ distribution is nearly time-invariant within $\tH$. Although individual triaxial orbits' periods typically vary by $\sim\!4\%$ between the first and third cycles, the population distributions of the first and third radial cycles agree to $\lesssim\!1\%$ in median and $\lesssim\!10\%$ in width.}, for the eight triaxial hosts with $T = 0, 1/3, 2/3, 1$ (left to right columns) and $s = 2/3, 1/3$ (top and bottom rows, respectively). In each panel, we compare the inner (red), middle (yellow), and outer (cyan) orbit distributions against their reference spherical counterparts, with the corresponding medians marked by the ticks along the bottom axis. As noted above, $\Trad$ is a function of $E$ and $|\mathbf{L}|$ alone in the spherical host; the period distributions therefore directly inherit the sharp truncations of the underlying energy bins\footnote{We have explicitly verified the spherical period distributions of the three energy bins analytically via direct quadrature integration; the per-orbit recovery is accurate to better than $10^{-3}$ for all $10^6$ orbits.}.

On a population level, the median orbital periods are insensitive to the host halo shape, agreeing with the spherical benchmark to within $1.5\%$ in every energy bin. However, the full distributions around those medians do markedly broaden. In the triaxial hosts, the sharply energy-truncated spherical distributions dissolve into extended tails. Specifically, the $16$--$84$ percentile width of the inner-bin period distribution grows by $\sim\!20\%$ for $s = 2/3$ and by $75$--$85\%$ for $s = 1/3$, with the central $95\%$ of inner-bin periods spreading over $5.7$--$10.3\Gyr$ while the spherical counterpart is confined to $6.6$--$8.7\Gyr$. Satellite orbits of the same energy thus no longer share a common radial period in a triaxial potential, and hence progressively dephase.

This measured broadening directly reflects the orientation dependence of $\Trad$ (\S\ref{sec:radii}); the individual period offsets average out on the population level, anchoring the medians, while their spread grows with the degree of asphericity and is fully present from the first radial cycle. The same orientation dependence also explains why flattening dominates over triaxiality in altering orbital properties. Specifically, $s$ sets the amplitude of the dominant distortion of the potential ($1 - s$, the largest axis-ratio deviation from unity), whereas $T$ merely reapportions that distortion azimuthally between the oblate and prolate limits. The number of surviving isolating integrals, by contrast, is controlled by $T$, with both $T = 0$ and $T = 1$ being axisymmetric, which is mostly relevant for chaos (\S\ref{sec:chaos}) but less so for the orbital dephasing quantified here.

%%%%%%%%%%%%%%%%%%%%%%%%%%%%%%%%%%%%%%%%%%
\section{Chaos and orbit reconstruction}
\label{sec:chaos}

The presence of chaotic orbits in triaxial potentials has long been established \citep[e.g.,][]{Schwarzschild1979, Merritt1996, Valluri1998}. However, the dynamical role of chaos in realistic hosts has so far been quantified chiefly for tidal streams and solar-vicinity halo stars in Milky Way-like potentials \citep[e.g.,][]{Maffione2015, PriceWhelan2016, Mestre2020}, with its impact on the satellite population remaining largely unexplored. In particular, the precise fraction of a cosmologically seeded satellite infall population that is chaotic, and the time-scale on which chaos erases the memory of the infall conditions relative to $\tH$ have yet to be quantified. These answers settle whether the spherical-modelling error is irreducible in nature due to exponentially diverging orbits, or a correctable modelling choice whose errors grow only linearly with time. We first present the chaos census (\S\ref{ssec:census}) and then quantify the position errors of spherical orbit reconstruction (\S\ref{ssec:divergence}), from an individual Milky Way satellite to the full population.

Here, we quantify chaos with Lyapunov exponents $\lamc$ that measure the exponential growth rate at which an orbit separates from an initially nearby orbit, so a chaotic orbit has $\lamc > 0$ and e-folds on the Lyapunov time $\tchaos = 1/\lamc$, whereas a regular and non-chaotic orbit has $\lamc = 0$. We measure $\lamc$ by integrating each orbit alongside a nearby companion \citep{Benettin1976}, for an identical subsample of $10^5$ orbits per host, large enough to determine the chaotic fraction of each energy bin to the $0.5\%$ level. We estimate $\lamc$ by the drift-corrected fit of \eref{eq:drift}, with the detection threshold calibrated on the integrable spherical control. Appendix~\ref{app:lyapunov} details the estimator and its validation.

%%%%%%%%%%%%%%%%%%%%%
\subsection{Physical importance of chaotic satellite orbits}
\label{ssec:census}

\begin{figure}
    \includegraphics[width=0.98\linewidth]{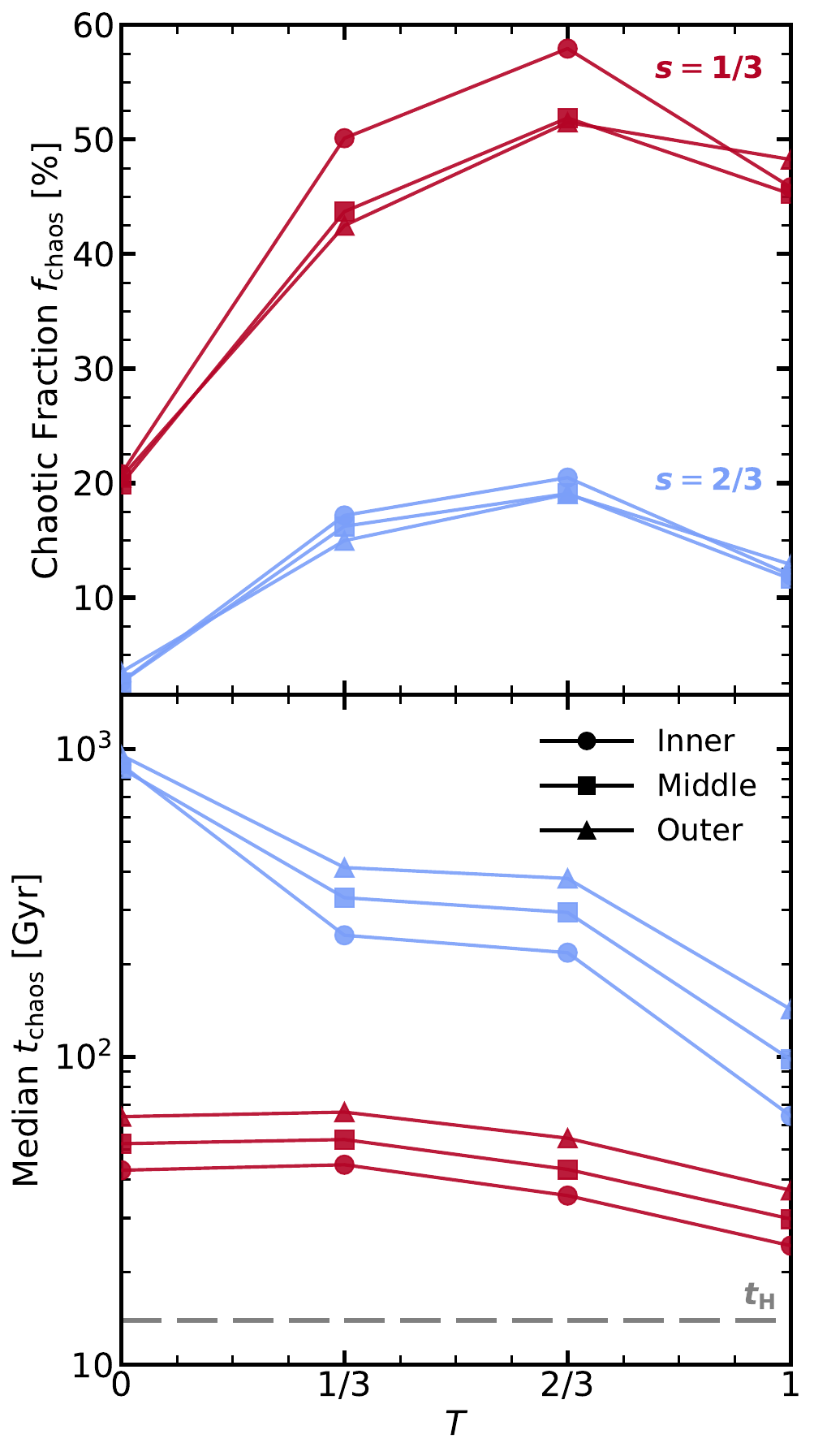}
    \caption{\textit{Top:} Chaotic orbit fraction $\fchaos$ as a function of triaxiality $T$ for the $s = 1/3$ (red) and $s = 2/3$ (blue) hosts; symbols distinguish the three orbital-energy bins. \textit{Bottom:} Median Lyapunov time $\tchaos$ of the chaotic orbits in each energy bin, against the Hubble time (grey dashed line).}
    \label{fig:census}
\end{figure}

\begin{table}
    \centering
    \caption{Classification of satellite orbits in each triaxial host. Each entry gives the percentage of all orbits in that class and, in parentheses, the percentage of those that are chaotic (see Appendix~\ref{app:lyapunov}). S-tubes (L-tubes) keep a fixed sense of circulation about the minor (major) axis throughout $1000\Gyr$, the baseline of our Lyapunov analysis, whereas box-like orbits do not; see \S\ref{ssec:census}.}
    \label{tab:families}
    \footnotesize
    \begin{tabular}{lccc}
        \hline
        $(T,\, s)$ & S-tube & L-tube & Box-like \\
        \hline\hline
        $(0,\, 2/3)$   & $100$ $(2.9)$      & --                 & --                \\
        $(1/3,\, 2/3)$ & $62.1$ $(9.3)$     & $29.3$ $(13.1)$    & $8.6$ $(73.2)$    \\
        $(2/3,\, 2/3)$ & $40.9$ $(13.3)$    & $50.8$ $(15.2)$    & $8.3$ $(74.0)$    \\
        $(1,\, 2/3)$   & --                 & $100$ $(12.1)$     & --                \\
        \hline
        $(0,\, 1/3)$   & $100$ $(20.2)$     & --                 & --                \\
        $(1/3,\, 1/3)$ & $54.2$ $(24.4)$    & $16.6$ $(18.3)$    & $29.3$ $(95.4)$   \\
        $(2/3,\, 1/3)$ & $32.2$ $(25.4)$    & $30.9$ $(28.5)$    & $37.0$ $(96.5)$   \\
        $(1,\, 1/3)$   & --                 & $100$ $(46.4)$     & --                \\
        \hline
    \end{tabular}
\end{table}

\fref{fig:census} shows the chaotic orbit fraction $\fchaos$ (top panel) and the median Lyapunov time $\tchaos$ of the chaotic orbits (bottom panel) as functions of triaxiality $T$, for the $s = 1/3$ (red) and $s = 2/3$ (blue) hosts, with the three orbital-energy bins as different symbols. The chaotic fraction $\fchaos$ reflects the underlying orbit families (\S\ref{sec:radii}), with box orbits rendered the most prone to chaos by their repeated passages through the innermost, most aspherical region of the potential \citep[cf.][]{Valluri2010}. In parallel, \tref{tab:families} provides orbit classification by their direction of circulation\footnote{\label{fn:boxlike}In the orbit classification of \citet{deZeeuw1985} for integrable triaxial potentials, box and tube orbits are both regular, distinguished by whether the orbit retains a fixed sense of circulation about a principal axis. Here, we apply the same circulation criterion operationally to every orbit, regular or chaotic, and label an orbit `box-like' if it never settles into a fixed sense of circulation over $\sim\!100$ radial periods \citep[cf.][]{Merritt1996}; this definition is distinct from the box family of regular orbits. In the axisymmetric hosts, all orbits are tubes about the symmetry axis, i.e., the minor (major) axis for $T = 0$ ($T = 1$).}. Box-like orbits are predominantly chaotic ($73$--$97\%$ in the four non-axisymmetric hosts), compared with $9$--$28\%$ of the tube orbits. As they make up only $\lesssim\!9\%$ of all orbits in the $s = 2/3$ hosts, box-like orbits account for only $32$--$40\%$ of the chaotic orbits there. In the $s = 1/3$ hosts, by contrast, box-like orbits make up $29$--$37\%$ of all orbits and $63$--$68\%$ of the chaotic ones; in the prolate axisymmetric host ($T = 1$, $s = 1/3$) the box family is absent altogether, and its substantial measured chaos is instead carried by long-axis-tube orbits perturbed within the meridional plane \citep{KandrupSiopis2003}. 

Interestingly, nearly all box-like orbits ($\geq\!99.8\%$) circulate about a principal axis for at least two radial periods before changing or losing their sense of circulation, i.e., they cross the boundaries (separatrices) between different orbital families \citep{deZeeuw1985}. Chaos is thus strongly clustered along these separatrices \citep[cf.][]{PriceWhelan2016}. We have also verified that resonant trapping is negligible for the tube orbits from a frequency analysis of the same orbits over $1000\Gyr$ with \textsc{naif} \citep{BeraldoeSilva2023}: the fraction of tubes with well-determined frequencies whose frequency ratios lie close to a ratio of small integers (e.g., 2:3) other than 1:1 is insignificant ($\lesssim\!4\%$) in all four non-axisymmetric hosts. As expected, $\fchaos$ is suppressed towards the axisymmetric limits ($T = 0$ or $1$), where the conserved axial angular momentum restores an isolating integral \citep[e.g.,][]{Schwarzschild1979, Valluri1998}. Quantitatively, $\fchaos$ peaks at intermediate triaxiality, reaching $52$--$58\%$ at $T = 2/3$ for $s = 1/3$ and $\sim\!20\%$ for $s = 2/3$, and falls to only $3$--$4\%$ in the most nearly integrable oblate $T = 0$, $s = 2/3$ host. Similar to the case of pericentre scatter (\fref{fig:cv}), $\fchaos$ is nearly independent of orbital energy.

Such chaotic fractions are broadly in line with literature measurements, which themselves span wide ranges with the system, the orbit population sampled, and the detection convention (Appendix~\ref{app:lyapunov}). In self-consistent triaxial $N$-body haloes, \citet{Valluri2010} classify $1$--$21\%$ of orbits as chaotic, from baryon-free haloes to those hosting a compact central baryonic component (their Table~2), while dense, cuspy triaxial stellar systems reach chaotic fractions above $75\%$ \citep{ZorziMuzzio2012}.

Chaos is therefore widespread, but at the same time exceedingly slow relative to both the halo crossing time $t_{\rm cross} \eee \Rvir/\Vvir \simeq 2.0\Gyr$ and $\tH$. The bottom panel of \fref{fig:census} shows the median Lyapunov time $\tchaos$ of the chaotic orbits, which over all energy bins is $30$--$60\Gyr$ for the $s = 1/3$ hosts, $110$--$340\Gyr$ for the triaxial $s = 2/3$ hosts, and $\sim\!900\Gyr$ for the nearly integrable oblate $s = 2/3$ endpoint. Within each host, $\tchaos$ decreases towards lower orbital energy; the inner-bin median is $24$--$45\Gyr$ for $s = 1/3$ and $64$--$250\Gyr$ for the triaxial $s = 2/3$ hosts. Note that per radial period, chaos is nearly energy-independent ($\tchaos/\Trad \simeq 3$--$6$ for $s = 1/3$ and $8$--$32$ for the triaxial $s = 2/3$ hosts), hence the trend in the bottom panel of \fref{fig:census} simply reflects the shorter radial periods of the more bound orbits. Even in the most chaotic host studied, a typical chaotic orbit's divergence grows by less than a factor of $e^{\tH/\tchaos} \simeq 1.6$ within a Hubble time. This inefficiency is consistent with the slow phase-space diffusion of chaotic orbits in simulated Milky Way-mass haloes reported by \citet{Maffione2015}. In the analytic triaxial Milky Way-like halo potential of \citet{PriceWhelan2016}, the majority of orbits likewise have chaotic time-scales of thousands of orbital periods ($\gtrsim\!1000\Gyr$). In self-consistent triaxial $N$-body systems, \citet{Voglis2002} showed that the detected chaotic components comprise $26$--$32\%$ of the mass, yet only $2$--$8\%$ of it can develop chaotic diffusion within a Hubble time, in broad agreement with our findings. Hence, even where $\fchaos$ is high and chaotic orbits are ubiquitous in realistic triaxial systems, across a wide range of central density slopes \citep[e.g.,][]{Merritt1996, Valluri1998, Valluri2010} and with or without a central massive object \citep[e.g.,][]{Gerhard1985, Udry1988}, chaos becomes dynamically important only after several $\tchaos$, i.e., after several e-folds of divergence.

The dephasing characterised in \S\ref{sec:radii} is therefore overwhelmingly \textit{regular}. The census also demonstrates that the bin-convergence blurring observed in the $s = 1/3$ row of \fref{fig:peridist} is unrelated to chaos. Its strength grows monotonically from $T = 0$ to $T = 1$ at $s = 1/3$, whereas $\fchaos$ (top panel of \fref{fig:census}) peaks at intermediate triaxiality and collapses at the axisymmetric endpoints; the two trends are plainly decoupled. In particular, the $T = 0$, $s = 1/3$ host is axisymmetric and has the smallest chaotic fraction of the $s = 1/3$ hosts, yet exhibits blurring comparable to the other $s = 1/3$ hosts. Similarly, the pericentre scatter, which increases monotonically with $T$ (\fref{fig:cv}), does not track the fraction of centrophilic orbits, which instead peaks at intermediate triaxiality (\tref{tab:families}). The pericentre scatter and its population-level imprints thus arise from the per-passage torques affecting all orbits, rather than chaos-driven evolution, within a Hubble time.

%%%%%%%%%%%%%%%%%%%%%
\subsection{Position errors of spherical orbit reconstruction}
\label{ssec:divergence}

Observational inferences of satellite orbital histories conventionally assume the host dark matter halo to be simply spherical, from the LMC \citep[e.g.,][]{Kallivayalil2013, GaiaHelmi2018, Patel2020}, through the Galactic classical dwarf spheroidals \citep[e.g.,][]{Fritz2018, Li2021, MartinezGarcia2026}, down to the ultra-faint satellites \citep[e.g.,][]{Simon2018, Battaglia2022, Pace2022}, and out to the few satellites beyond the Milky Way with measured proper motions \citep[e.g.,][]{Patel2017, vanderMarel2019, Sohn2020}, all obtained by direct backward integration from their measured present-day phase-space coordinates. However, as demonstrated in \S\ref{sec:radii}, orbit integration from \textit{identical infall conditions} yields diverse orbital properties and \textit{diverging orbital trajectories} that depend sensitively on the exact aspherical shape of the host.

\begin{figure*}
    \includegraphics[width=0.95\linewidth]{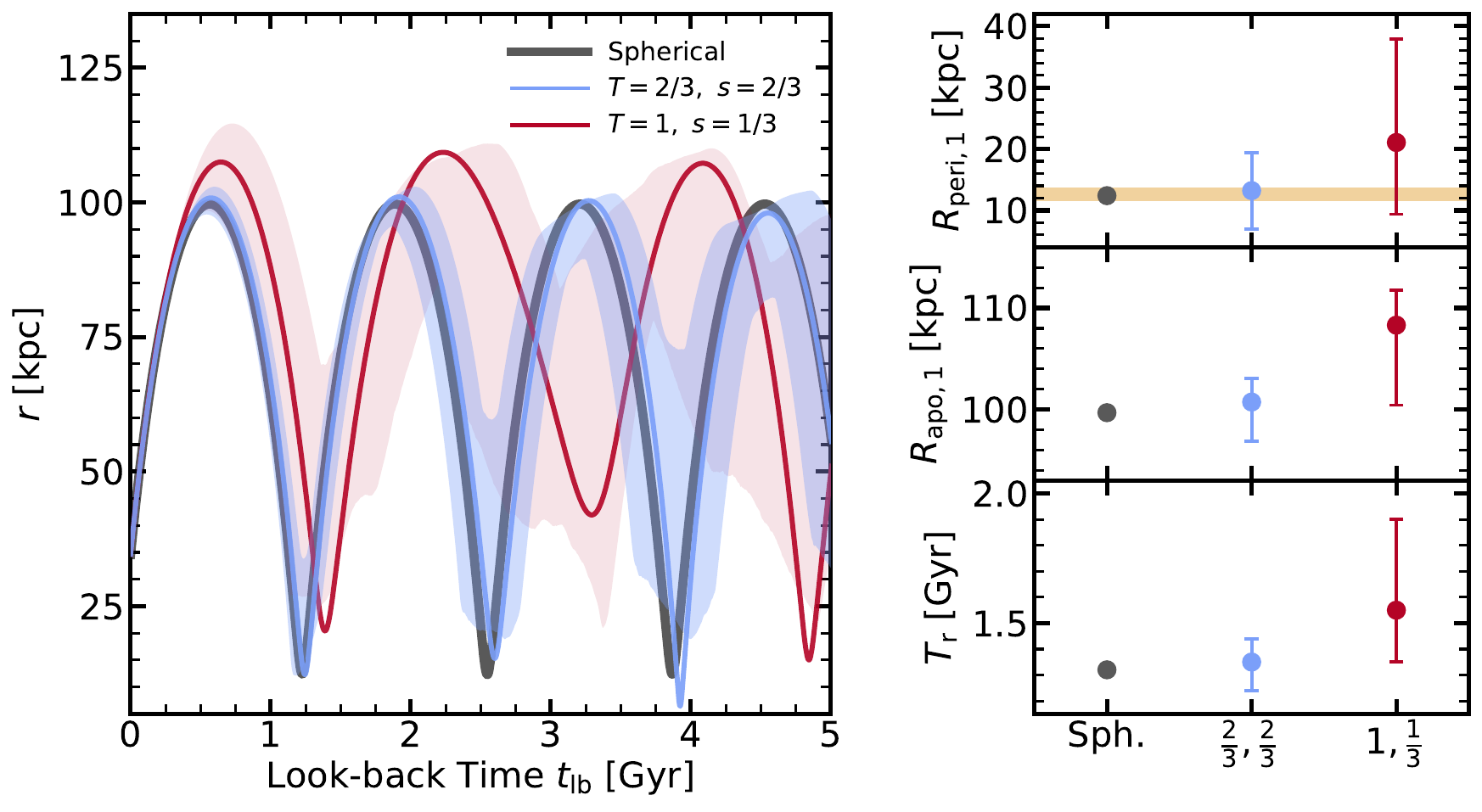}
    \caption{Backward orbital integration of the ultra-faint dwarf satellite Triangulum~II from its measured phase-space coordinates \citep{Pace2022}, in the full Milky Way model of \citet{McMillan2017} (grey) and in two hosts differing from it \textit{only} in the shape of the dark matter halo (colour-coded as indicated), sampling the unknown orientation of the halo principal axes relative to the disc plane with $10^4$ random realisations. \textit{Left:} Galactocentric distance as a function of look-back time $t_{\rm lb}$, with the $16$--$84$ percentile envelope over all realisations (colour-shaded) and, for each triaxial host, the single realisation lying closest to the ensemble median over the interval shown. \textit{Right:} The three subpanels compare, from top to bottom, the \textit{single most recent} pericentre $R_{\rm peri,1}$, apocentre $R_{\rm apo,1}$, and radial orbital period $\Trad$ recovered by each backward integration; symbols and error bars give the median and $16$--$84$ percentile \textit{across the $10^4$ halo-orientation realisations}, not the variation over successive passages of any single orbit. The horizontal ribbon in the top subpanel marks the published $\rperi = 12.6 \pm 1.1\kpc$ of \citet{Pace2022} (their no-LMC model).}
    \label{fig:casestudy}
\end{figure*}

We now quantify such orbital divergence at the single-object level, by backward integration from the observed present-day phase-space coordinates of a Milky Way satellite (\fref{fig:casestudy}), and generalise it to the full satellite population at the close of this section. This absolute positional uncertainty is especially consequential for perturbers massive enough to co-shape the host potential themselves, and thereby the orbits of all substructures within it; the most massive Milky Way satellite, the LMC, is the clearest such case. To date, the LMC's past orbit remains debated even at the level of first versus second passage \citep[e.g.,][]{Vasiliev2023, Vasiliev2024, Lucchini2025}. That distinction can itself hinge on the assumed halo shape; for the same LMC model, \citet{Sheng2024} recover a second-passage history in a spherical Milky Way halo but a first-infall history in an oblate halo with $q = 0.7$. In the case study below, we thus do not model the ill-constrained LMC component to isolate the role of the halo shape, on which the reconstructed orbit of the LMC itself depends.

As a case study set in the context of Milky Way dwarf satellite orbits, we now replace the generic hosts of \S\ref{sec:methods} by the best-fitting Milky Way model parameters of \citet{McMillan2017} as adopted in \citet{Pace2022}, and construct the triaxial counterparts of the same spherical halo as in \fref{fig:shapes} following the identical prescription of \S\ref{sec:methods}. The \citet{McMillan2017} model comprises axisymmetric baryonic components (flattened bulge, thin and thick stellar discs, atomic and molecular gas discs) and a spherical NFW dark matter halo. The baryonic components follow \citet{McMillan2017} verbatim, so the spherical benchmark now \textit{is} \citet{McMillan2017} and the triaxial members differ from this new fiducial setup only in the host halo shape. The satellite's phase-space coordinates are observationally fixed relative to the disc, leaving the halo orientation as the remaining free parameter; we uniformly sample the relative orientation with $10^4$ realisations per host. Integrating the measured coordinates backward over the rewind span of \citet{Pace2022}, our spherical orbits recover all their published pericentres and apocentres to the $\leq 3\%$ level. Any resulting orbital deviation is thus cleanly attributable to halo shape and orientation alone.

\fref{fig:casestudy} compares the backward-integrated radial orbits of Triangulum~II (Tri~II), in the spherical benchmark (grey) and in two selected triaxial hosts (blue for $T = 2/3$, $s = 2/3$; red for $T = 1$, $s = 1/3$). Tri~II has its most recent pericentric passage $1.23\Gyr$ ago in the spherical host, with its most recent pericentre distance, hereafter $R_{\rm peri,1}$, among the more precisely determined in the sample of \citet{Pace2022}. The left-hand panel compares its Galactocentric distance in each host, with the $16$--$84$ percentile envelope (colour-shaded) over halo orientations and one statistically representative realisation shown for either triaxial case. The right-hand panels compare the recovered $R_{\rm peri,1}$, the analogous most recent apocentre $R_{\rm apo,1}$, and the radial period $\Trad$ against the published values. We emphasise that each realisation contributes a single number per panel, the most recent value that an observational backward integration would infer; the error bars therefore quantify the spread of that one inference across halo orientations, not passage-to-passage variation along an orbit.

The spherical host places that passage at $12.4\kpc$, whereas $R_{\rm peri,1}$ physically spans $13.2\,(7.0$--$19.4)\kpc$ in the cosmologically representative triaxial host and $21.1\,(9.4$--$37.9)\kpc$ in the flattened one ($16$--$84$ percentiles), an uncertainty of $47\%$ and $68\%$ (half the $16$--$84$ range) induced by halo shape alone. If the host is sufficiently triaxial, the spherical host assumption does significantly bias the median of the recovered pericentre distances, in a way that depends on the satellite's orbital energy (cf. \fref{fig:peridist}). Moreover, for any fixed halo orientation, the orbit does not repeat a fixed pericentre but alternates between deeper and shallower encounters, more strongly so in the flattened host; this is precisely the per-orbit pericentre scatter of \S\ref{ssec:phase_mixing} manifested in a single observed satellite. The apocentre is comparatively robust, with its median shifting by only $+1.1\%$ ($+8.7\%$) in the typical (flattened) host and an uncertainty of $3.1\%$ ($5.2\%$); the ratio $R_{\rm peri,1}/R_{\rm apo,1}$ (not shown) correspondingly inherits the pericentre's scatter and bias, with median offsets of $+5\%$ and $+65\%$ in the two triaxial hosts.

Two further dwarfs, integrated identically but not shown, confirm that the size of this host-shape-driven scatter and bias grows with the time elapsed since the satellite's last pericentric passage, since $|\Delta\mathbf{x}|$ grows with time. Coma Berenices has just passed its pericentre, $0.03\,\Trad$ ago, and its inferred $R_{\rm peri,1} = 42.5\kpc$ is reproduced to within $0.2\%$ in both triaxial hosts; Sculptor, $0.25\,\Trad$ past its most recent passage, already scatters by $5\%$ and $13\%$ in the two triaxial hosts. Together with Tri~II at $0.93\,\Trad$ past its most recent passage, the three trace the growth of the shape term across a full radial cycle. The published precision of $R_{\rm peri,1}$ is thus most robust against unknown host shape and relative orientation only when the passage just occurred. Even with a reliable $R_{\rm peri,1}$ as in Coma Berenices, its radial period varies by $7\%$ and $20\%$ across orientations in the two triaxial hosts. Hence, robustness in recovered $R_{\rm peri,1}$ does not guarantee that in other orbital parameters, and vice versa. This shape-induced error moreover grows with look-back time, so that the recovery of earlier passages becomes prohibitively unreliable. \citet{Pace2022} already restrict the reported orbital attributes to the most recent passage on the grounds that earlier ones are poorly constrained even under the spherical host assumption \citep[e.g.,][]{DSouza2022}. The shape-induced uncertainty tightens that restriction further.

Relative to the quoted uncertainties in published satellite orbits, `error' sourced by our ignorance of the underlying host shape and orientation is not subdominant. For Tri~II the shape-induced scatter of $\sim\!47\%$ at cosmologically typical triaxiality greatly exceeds both the quoted $9\%$ uncertainty in \citet{Pace2022} and their LMC-induced shift of $3.2\%$. Across their full sample of $46$ dwarfs, the LMC-induced $3$--$33\%$ ($16$--$84$ range) displacement in $R_{\rm peri,1}$ is comparable to the $5$--$23\%$ per-pericentre scatter sourced by halo shape alone (\fref{fig:cv}). The comparison is a notable one, because the LMC is now routinely included in orbit reconstructions while the halo shape is rarely assessed and should be incorporated into the error budget of backward-integrated satellite orbits \citep[cf.][]{DSouza2022}. Furthermore, the host-shape- and LMC-induced uncertainties are coupled, as the LMC constraints per se are conditional on the assumed Milky Way potential and the halo shape \citep[e.g.,][]{Petersen2021, Vasiliev2021}. Given that the LMC trajectories adopted in such reconstructions, including that of \citet{Pace2022}, are often computed in a spherical Milky Way halo, tighter LMC constraints alone cannot fully lift this degeneracy and require joint inference with the halo shape.

On the population level, we now quantify, using the full infall population (\S\ref{ssec:population}) in our fiducial hosts (\S\ref{ssec:haloes}), the unaccounted-for `error' that observational inferences incur from the spherical host assumption even when the host mass profile is known exactly. In our static hosts with exact time reversibility, we measure the reconstruction error by re-analysing the forward-integrated suite of \S\ref{sec:radii}. By pairing each orbit in a triaxial host with its spherical-benchmark `twin', we explicitly track their 3D separation $|\Delta\mathbf{x}(t)|$ over time. The median positional deviation crosses $0.1\,\Rvir$ ($\simeq\!26\kpc$) within $1.9$--$2.2\Gyr$ for the flattened $s = 1/3$ hosts and $3.4$--$3.9\Gyr$ for the typical $s = 2/3$ shapes, and reaches $0.85$--$0.95\,\Rvir$ ($\sim\!230$--$250\kpc$) and $0.37$--$0.41\,\Rvir$ ($\sim\!100\kpc$), respectively, by $\tH$. \textit{A spherically reconstructed orbit is positionally reliable for only $\sim\!1$--$2\Gyr$}; already by $t = 1\Gyr$ ($3\Gyr$) the median deviation is $2.5$--$7.3\kpc$ ($18$--$59\kpc$), and the $16$--$84$ orbit-to-orbit range spans a factor of $\sim\!3$ about the median at all times, so the error of an individual satellite depends sensitively on its exact infall or present-day phase-space coordinates. The deviation predominantly reflects a change of orbital orientation rather than of orbit size. The spherical twin conserves its orbital plane by symmetry, whereas in a triaxial host the orbital plane naturally precesses away secularly---by a median of $15.7\degr$ ($s = 2/3$) and $44.3\degr$ ($s = 1/3$) within $\tH$. Also, $|\Delta\mathbf{x}(t)|$ is nearly identical for all four triaxialities at fixed $s$ and for orbits of all energies, pericentre distances, and chaotic or regular character. Similar to the pericentre scatter, the measured orbital-plane precession is paced by the radial period $\Trad$, so that more bound orbits precess faster in physical time; this is offset by their smaller orbital radii, giving a nearly energy-independent $|\Delta\mathbf{x}(t)|$ in physical time. Hence, the statistical reconstruction error is set by the host flattening $s$ and the orientation of the orbit relative to the host's principal axes, with the orbit's own properties entering only at second order.

%%%%%%%%%%%%%%%%%%%%%%%%%%%%%%%%%%%%%%%%%%
\section{Summary, Implications, and Conclusions}
\label{sec:conclusions}

Satellite orbits are almost universally modelled with spherical host dark matter haloes, whereas these haloes are generically triaxial. The error incurred by this geometric simplification has been systematically quantified at neither the orbit-by-orbit nor the population level for cosmologically infalling satellites, with the halo shape systematically varied at fixed mass profile. In this work, we integrate $10^6$ satellites drawn from the cosmological infall distribution of \citet{Li2020} through nine static NFW hosts of identical virial mass and matched spherically averaged mass profile, spanning triaxialities $T = 0$--$1$ at sphericities $s = 1/3$ and $2/3$, and trace the observational consequences down to individual Milky Way dwarfs within a \citet{McMillan2017}-calibrated case study. Our main conclusions are as follows:
\begin{itemize}
    \item At fixed virial mass and spherically averaged mass profile, the host halo shape leaves the population medians of pericentre, apocentre, pericentre-to-apocentre ratio, and radial period essentially unchanged ($\lesssim 2\%$ for cosmologically typical shapes; \frefs{fig:peridist} and \ref{fig:periods}). Spherical orbit modelling is unbiased on average, consistent with the cosmological-simulation measurement of \citet{Smith2022}.
    \item Halo triaxiality instead \textit{dephases} satellite orbits; successive pericentres of an individual orbit scatter by $5$--$23\%$, a fraction that grows with triaxiality and flattening but is independent of orbital energy (\fref{fig:cv}). The apocentre, by contrast, is pinned by the conserved orbital energy and scatters at only the $\lesssim\!1\%$ level in every host. The tight period--energy linkage of the spherical host correspondingly loosens, with the period distribution at fixed energy developing extended tails, its inner-bin $16$--$84$ width growing by $\sim\!20\%$ and $75$--$85\%$ for hosts at the two flattenings (\fref{fig:periods}).
    \item The minimum pericentre reached by infalling satellites within a Hubble time $\tH$ is systematically deeper than the spherical prediction, by $5$--$12\%$ on average for the most bound orbits and by up to $21\%$ in strongly flattened hosts, where the distribution moreover piles up in a localised peak at $\simeq\!0.1\,\Rvir$, up to $\sim\!30\%$ in amplitude above the spherical counterpart (\fref{fig:minperi}).
    \item The satellite orbit dephasing within $\tH$ is overwhelmingly regular, not chaotic. Drift-corrected Lyapunov exponents show that although up to $\sim\!58\%$ of orbits are chaotic in the most aspherical hosts, the median Lyapunov time of the chaotic orbits is $30$--$900\Gyr$, far exceeding the Hubble time (\fref{fig:census}). Chaos is insignificant for orbit reconstruction and most population-level statistics, except for the \textit{minimum pericentre distances} that directly tie to tidal mass loss, where chaotic orbits show the largest passage-to-passage scatter and plunge deepest towards the host centre.
    \item For the Milky Way dwarfs, the shape-sourced reconstruction error is already competitive with, and can dominate, the published error budgets. Backward integrated in the \citet{McMillan2017}-calibrated host potential, Triangulum~II has its recovered pericentre uncertain at the $\sim\!50\%$ level in a cosmologically typical triaxial halo (\fref{fig:casestudy}), five times its measurement uncertainty and more than an order of magnitude above its LMC-induced shift quoted in \citet{Pace2022}. Underlying this, an orbit reconstructed in a spherical potential in lieu of a truly triaxial host accumulates a median position error of $0.1\,\Rvir$ within only $2$--$4\Gyr$, driven by the secular precession of the orbital plane; the unknown orientation of the halo contributes an uncertainty comparable to that of its unknown shape (\S\ref{ssec:divergence}).
\end{itemize}

An immediate implication of these results concerns the numerical convergence of simulated subhaloes. \citet{Chiang2026a} recently demonstrated that properly modelling subhalo tidal evolution requires resolving the minimum tidal radius it has experienced since infall, set precisely at the deepest pericentric passage $R^{\rm min}_{\rm peri}$. With $R^{\rm min}_{\rm peri}$ computed under the spherical host assumption, about half of the subhaloes in a typical cosmological simulation are found to fail this criterion. In particular, subhaloes with $R^{\rm min}_{\rm peri}\leq 0.2\Rvir$ are nearly all force-unresolved (see Fig.~11 therein). However, halo triaxiality further shifts the entire $R^{\rm min}_{\rm peri}$ distribution to smaller radii, reducing its median by up to $21\%$ for the most bound energy bin, and, in strongly flattened hosts, creates a localised peak at $\simeq\!0.1\,\Rvir$ (\fref{fig:minperi}) sourced predominantly by chaotic orbits (\S\ref{ssec:minperi}). The force-unresolved fraction quoted in \citet{Chiang2026a} is therefore an optimistic lower bound.

Similarly, tidal stripping as set by $R^{\rm min}_{\rm peri}$ is also impacted in semi-analytic models. Semi-analytic frameworks that explicitly integrate subhalo orbits, from the pioneering models of \citet{Taylor2001} and \citet{Zentner2005} to the modern \textsc{SatGen} \citep{Jiang2021, Green2022, Monzon2026} and \textsc{Galacticus} \citep{Benson2012, Du2024}, uniformly assume a spherical host halo. Importantly, the tidal mass loss is set primarily by the minimum tidal radius a subhalo has experienced, and thus by its $R^{\rm min}_{\rm peri}$ \citep[e.g.,][]{Jiang2016, Stucker2023, Errani2024}. With the triaxiality-induced scatter and systematic deepening of $R^{\rm min}_{\rm peri}$ (\S\ref{ssec:minperi}) left unmodelled, these spherical frameworks are therefore expected to predict biased individual subhalo bound masses and radial distribution of subhalo abundance, with underestimated population-level scatter. The energy independence of the fractional pericentre scatter suggests a compact practical remedy: a shape-dependent `dephasing kernel' that reshuffles the angular-momentum magnitude of a spherically computed orbit at each pericentric passage, with an amplitude that grows with decreasing pericentric distance (\S\ref{ssec:minperi}), which could bring semi-analytic subhalo models and orbit-reconstruction pipelines most of the benefit of triaxial modelling at negligible cost, an implementation we leave for future work.

On the observational front, orbit histories of individual Milky Way satellites are obtained by direct backward integration of their present-day phase-space coordinates in an assumed, usually spherical or mildly flattened, Galactic potential \citep[e.g.,][]{Fritz2018, Battaglia2022, Pace2022}. The most recent such catalogue by \citet{MartinezGarcia2026} samples six time-evolving Milky Way--LMC potentials, of which one carries a triaxial halo; the halo triaxiality per se, however, was not varied in isolation but jointly with other model parameters. Such reconstructions are already known to be sensitive to the assumed time-dependence of the potential; using simulated Milky Way-mass hosts, \citet{DSouza2022} show that the host's mass growth and the recent accretion of the LMC \citep[cf.][]{GaravitoCamargo2019, Conroy2021, Cavieres2025} induce substantial errors in the recovered pericentres and infall times, failing to recover the penultimate pericentric distance to within $30\%$ in $44\%$ of their cases \citep[see also][]{Santistevan2023, Santistevan2024}. However, none of these error budgets isolates and quantifies the impact of the host halo shape. As demonstrated in \S\ref{ssec:divergence}, at cosmologically typical triaxiality, the reconstructed position is statistically off by $\sim\!0.1\,\Rvir$ within $2$--$4\Gyr$ of look-back time, and the pericentre recovered for a satellite rewound through a full radial period in a host of unknown orientation is uncertain at the $\sim\!50\%$ level (\fref{fig:casestudy}). 

More broadly, present-day Milky Way satellites span a broad distribution of infall times \citep[e.g.,][]{Fillingham2019, Santistevan2023}, and their current orbital energies and angular momenta retain progressively less memory of their infall values with each orbit \citep[e.g.,][]{Santistevan2024, Necib2026}, owing to the continuous growth of the host and, for massive satellites, dynamical friction \citep[e.g.,][]{vandenBosch2016, Santistevan2023, Santistevan2024}. We do not model these effects, and by construction our infall orbits, launched at $\Rvir$, do not include the innermost observed satellites with $R_{\rm apo,1}$ well within $\Rvir$. Nevertheless, our key results are nearly independent of orbital energy: the per-passage pericentre scatter (\fref{fig:cv}) and the $R^{\rm min}_{\rm peri}$ reduction at a fixed number of pericentric passages (\S\ref{ssec:minperi}), as well as the reconstruction error at a fixed look-back time (\S\ref{ssec:divergence}). The host-shape imprint on a present-day satellite is therefore primarily set by the number of pericentric passages completed since infall, and its reconstruction error by the look-back time. Early-accreted satellites, which today typically occupy close-in orbits with short radial periods and have thus completed the most passages \citep[e.g.,][]{Rocha2012, Fillingham2019, Santistevan2023}, are expected to be the most strongly affected, both in their minimum pericentres and in the recovery of their orbital histories. If restricted to look-back times of $\lesssim\!1$--$2\Gyr$, however, orbital reconstructions remain robust against the unknown halo shape, irrespective of the infall time; Coma Berenices, having just passed its pericentre, has its $R_{\rm peri,1}$ reproduced to within $0.2\%$ (\S\ref{ssec:divergence}), although it was accreted about as early as Triangulum~II ($\simeq\!10\Gyr$ ago; \citealp{Fillingham2019}), whose halo-shape-induced uncertainty in $R_{\rm peri,1}$ is at the $\sim\!50\%$ level.

As \textit{Gaia} precision advances, this halo-shape systematic could soon dominate the uncertainties on inferred orbital histories. In parallel, the Galactic halo shape and absolute orientation are steadily being charted by ever more detailed characterisation of stellar streams. The same aspherical torques that dephase satellite orbits imprint a coherent, secular precession on stellar streams \citep[e.g.,][]{Erkal2016}, making streams the complementary counterpart of the very effect quantified in this work. First stream-based measurements of the Galactic halo shape, tilt, and triaxiality are now becoming available \citep[e.g.,][]{Vasiliev2021, Woudenberg2024, Nibauer2025}, and the rapidly growing stream census from wide-area photometric surveys, together with dedicated stream spectroscopy such as the upcoming Via project \citep{Via2026}, promises an ensemble mapping of the Galactic halo shape and orientation. Such constraints on the halo shape and orientation are precisely what is required to narrow the shape-induced uncertainties in reconstructed satellite orbital histories.

%%%%%%%%%%%%%%%%%%%%%%%%%%%%%%%%%%%%%%%%%%
\section*{Acknowledgements}
We thank the anonymous referee for a detailed and constructive report that improved the quality of this paper. BC and FvdB are supported by the National Science Foundation (NSF) through grants AST-2307280 and AST-2407063. We use \texttt{NumPy} \citep{numpy} and \texttt{SciPy} \citep{scipy} for data analysis, and \texttt{Matplotlib} \citep{matplotlib} for data visualisation. We also use \textsc{galpy} v1.11.0 \citep{Bovy2015} for orbit integration.

%%%%%%%%%%%%%%%%%%%%%%%%%%%%%%%%%%%%%%%%%%
\section*{Data Availability}
The data underlying this article will be shared on reasonable request to the corresponding author.

%%%%%%%%%%%%%%%%%%%%%%%%%%%%%%%%%%%%%%%%%%
\bibliographystyle{mnras}
\bibliography{bibliography}

@ARTICLE{Smith2022,
       author = {{Smith}, Rory and {Calder{\'o}n-Castillo}, Paula and {Shin}, Jihye and {Raouf}, Mojtaba and {Ko}, Jongwan},
        title = "{The First Fall is the Hardest: The Importance of Peculiar Galaxy Dynamics at Infall Time for Tidal Stripping Acting at the Centers of Groups and Clusters}",
      journal = {\aj},
         year = 2022,
        month = sep,
       volume = {164},
       number = {3},
          eid = {95},
        pages = {95},
          doi = {10.3847/1538-3881/ac8053},
archivePrefix = {arXiv},
       eprint = {2207.05099},
 primaryClass = {astro-ph.GA},
       adsurl = {https://ui.adsabs.harvard.edu/abs/2022AJ....164...95S}
}

@ARTICLE{vandenBosch2018,
       author = {{van den Bosch}, Frank C. and {Ogiya}, Go and {Hahn}, Oliver and {Burkert}, Andreas},
        title = "{Disruption of dark matter substructure: fact or fiction?}",
      journal = {\mnras},
         year = 2018,
        month = mar,
       volume = {474},
       number = {3},
        pages = {3043-3066},
          doi = {10.1093/mnras/stx2956},
archivePrefix = {arXiv},
       eprint = {1711.05276},
 primaryClass = {astro-ph.GA},
       adsurl = {https://ui.adsabs.harvard.edu/abs/2018MNRAS.474.3043V}
}

@BOOK{Mo2010,
       author = {{Mo}, Houjun and {van den Bosch}, Frank C. and {White}, Simon},
        title = "{Galaxy Formation and Evolution}",
         year = 2010,
    publisher = {Cambridge University Press},
       adsurl = {https://ui.adsabs.harvard.edu/abs/2010gfe..book.....M}
}

@ARTICLE{Zentner2005,
       author = {{Zentner}, Andrew R. and {Berlind}, Andreas A. and {Bullock}, James S. and {Kravtsov}, Andrey V. and {Wechsler}, Risa H.},
        title = "{The Physics of Galaxy Clustering. I. A Model for Subhalo Populations}",
      journal = {\apj},
         year = 2005,
        month = may,
       volume = {624},
       number = {2},
        pages = {505-525},
          doi = {10.1086/428898},
archivePrefix = {arXiv},
       eprint = {astro-ph/0411586},
 primaryClass = {astro-ph},
       adsurl = {https://ui.adsabs.harvard.edu/abs/2005ApJ...624..505Z}
}

@ARTICLE{vandenBosch1999,
       author = {{van den Bosch}, Frank C. and {Lewis}, Geraint F. and {Lake}, George and {Stadel}, Joachim},
        title = "{Substructure in Dark Halos: Orbital Eccentricities and Dynamical Friction}",
      journal = {\apj},
         year = 1999,
        month = apr,
       volume = {515},
       number = {1},
        pages = {50-68},
          doi = {10.1086/307023},
archivePrefix = {arXiv},
       eprint = {astro-ph/9811229},
 primaryClass = {astro-ph},
       adsurl = {https://ui.adsabs.harvard.edu/abs/1999ApJ...515...50V}
}

@ARTICLE{Navarro1997,
       author = {{Navarro}, Julio F. and {Frenk}, Carlos S. and {White}, Simon D.~M.},
        title = "{A Universal Density Profile from Hierarchical Clustering}",
      journal = {\apj},
         year = 1997,
        month = dec,
       volume = {490},
       number = {2},
        pages = {493-508},
          doi = {10.1086/304888},
archivePrefix = {arXiv},
       eprint = {astro-ph/9611107},
 primaryClass = {astro-ph},
       adsurl = {https://ui.adsabs.harvard.edu/abs/1997ApJ...490..493N}
}

@ARTICLE{Franx1991,
       author = {{Franx}, Marijn and {Illingworth}, Garth and {de Zeeuw}, Tim},
        title = "{The Ordered Nature of Elliptical Galaxies: Implications for Their Intrinsic Angular Momenta and Shapes}",
      journal = {\apj},
         year = 1991,
        month = dec,
       volume = {383},
        pages = {112},
          doi = {10.1086/170769},
       adsurl = {https://ui.adsabs.harvard.edu/abs/1991ApJ...383..112F}
}

@ARTICLE{Bailin2005,
       author = {{Bailin}, Jeremy and {Steinmetz}, Matthias},
        title = "{Internal and External Alignment of the Shapes and Angular Momenta of {\ensuremath{\Lambda}}CDM Halos}",
      journal = {\apj},
         year = 2005,
        month = jul,
       volume = {627},
       number = {2},
        pages = {647-665},
          doi = {10.1086/430397},
archivePrefix = {arXiv},
       eprint = {astro-ph/0408163},
 primaryClass = {astro-ph},
       adsurl = {https://ui.adsabs.harvard.edu/abs/2005ApJ...627..647B}
}

@ARTICLE{Allgood2006,
       author = {{Allgood}, Brandon and {Flores}, Ricardo A. and {Primack}, Joel R. and {Kravtsov}, Andrey V. and {Wechsler}, Risa H. and {Faltenbacher}, Andreas and {Bullock}, James S.},
        title = "{The shape of dark matter haloes: dependence on mass, redshift, radius and formation}",
      journal = {\mnras},
         year = 2006,
        month = apr,
       volume = {367},
       number = {4},
        pages = {1781-1796},
          doi = {10.1111/j.1365-2966.2006.10094.x},
archivePrefix = {arXiv},
       eprint = {astro-ph/0508497},
 primaryClass = {astro-ph},
       adsurl = {https://ui.adsabs.harvard.edu/abs/2006MNRAS.367.1781A}
}

@ARTICLE{Bett2007,
       author = {{Bett}, Philip and {Eke}, Vincent and {Frenk}, Carlos S. and {Jenkins}, Adrian and {Helly}, John and {Navarro}, Julio},
        title = "{The spin and shape of dark matter haloes in the Millennium simulation of a {\ensuremath{\Lambda}} cold dark matter universe}",
      journal = {\mnras},
         year = 2007,
        month = mar,
       volume = {376},
       number = {1},
        pages = {215-232},
          doi = {10.1111/j.1365-2966.2007.11432.x},
archivePrefix = {arXiv},
       eprint = {astro-ph/0608607},
 primaryClass = {astro-ph},
       adsurl = {https://ui.adsabs.harvard.edu/abs/2007MNRAS.376..215B}
}

@ARTICLE{VeraCiro2011,
       author = {{Vera-Ciro}, Carlos A. and {Sales}, Laura V. and {Helmi}, Amina and {Frenk}, Carlos S. and {Navarro}, Julio F. and {Springel}, Volker and {Vogelsberger}, Mark and {White}, Simon D.~M.},
        title = "{The shape of dark matter haloes in the Aquarius simulations: evolution and memory}",
      journal = {\mnras},
         year = 2011,
        month = sep,
       volume = {416},
       number = {2},
        pages = {1377-1391},
          doi = {10.1111/j.1365-2966.2011.19134.x},
archivePrefix = {arXiv},
       eprint = {1104.1566},
 primaryClass = {astro-ph.CO},
       adsurl = {https://ui.adsabs.harvard.edu/abs/2011MNRAS.416.1377V}
}

@ARTICLE{Chua2019,
       author = {{Chua}, Kun Ting Eddie and {Pillepich}, Annalisa and {Vogelsberger}, Mark and {Hernquist}, Lars},
        title = "{Shape of dark matter haloes in the Illustris simulation: effects of baryons}",
      journal = {\mnras},
         year = 2019,
        month = mar,
       volume = {484},
       number = {1},
        pages = {476-493},
          doi = {10.1093/mnras/sty3531},
archivePrefix = {arXiv},
       eprint = {1809.07255},
 primaryClass = {astro-ph.GA},
       adsurl = {https://ui.adsabs.harvard.edu/abs/2019MNRAS.484..476C}
}

@ARTICLE{VegaFerrero2017,
       author = {{Vega-Ferrero}, Jes{\'u}s and {Yepes}, Gustavo and {Gottl{\"o}ber}, Stefan},
        title = "{On the shape of dark matter haloes from MultiDark Planck simulations}",
      journal = {\mnras},
         year = 2017,
        month = may,
       volume = {467},
       number = {3},
        pages = {3226-3238},
          doi = {10.1093/mnras/stx282},
archivePrefix = {arXiv},
       eprint = {1603.02256},
 primaryClass = {astro-ph.CO},
       adsurl = {https://ui.adsabs.harvard.edu/abs/2017MNRAS.467.3226V}
}

@ARTICLE{Prada2019,
       author = {{Prada}, Jesus and {Forero-Romero}, Jaime E. and {Grand}, Robert J.~J. and {Pakmor}, R{\"u}diger and {Springel}, Volker},
        title = "{Dark matter halo shapes in the Auriga simulations}",
      journal = {\mnras},
         year = 2019,
        month = dec,
       volume = {490},
       number = {4},
        pages = {4877-4888},
          doi = {10.1093/mnras/stz2873},
archivePrefix = {arXiv},
       eprint = {1910.04045},
 primaryClass = {astro-ph.GA},
       adsurl = {https://ui.adsabs.harvard.edu/abs/2019MNRAS.490.4877P}
}

@ARTICLE{Despali2014,
       author = {{Despali}, Giulia and {Giocoli}, Carlo and {Tormen}, Giuseppe},
        title = "{Some like it triaxial: the universality of dark matter halo shapes and their evolution along the cosmic time}",
      journal = {\mnras},
         year = 2014,
        month = oct,
       volume = {443},
       number = {4},
        pages = {3208-3217},
          doi = {10.1093/mnras/stu1393},
archivePrefix = {arXiv},
       eprint = {1404.6527},
 primaryClass = {astro-ph.CO},
       adsurl = {https://ui.adsabs.harvard.edu/abs/2014MNRAS.443.3208D}
}

@ARTICLE{Maccio2008,
       author = {{Macci{\`o}}, Andrea V. and {Dutton}, Aaron A. and {van den Bosch}, Frank C.},
        title = "{Concentration, spin and shape of dark matter haloes as a function of the cosmological model: WMAP1, WMAP3 and WMAP5 results}",
      journal = {\mnras},
         year = 2008,
        month = dec,
       volume = {391},
       number = {4},
        pages = {1940-1954},
          doi = {10.1111/j.1365-2966.2008.14029.x},
archivePrefix = {arXiv},
       eprint = {0805.1926},
 primaryClass = {astro-ph},
       adsurl = {https://ui.adsabs.harvard.edu/abs/2008MNRAS.391.1940M}
}

@ARTICLE{Emami2021,
       author = {{Emami}, Razieh and {Genel}, Shy and {Hernquist}, Lars and {Alcock}, Charles and {Bose}, Sownak and {Weinberger}, Rainer and {Vogelsberger}, Mark and {Marinacci}, Federico and {Loeb}, Abraham and {Torrey}, Paul and {Forbes}, John C.},
        title = "{Morphological Types of DM Halos in Milky Way-like Galaxies in the TNG50 Simulation: Simple, Twisted, or Stretched}",
      journal = {\apj},
         year = 2021,
        month = may,
       volume = {913},
       number = {1},
          eid = {36},
        pages = {36},
          doi = {10.3847/1538-4357/abf147},
archivePrefix = {arXiv},
       eprint = {2009.09220},
 primaryClass = {astro-ph.GA},
       adsurl = {https://ui.adsabs.harvard.edu/abs/2021ApJ...913...36E}
}

@ARTICLE{Bryan1998,
       author = {{Bryan}, Greg L. and {Norman}, Michael L.},
        title = "{Statistical Properties of X-Ray Clusters: Analytic and Numerical Comparisons}",
      journal = {\apj},
         year = 1998,
        month = mar,
       volume = {495},
       number = {1},
        pages = {80-99},
          doi = {10.1086/305262},
archivePrefix = {arXiv},
       eprint = {astro-ph/9710107},
 primaryClass = {astro-ph},
       adsurl = {https://ui.adsabs.harvard.edu/abs/1998ApJ...495...80B}
}

@ARTICLE{Balogh2000,
       author = {{Balogh}, Michael L. and {Navarro}, Julio F. and {Morris}, Simon L.},
        title = "{The Origin of Star Formation Gradients in Rich Galaxy Clusters}",
      journal = {\apj},
         year = 2000,
        month = sep,
       volume = {540},
       number = {1},
        pages = {113-121},
          doi = {10.1086/309323},
archivePrefix = {arXiv},
       eprint = {astro-ph/0004078},
 primaryClass = {astro-ph},
       adsurl = {https://ui.adsabs.harvard.edu/abs/2000ApJ...540..113B}
}

@ARTICLE{Gunn1972,
       author = {{Gunn}, James E. and {Gott}, J. Richard, III},
        title = "{On the Infall of Matter Into Clusters of Galaxies and Some Effects on Their Evolution}",
      journal = {\apj},
         year = 1972,
        month = aug,
       volume = {176},
        pages = {1},
          doi = {10.1086/151605},
       adsurl = {https://ui.adsabs.harvard.edu/abs/1972ApJ...176....1G}
}

@BOOK{Binney1987,
       author = {{Binney}, James and {Tremaine}, Scott},
        title = "{Galactic dynamics}",
         year = 1987,
    publisher = {Princeton University Press},
       adsurl = {https://ui.adsabs.harvard.edu/abs/1987gady.book.....B}
}

@ARTICLE{Tormen1997,
       author = {{Tormen}, Giuseppe},
        title = "{The rise and fall of satellites in galaxy clusters}",
      journal = {\mnras},
         year = 1997,
        month = sep,
       volume = {290},
       number = {3},
        pages = {411-421},
          doi = {10.1093/mnras/290.3.411},
archivePrefix = {arXiv},
       eprint = {astro-ph/9611078},
 primaryClass = {astro-ph},
       adsurl = {https://ui.adsabs.harvard.edu/abs/1997MNRAS.290..411T}
}

@ARTICLE{Benson2005,
       author = {{Benson}, A.~J.},
        title = "{Orbital parameters of infalling dark matter substructures}",
      journal = {\mnras},
         year = 2005,
        month = apr,
       volume = {358},
       number = {2},
        pages = {551-562},
          doi = {10.1111/j.1365-2966.2005.08788.x},
archivePrefix = {arXiv},
       eprint = {astro-ph/0407428},
 primaryClass = {astro-ph},
       adsurl = {https://ui.adsabs.harvard.edu/abs/2005MNRAS.358..551B}
}

@ARTICLE{Khochfar2006,
       author = {{Khochfar}, S. and {Burkert}, A.},
        title = "{Orbital parameters of merging dark matter halos}",
      journal = {\aap},
         year = 2006,
        month = jan,
       volume = {445},
       number = {2},
        pages = {403-412},
          doi = {10.1051/0004-6361:20053241},
archivePrefix = {arXiv},
       eprint = {astro-ph/0309611},
 primaryClass = {astro-ph},
       adsurl = {https://ui.adsabs.harvard.edu/abs/2006A&A...445..403K}
}

@ARTICLE{Wetzel2011,
       author = {{Wetzel}, Andrew R.},
        title = "{On the orbits of infalling satellite haloes}",
      journal = {\mnras},
         year = 2011,
        month = mar,
       volume = {412},
       number = {1},
        pages = {49-58},
          doi = {10.1111/j.1365-2966.2010.17877.x},
archivePrefix = {arXiv},
       eprint = {1001.4792},
 primaryClass = {astro-ph.CO},
       adsurl = {https://ui.adsabs.harvard.edu/abs/2011MNRAS.412...49W}
}

@ARTICLE{Jiang2015,
       author = {{Jiang}, Lilian and {Cole}, Shaun and {Sawala}, Till and {Frenk}, Carlos S.},
        title = "{Orbital parameters of infalling satellite haloes in the hierarchical {\ensuremath{\Lambda}}CDM model}",
      journal = {\mnras},
         year = 2015,
        month = apr,
       volume = {448},
       number = {2},
        pages = {1674-1686},
          doi = {10.1093/mnras/stv053},
archivePrefix = {arXiv},
       eprint = {1409.1179},
 primaryClass = {astro-ph.CO},
       adsurl = {https://ui.adsabs.harvard.edu/abs/2015MNRAS.448.1674J}
}

@ARTICLE{DiemerKravtsov2014,
       author = {{Diemer}, Benedikt and {Kravtsov}, Andrey V.},
        title = "{Dependence of the Outer Density Profiles of Halos on Their Mass Accretion Rate}",
      journal = {\apj},
         year = 2014,
        month = jul,
       volume = {789},
       number = {1},
          eid = {1},
        pages = {1},
          doi = {10.1088/0004-637X/789/1/1},
archivePrefix = {arXiv},
       eprint = {1401.1216},
 primaryClass = {astro-ph.CO},
       adsurl = {https://ui.adsabs.harvard.edu/abs/2014ApJ...789....1D}
}

@ARTICLE{Adhikari2014,
       author = {{Adhikari}, Susmita and {Dalal}, Neal and {Chamberlain}, Robert T.},
        title = "{Splashback in accreting dark matter halos}",
      journal = {\jcap},
         year = 2014,
        month = nov,
       volume = {2014},
       number = {11},
        pages = {019-019},
          doi = {10.1088/1475-7516/2014/11/019},
archivePrefix = {arXiv},
       eprint = {1409.4482},
 primaryClass = {astro-ph.CO},
       adsurl = {https://ui.adsabs.harvard.edu/abs/2014JCAP...11..019A}
}

@ARTICLE{More2015,
       author = {{More}, Surhud and {Diemer}, Benedikt and {Kravtsov}, Andrey V.},
        title = "{The Splashback Radius as a Physical Halo Boundary and the Growth of Halo Mass}",
      journal = {\apj},
         year = 2015,
        month = sep,
       volume = {810},
       number = {1},
          eid = {36},
        pages = {36},
          doi = {10.1088/0004-637X/810/1/36},
archivePrefix = {arXiv},
       eprint = {1504.05591},
 primaryClass = {astro-ph.CO},
       adsurl = {https://ui.adsabs.harvard.edu/abs/2015ApJ...810...36M}
}

@ARTICLE{Diemer2017,
       author = {{Diemer}, Benedikt and {Mansfield}, Philip and {Kravtsov}, Andrey V. and {More}, Surhud},
        title = "{The Splashback Radius of Halos from Particle Dynamics. II. Dependence on Mass, Accretion Rate, Redshift, and Cosmology}",
      journal = {\apj},
         year = 2017,
        month = jul,
       volume = {843},
       number = {2},
          eid = {140},
        pages = {140},
          doi = {10.3847/1538-4357/aa79ab},
archivePrefix = {arXiv},
       eprint = {1703.09716},
 primaryClass = {astro-ph.CO},
       adsurl = {https://ui.adsabs.harvard.edu/abs/2017ApJ...843..140D}
}

@ARTICLE{vandenBosch2017,
       author = {{van den Bosch}, Frank C.},
        title = "{Dissecting the evolution of dark matter subhaloes in the Bolshoi simulation}",
      journal = {\mnras},
         year = 2017,
        month = jun,
       volume = {468},
       number = {1},
        pages = {885-909},
          doi = {10.1093/mnras/stx520},
archivePrefix = {arXiv},
       eprint = {1611.02657},
 primaryClass = {astro-ph.GA},
       adsurl = {https://ui.adsabs.harvard.edu/abs/2017MNRAS.468..885V}
}

@ARTICLE{Bovy2015,
       author = {{Bovy}, Jo},
        title = "{galpy: A python Library for Galactic Dynamics}",
      journal = {\apjs},
         year = 2015,
        month = feb,
       volume = {216},
       number = {2},
          eid = {29},
        pages = {29},
          doi = {10.1088/0067-0049/216/2/29},
archivePrefix = {arXiv},
       eprint = {1412.3451},
 primaryClass = {astro-ph.GA},
       adsurl = {https://ui.adsabs.harvard.edu/abs/2015ApJS..216...29B}
}

@ARTICLE{Li2020,
       author = {{Li}, Zhao-Zhou and {Zhao}, Dong-Hai and {Jing}, Y.~P. and {Han}, Jiaxin and {Dong}, Fu-Yu},
        title = "{Orbital Distribution of Infalling Satellite Halos across Cosmic Time}",
      journal = {\apj},
         year = 2020,
        month = dec,
       volume = {905},
       number = {2},
          eid = {177},
        pages = {177},
          doi = {10.3847/1538-4357/abc481},
archivePrefix = {arXiv},
       eprint = {2008.05710},
 primaryClass = {astro-ph.CO},
       adsurl = {https://ui.adsabs.harvard.edu/abs/2020ApJ...905..177L}
}

@ARTICLE{GaiaCollaboration2016,
	author = {{Gaia Collaboration} and {Prusti}, T. and {de Bruijne}, J.~H.~J. and {Brown}, A.~G.~A. and {Vallenari}, A. and {Babusiaux}, C. and {Bailer-Jones}, C.~A.~L. and {Bastian}, U. and {Biermann}, M. and {Evans}, D.~W. and {Eyer}, L. and {Jansen}, F. and {Jordi}, C. and {Klioner}, S.~A. and {Lammers}, U. and {Lindegren}, L. and {Luri}, X. and {Mignard}, F. and {Milligan}, D.~J. and {Panem}, C. and {Poinsignon}, V. and {Pourbaix}, D. and {Randich}, S. and {Sarri}, G. and {Sartoretti}, P. and {Siddiqui}, H.~I. and {Soubiran}, C. and {Valette}, V. and {van Leeuwen}, F. and {Walton}, N.~A. and {Aerts}, C. and {Arenou}, F. and {Cropper}, M. and {Drimmel}, R. and {H{\o}g}, E. and {Katz}, D. and {Lattanzi}, M.~G. and {O'Mullane}, W. and {Grebel}, E.~K. and {Holland}, A.~D. and {Huc}, C. and {Passot}, X. and {Bramante}, L. and {Cacciari}, C. and {Casta{\~n}eda}, J. and {Chaoul}, L. and {Cheek}, N. and {De Angeli}, F. and {Fabricius}, C. and {Guerra}, R. and {Hern{\'a}ndez}, J. and {Jean-Antoine-Piccolo}, A. and {Masana}, E. and {Messineo}, R. and {Mowlavi}, N. and {Nienartowicz}, K. and {Ord{\'o}{\~n}ez-Blanco}, D. and {Panuzzo}, P. and {Portell}, J. and {Richards}, P.~J. and {Riello}, M. and {Seabroke}, G.~M. and {Tanga}, P. and {Th{\'e}venin}, F. and {Torra}, J. and {Els}, S.~G. and {Gracia-Abril}, G. and {Comoretto}, G. and {Garcia-Reinaldos}, M. and {Lock}, T. and {Mercier}, E. and {Altmann}, M. and {Andrae}, R. and {Astraatmadja}, T.~L. and {Bellas-Velidis}, I. and {Benson}, K. and {Berthier}, J. and {Blomme}, R. and {Busso}, G. and {Carry}, B. and {Cellino}, A. and {Clementini}, G. and {Cowell}, S. and {Creevey}, O. and {Cuypers}, J. and {Davidson}, M. and {De Ridder}, J. and {de Torres}, A. and {Delchambre}, L. and {Dell'Oro}, A. and {Ducourant}, C. and {Fr{\'e}mat}, Y. and {Garc{\'\i}a-Torres}, M. and {Gosset}, E. and {Halbwachs}, J. -L. and {Hambly}, N.~C. and {Harrison}, D.~L. and {Hauser}, M. and {Hestroffer}, D. and {Hodgkin}, S.~T. and {Huckle}, H.~E. and {Hutton}, A. and {Jasniewicz}, G. and {Jordan}, S. and {Kontizas}, M. and {Korn}, A.~J. and {Lanzafame}, A.~C. and {Manteiga}, M. and {Moitinho}, A. and {Muinonen}, K. and {Osinde}, J. and {Pancino}, E. and {Pauwels}, T. and {Petit}, J. -M. and {Recio-Blanco}, A. and {Robin}, A.~C. and {Sarro}, L.~M. and {Siopis}, C. and {Smith}, M. and {Smith}, K.~W. and {Sozzetti}, A. and {Thuillot}, W. and {van Reeven}, W. and {Viala}, Y. and {Abbas}, U. and {Abreu Aramburu}, A. and {Accart}, S. and {Aguado}, J.~J. and {Allan}, P.~M. and {Allasia}, W. and {Altavilla}, G. and {{\'A}lvarez}, M.~A. and {Alves}, J. and {Anderson}, R.~I. and {Andrei}, A.~H. and {Anglada Varela}, E. and {Antiche}, E. and {Antoja}, T. and {Ant{\'o}n}, S. and {Arcay}, B. and {Atzei}, A. and {Ayache}, L. and {Bach}, N. and {Baker}, S.~G. and {Balaguer-N{\'u}{\~n}ez}, L. and {Barache}, C. and {Barata}, C. and {Barbier}, A. and {Barblan}, F. and {Baroni}, M. and {Barrado y Navascu{\'e}s}, D. and {Barros}, M. and {Barstow}, M.~A. and {Becciani}, U. and {Bellazzini}, M. and {Bellei}, G. and {Bello Garc{\'\i}a}, A. and {Belokurov}, V. and {Bendjoya}, P. and {Berihuete}, A. and {Bianchi}, L. and {Bienaym{\'e}}, O. and {Billebaud}, F. and {Blagorodnova}, N. and {Blanco-Cuaresma}, S. and {Boch}, T. and {Bombrun}, A. and {Borrachero}, R. and {Bouquillon}, S. and {Bourda}, G. and {Bouy}, H. and {Bragaglia}, A. and {Breddels}, M.~A. and {Brouillet}, N. and {Br{\"u}semeister}, T. and {Bucciarelli}, B. and {Budnik}, F. and {Burgess}, P. and {Burgon}, R. and {Burlacu}, A. and {Busonero}, D. and {Buzzi}, R. and {Caffau}, E. and {Cambras}, J. and {Campbell}, H. and {Cancelliere}, R. and {Cantat-Gaudin}, T. and {Carlucci}, T. and {Carrasco}, J.~M. and {Castellani}, M. and {Charlot}, P. and {Charnas}, J. and {Charvet}, P. and {Chassat}, F. and {Chiavassa}, A. and {Clotet}, M. and {Cocozza}, G. and {Collins}, R.~S. and {Collins}, P. and {Costigan}, G. and {Crifo}, F. and {Cross}, N.~J.~G. and {Crosta}, M. and {Crowley}, C. and {Dafonte}, C. and {Damerdji}, Y. and {Dapergolas}, A. and {David}, P. and {David}, M. and {De Cat}, P. and {de Felice}, F. and {de Laverny}, P. and {De Luise}, F. and {De March}, R. and {de Martino}, D. and {de Souza}, R. and {Debosscher}, J. and {del Pozo}, E. and {Delbo}, M. and {Delgado}, A. and {Delgado}, H.~E. and {di Marco}, F. and {Di Matteo}, P. and {Diakite}, S. and {Distefano}, E. and {Dolding}, C. and {Dos Anjos}, S. and {Drazinos}, P. and {Dur{\'a}n}, J. and {Dzigan}, Y. and {Ecale}, E. and {Edvardsson}, B. and {Enke}, H. and {Erdmann}, M. and {Escolar}, D. and {Espina}, M. and {Evans}, N.~W. and {Eynard Bontemps}, G. and {Fabre}, C. and {Fabrizio}, M. and {Faigler}, S. and {Falc{\~a}o}, A.~J. and {Farr{\`a}s Casas}, M. and {Faye}, F. and {Federici}, L. and {Fedorets}, G. and {Fern{\'a}ndez-Hern{\'a}ndez}, J. and {Fernique}, P. and {Fienga}, A. and {Figueras}, F. and {Filippi}, F. and {Findeisen}, K. and {Fonti}, A. and {Fouesneau}, M. and {Fraile}, E. and {Fraser}, M. and {Fuchs}, J. and {Furnell}, R. and {Gai}, M. and {Galleti}, S. and {Galluccio}, L. and {Garabato}, D. and {Garc{\'\i}a-Sedano}, F. and {Gar{\'e}}, P. and {Garofalo}, A. and {Garralda}, N. and {Gavras}, P. and {Gerssen}, J. and {Geyer}, R. and {Gilmore}, G. and {Girona}, S. and {Giuffrida}, G. and {Gomes}, M. and {Gonz{\'a}lez-Marcos}, A. and {Gonz{\'a}lez-N{\'u}{\~n}ez}, J. and {Gonz{\'a}lez-Vidal}, J.~J. and {Granvik}, M. and {Guerrier}, A. and {Guillout}, P. and {Guiraud}, J. and {G{\'u}rpide}, A. and {Guti{\'e}rrez-S{\'a}nchez}, R. and {Guy}, L.~P. and {Haigron}, R. and {Hatzidimitriou}, D. and {Haywood}, M. and {Heiter}, U. and {Helmi}, A. and {Hobbs}, D. and {Hofmann}, W. and {Holl}, B. and {Holland}, G. and {Hunt}, J.~A.~S. and {Hypki}, A. and {Icardi}, V. and {Irwin}, M. and {Jevardat de Fombelle}, G. and {Jofr{\'e}}, P. and {Jonker}, P.~G. and {Jorissen}, A. and {Julbe}, F. and {Karampelas}, A. and {Kochoska}, A. and {Kohley}, R. and {Kolenberg}, K. and {Kontizas}, E. and {Koposov}, S.~E. and {Kordopatis}, G. and {Koubsky}, P. and {Kowalczyk}, A. and {Krone-Martins}, A. and {Kudryashova}, M. and {Kull}, I. and {Bachchan}, R.~K. and {Lacoste-Seris}, F. and {Lanza}, A.~F. and {Lavigne}, J. -B. and {Le Poncin-Lafitte}, C. and {Lebreton}, Y. and {Lebzelter}, T. and {Leccia}, S. and {Leclerc}, N. and {Lecoeur-Taibi}, I. and {Lemaitre}, V. and {Lenhardt}, H. and {Leroux}, F. and {Liao}, S. and {Licata}, E. and {Lindstr{\o}m}, H.~E.~P. and {Lister}, T.~A. and {Livanou}, E. and {Lobel}, A. and {L{\"o}ffler}, W. and {L{\'o}pez}, M. and {Lopez-Lozano}, A. and {Lorenz}, D. and {Loureiro}, T. and {MacDonald}, I. and {Magalh{\~a}es Fernandes}, T. and {Managau}, S. and {Mann}, R.~G. and {Mantelet}, G. and {Marchal}, O. and {Marchant}, J.~M. and {Marconi}, M. and {Marie}, J. and {Marinoni}, S. and {Marrese}, P.~M. and {Marschalk{\'o}}, G. and {Marshall}, D.~J. and {Mart{\'\i}n-Fleitas}, J.~M. and {Martino}, M. and {Mary}, N. and {Matijevi{\v{c}}}, G. and {Mazeh}, T. and {McMillan}, P.~J. and {Messina}, S. and {Mestre}, A. and {Michalik}, D. and {Millar}, N.~R. and {Miranda}, B.~M.~H. and {Molina}, D. and {Molinaro}, R. and {Molinaro}, M. and {Moln{\'a}r}, L. and {Moniez}, M. and {Montegriffo}, P. and {Monteiro}, D. and {Mor}, R. and {Mora}, A. and {Morbidelli}, R. and {Morel}, T. and {Morgenthaler}, S. and {Morley}, T. and {Morris}, D. and {Mulone}, A.~F. and {Muraveva}, T. and {Musella}, I. and {Narbonne}, J. and {Nelemans}, G. and {Nicastro}, L. and {Noval}, L. and {Ord{\'e}novic}, C. and {Ordieres-Mer{\'e}}, J. and {Osborne}, P. and {Pagani}, C. and {Pagano}, I. and {Pailler}, F. and {Palacin}, H. and {Palaversa}, L. and {Parsons}, P. and {Paulsen}, T. and {Pecoraro}, M. and {Pedrosa}, R. and {Pentik{\"a}inen}, H. and {Pereira}, J. and {Pichon}, B. and {Piersimoni}, A.~M. and {Pineau}, F. -X. and {Plachy}, E. and {Plum}, G. and {Poujoulet}, E. and {Pr{\v{s}}a}, A. and {Pulone}, L. and {Ragaini}, S. and {Rago}, S. and {Rambaux}, N. and {Ramos-Lerate}, M. and {Ranalli}, P. and {Rauw}, G. and {Read}, A. and {Regibo}, S. and {Renk}, F. and {Reyl{\'e}}, C. and {Ribeiro}, R.~A. and {Rimoldini}, L. and {Ripepi}, V. and {Riva}, A. and {Rixon}, G. and {Roelens}, M. and {Romero-G{\'o}mez}, M. and {Rowell}, N. and {Royer}, F. and {Rudolph}, A. and {Ruiz-Dern}, L. and {Sadowski}, G. and {Sagrist{\`a} Sell{\'e}s}, T. and {Sahlmann}, J. and {Salgado}, J. and {Salguero}, E. and {Sarasso}, M. and {Savietto}, H. and {Schnorhk}, A. and {Schultheis}, M. and {Sciacca}, E. and {Segol}, M. and {Segovia}, J.~C. and {Segransan}, D. and {Serpell}, E. and {Shih}, I. -C. and {Smareglia}, R. and {Smart}, R.~L. and {Smith}, C. and {Solano}, E. and {Solitro}, F. and {Sordo}, R. and {Soria Nieto}, S. and {Souchay}, J. and {Spagna}, A. and {Spoto}, F. and {Stampa}, U. and {Steele}, I.~A. and {Steidelm{\"u}ller}, H. and {Stephenson}, C.~A. and {Stoev}, H. and {Suess}, F.~F. and {S{\"u}veges}, M. and {Surdej}, J. and {Szabados}, L. and {Szegedi-Elek}, E. and {Tapiador}, D. and {Taris}, F. and {Tauran}, G. and {Taylor}, M.~B. and {Teixeira}, R. and {Terrett}, D. and {Tingley}, B. and {Trager}, S.~C. and {Turon}, C. and {Ulla}, A. and {Utrilla}, E. and {Valentini}, G. and {van Elteren}, A. and {Van Hemelryck}, E. and {van Leeuwen}, M. and {Varadi}, M. and {Vecchiato}, A. and {Veljanoski}, J. and {Via}, T. and {Vicente}, D. and {Vogt}, S. and {Voss}, H. and {Votruba}, V. and {Voutsinas}, S. and {Walmsley}, G. and {Weiler}, M. and {Weingrill}, K. and {Werner}, D. and {Wevers}, T. and {Whitehead}, G. and {Wyrzykowski}, {\L}. and {Yoldas}, A. and {{\v{Z}}erjal}, M. and {Zucker}, S. and {Zurbach}, C. and {Zwitter}, T. and {Alecu}, A. and {Allen}, M. and {Allende Prieto}, C. and {Amorim}, A. and {Anglada-Escud{\'e}}, G. and {Arsenijevic}, V. and {Azaz}, S. and {Balm}, P. and {Beck}, M. and {Bernstein}, H. -H. and {Bigot}, L. and {Bijaoui}, A. and {Blasco}, C. and {Bonfigli}, M. and {Bono}, G. and {Boudreault}, S. and {Bressan}, A. and {Brown}, S. and {Brunet}, P. -M. and {Bunclark}, P. and {Buonanno}, R. and {Butkevich}, A.~G. and {Carret}, C. and {Carrion}, C. and {Chemin}, L. and {Ch{\'e}reau}, F. and {Corcione}, L. and {Darmigny}, E. and {de Boer}, K.~S. and {de Teodoro}, P. and {de Zeeuw}, P.~T. and {Delle Luche}, C. and {Domingues}, C.~D. and {Dubath}, P. and {Fodor}, F. and {Fr{\'e}zouls}, B. and {Fries}, A. and {Fustes}, D. and {Fyfe}, D. and {Gallardo}, E. and {Gallegos}, J. and {Gardiol}, D. and {Gebran}, M. and {Gomboc}, A. and {G{\'o}mez}, A. and {Grux}, E. and {Gueguen}, A. and {Heyrovsky}, A. and {Hoar}, J. and {Iannicola}, G. and {Isasi Parache}, Y. and {Janotto}, A. -M. and {Joliet}, E. and {Jonckheere}, A. and {Keil}, R. and {Kim}, D. -W. and {Klagyivik}, P. and {Klar}, J. and {Knude}, J. and {Kochukhov}, O. and {Kolka}, I. and {Kos}, J. and {Kutka}, A. and {Lainey}, V. and {LeBouquin}, D. and {Liu}, C. and {Loreggia}, D. and {Makarov}, V.~V. and {Marseille}, M.~G. and {Martayan}, C. and {Martinez-Rubi}, O. and {Massart}, B. and {Meynadier}, F. and {Mignot}, S. and {Munari}, U. and {Nguyen}, A. -T. and {Nordlander}, T. and {Ocvirk}, P. and {O'Flaherty}, K.~S. and {Olias Sanz}, A. and {Ortiz}, P. and {Osorio}, J. and {Oszkiewicz}, D. and {Ouzounis}, A. and {Palmer}, M. and {Park}, P. and {Pasquato}, E. and {Peltzer}, C. and {Peralta}, J. and {P{\'e}turaud}, F. and {Pieniluoma}, T. and {Pigozzi}, E. and {Poels}, J. and {Prat}, G. and {Prod'homme}, T. and {Raison}, F. and {Rebordao}, J.~M. and {Risquez}, D. and {Rocca-Volmerange}, B. and {Rosen}, S. and {Ruiz-Fuertes}, M.~I. and {Russo}, F. and {Sembay}, S. and {Serraller Vizcaino}, I. and {Short}, A. and {Siebert}, A. and {Silva}, H. and {Sinachopoulos}, D. and {Slezak}, E. and {Soffel}, M. and {Sosnowska}, D. and {Strai{\v{z}}ys}, V. and {ter Linden}, M. and {Terrell}, D. and {Theil}, S. and {Tiede}, C. and {Troisi}, L. and {Tsalmantza}, P. and {Tur}, D. and {Vaccari}, M. and {Vachier}, F. and {Valles}, P. and {Van Hamme}, W. and {Veltz}, L. and {Virtanen}, J. and {Wallut}, J. -M. and {Wichmann}, R. and {Wilkinson}, M.~I. and {Ziaeepour}, H. and {Zschocke}, S.},
	title = "{The Gaia mission}",
	journal = {Astron. Astrophys.},
	year = 2016,
	volume = {595},
	eid = {A1},
	pages = {A1},
	doi = {10.1051/0004-6361/201629272},
	archivePrefix = {arXiv},
	eprint = {1609.04153},
	primaryClass = {astro-ph.IM},
	adsurl = {https://ui.adsabs.harvard.edu/abs/2016A&A...595A...1G}
}

@ARTICLE{Fritz2018,
	author = {{Fritz}, T.~K. and {Battaglia}, G. and {Pawlowski}, M.~S. and {Kallivayalil}, N. and {van der Marel}, R. and {Sohn}, S.~T. and {Brook}, C. and {Besla}, G.},
	title = "{Gaia DR2 proper motions of dwarf galaxies within 420 kpc. Orbits, Milky Way mass, tidal influences, planar alignments, and group infall}",
	journal = {Astron. Astrophys.},
	year = 2018,
	volume = {619},
	eid = {A103},
	pages = {A103},
	doi = {10.1051/0004-6361/201833343},
	archivePrefix = {arXiv},
	eprint = {1805.00908},
	primaryClass = {astro-ph.GA},
	adsurl = {https://ui.adsabs.harvard.edu/abs/2018A&A...619A.103F}
}

@ARTICLE{Pace2022,
	author = {{Pace}, Andrew B. and {Erkal}, Denis and {Li}, Ting S.},
	title = "{Proper Motions, Orbits, and Tidal Influences of Milky Way Dwarf Spheroidal Galaxies}",
	journal = {\apj},
	year = 2022,
	month = dec,
	volume = {940},
	number = {2},
	eid = {136},
	pages = {136},
	doi = {10.3847/1538-4357/ac997b},
	archivePrefix = {arXiv},
	eprint = {2205.05699},
	primaryClass = {astro-ph.GA},
	adsurl = {https://ui.adsabs.harvard.edu/abs/2022ApJ...940..136P}
}

@ARTICLE{Battaglia2022,
	author = {{Battaglia}, G. and {Taibi}, S. and {Thomas}, G.~F. and {Fritz}, T.~K.},
	title = "{Gaia early DR3 systemic motions of Local Group dwarf galaxies and orbital properties with a massive Large Magellanic Cloud}",
	journal = {Astron. Astrophys.},
	year = 2022,
	volume = {657},
	eid = {A54},
	pages = {A54},
	doi = {10.1051/0004-6361/202141528},
	archivePrefix = {arXiv},
	eprint = {2106.08819},
	primaryClass = {astro-ph.GA},
	adsurl = {https://ui.adsabs.harvard.edu/abs/2022A&A...657A..54B}
}

@ARTICLE{Jiang2021,
	author = {{Jiang}, Fangzhou and {Dekel}, Avishai and {Freundlich}, Jonathan and {van den Bosch}, Frank C. and {Green}, Sheridan B. and {Hopkins}, Philip F. and {Benson}, Andrew and {Du}, Xiaolong},
	title = "{SatGen: a semi-analytical satellite galaxy generator - I. The model and its application to Local-Group satellite statistics}",
	journal = {\mnras},
	year = 2021,
	month = mar,
	volume = {502},
	number = {1},
	pages = {621-641},
	doi = {10.1093/mnras/staa4034},
	archivePrefix = {arXiv},
	eprint = {2005.05974},
	primaryClass = {astro-ph.GA},
	adsurl = {https://ui.adsabs.harvard.edu/abs/2021MNRAS.502..621J}
}

@ARTICLE{Jiang2016,
   author = {{Jiang}, F. and {van den Bosch}, F.~C.},
    title = "{Statistics of dark matter substructure - I. Model and universal fitting functions}",
  journal = {\mnras},
     year = 2016,
    month = may,
   volume = 458,
    pages = {2848-2869},
      doi = {10.1093/mnras/stw439},
   adsurl = {http://adsabs.harvard.edu/abs/2016MNRAS.458.2848J}
}

@BOOK{Binney2008,
	author = {{Binney}, James and {Tremaine}, Scott},
	title = "{Galactic Dynamics: Second Edition}",
	address = {Princeton, N.J},
	isbn = {9780691130262},
	publisher = {Princeton University Press},
	year = {2008}
}

@ARTICLE{Schwarzschild1979,
	author = {{Schwarzschild}, M.},
	title = "{A numerical model for a triaxial stellar system in dynamical equilibrium.}",
	journal = {\apj},
	year = 1979,
	month = aug,
	volume = {232},
	pages = {236-247},
	doi = {10.1086/157282},
	adsurl = {https://ui.adsabs.harvard.edu/abs/1979ApJ...232..236S}
}

@ARTICLE{Chiang2025,
       author = {{Chiang}, Barry T. and {van den Bosch}, Frank C. and {Schive}, Hsi-Yu},
        title = "{The tidal evolution of anisotropic subhaloes: a new pathway to creating isotropic and cored satellites}",
      journal = {\mnras},
         year = 2025,
        month = nov,
       volume = {544},
       number = {1},
        pages = {36-52},
          doi = {10.1093/mnras/staf1639},
archivePrefix = {arXiv},
       eprint = {2411.03192},
 primaryClass = {astro-ph.GA},
       adsurl = {https://ui.adsabs.harvard.edu/abs/2025MNRAS.544...36C}
}

@ARTICLE{Chiang2026a,
       author = {{Chiang}, Barry T. and {van den Bosch}, Frank C. and {Schive}, Hsi-Yu},
        title = "{Universal numerical convergence criteria for subhalo tidal evolution}",
      journal = {The Open Journal of Astrophysics},
         year = 2026,
        month = jan,
       volume = {9},
        pages = {55367},
          doi = {10.33232/001c.155367},
archivePrefix = {arXiv},
       eprint = {2510.26901},
 primaryClass = {astro-ph.CO},
       adsurl = {https://ui.adsabs.harvard.edu/abs/2026OJAp....955367C}
}

@ARTICLE{Benettin1976,
       author = {{Benettin}, Giancarlo and {Galgani}, Luigi and {Strelcyn}, Jean-Marie},
        title = "{Kolmogorov entropy and numerical experiments}",
      journal = {\pra},
         year = 1976,
        month = dec,
       volume = {14},
       number = {6},
        pages = {2338-2345},
          doi = {10.1103/PhysRevA.14.2338},
       adsurl = {https://ui.adsabs.harvard.edu/abs/1976PhRvA..14.2338B}
}

@ARTICLE{Merritt1996,
       author = {{Merritt}, David and {Valluri}, Monica},
        title = "{Chaos and Mixing in Triaxial Stellar Systems}",
      journal = {\apj},
         year = 1996,
        month = nov,
       volume = {471},
        pages = {82},
          doi = {10.1086/177955},
archivePrefix = {arXiv},
       eprint = {astro-ph/9602079},
 primaryClass = {astro-ph},
       adsurl = {https://ui.adsabs.harvard.edu/abs/1996ApJ...471...82M}
}

@ARTICLE{Valluri1998,
       author = {{Valluri}, Monica and {Merritt}, David},
        title = "{Regular and Chaotic Dynamics of Triaxial Stellar Systems}",
      journal = {\apj},
         year = 1998,
        month = oct,
       volume = {506},
       number = {2},
        pages = {686-711},
          doi = {10.1086/306269},
archivePrefix = {arXiv},
       eprint = {astro-ph/9801041},
 primaryClass = {astro-ph},
       adsurl = {https://ui.adsabs.harvard.edu/abs/1998ApJ...506..686V}
}

@ARTICLE{Benson2012,
       author = {{Benson}, Andrew J.},
        title = "{G ALACTICUS: A semi-analytic model of galaxy formation}",
      journal = {\na},
         year = 2012,
        month = feb,
       volume = {17},
       number = {2},
        pages = {175-197},
          doi = {10.1016/j.newast.2011.07.004},
archivePrefix = {arXiv},
       eprint = {1008.1786},
 primaryClass = {astro-ph.CO},
       adsurl = {https://ui.adsabs.harvard.edu/abs/2012NewA...17..175B}
}

@ARTICLE{Du2024,
       author = {{Du}, Xiaolong and {Benson}, Andrew and {Zeng}, Zhichao Carton and {Treu}, Tommaso and {Peter}, Annika H.~G. and {Mace}, Charlie and {Jiang}, Fangzhou and {Yang}, Shengqi and {Gannon}, Charles and {Gilman}, Daniel and {Nierenberg}, Anna. M. and {Nadler}, Ethan O.},
        title = "{Tidal evolution of cored and cuspy dark matter halos}",
      journal = {\prd},
         year = 2024,
        month = jul,
       volume = {110},
       number = {2},
          eid = {023019},
        pages = {023019},
          doi = {10.1103/PhysRevD.110.023019},
archivePrefix = {arXiv},
       eprint = {2403.09597},
 primaryClass = {astro-ph.GA},
       adsurl = {https://ui.adsabs.harvard.edu/abs/2024PhRvD.110b3019D}
}

@ARTICLE{Miller2020,
       author = {{Miller}, Tim B. and {van den Bosch}, Frank C. and {Green}, Sheridan B. and {Ogiya}, Go},
        title = "{Dynamical self-friction: how mass loss slows you down}",
      journal = {\mnras},
         year = 2020,
        month = jul,
       volume = {495},
       number = {4},
        pages = {4496-4507},
          doi = {10.1093/mnras/staa1450},
archivePrefix = {arXiv},
       eprint = {2001.06489},
 primaryClass = {astro-ph.GA},
       adsurl = {https://ui.adsabs.harvard.edu/abs/2020MNRAS.495.4496M}
}

@ARTICLE{Klypin2016,
       author = {{Klypin}, Anatoly and {Yepes}, Gustavo and {Gottl{\"o}ber}, Stefan and {Prada}, Francisco and {He{\ss}}, Steffen},
        title = "{MultiDark simulations: the story of dark matter halo concentrations and density profiles}",
      journal = {\mnras},
         year = 2016,
        month = apr,
       volume = {457},
       number = {4},
        pages = {4340-4359},
          doi = {10.1093/mnras/stw248},
archivePrefix = {arXiv},
       eprint = {1411.4001},
 primaryClass = {astro-ph.CO},
       adsurl = {https://ui.adsabs.harvard.edu/abs/2016MNRAS.457.4340K}
}

@ARTICLE{Gao2004,
       author = {{Gao}, L. and {White}, S.~D.~M. and {Jenkins}, A. and {Stoehr}, F. and {Springel}, V.},
        title = "{The subhalo populations of {\ensuremath{\Lambda}}CDM dark haloes}",
      journal = {\mnras},
         year = 2004,
        month = dec,
       volume = {355},
       number = {3},
        pages = {819-834},
          doi = {10.1111/j.1365-2966.2004.08360.x},
archivePrefix = {arXiv},
       eprint = {astro-ph/0404589},
 primaryClass = {astro-ph},
       adsurl = {https://ui.adsabs.harvard.edu/abs/2004MNRAS.355..819G}
}

@ARTICLE{Grand2017,
       author = {{Grand}, Robert J.~J. and {G{\'o}mez}, Facundo A. and {Marinacci}, Federico and {Pakmor}, R{\"u}diger and {Springel}, Volker and {Campbell}, David J.~R. and {Frenk}, Carlos S. and {Jenkins}, Adrian and {White}, Simon D.~M.},
        title = "{The Auriga Project: the properties and formation mechanisms of disc galaxies across cosmic time}",
      journal = {\mnras},
         year = 2017,
        month = may,
       volume = {467},
       number = {1},
        pages = {179-207},
          doi = {10.1093/mnras/stx071},
archivePrefix = {arXiv},
       eprint = {1610.01159},
 primaryClass = {astro-ph.GA},
       adsurl = {https://ui.adsabs.harvard.edu/abs/2017MNRAS.467..179G}
}

@ARTICLE{Vogelsberger2014,
       author = {{Vogelsberger}, Mark and {Genel}, Shy and {Springel}, Volker and {Torrey}, Paul and {Sijacki}, Debora and {Xu}, Dandan and {Snyder}, Greg and {Nelson}, Dylan and {Hernquist}, Lars},
        title = "{Introducing the Illustris Project: simulating the coevolution of dark and visible matter in the Universe}",
      journal = {\mnras},
         year = 2014,
        month = oct,
       volume = {444},
       number = {2},
        pages = {1518-1547},
          doi = {10.1093/mnras/stu1536},
archivePrefix = {arXiv},
       eprint = {1405.2921},
 primaryClass = {astro-ph.CO},
       adsurl = {https://ui.adsabs.harvard.edu/abs/2014MNRAS.444.1518V}
}

@ARTICLE{Pillepich2019,
       author = {{Pillepich}, Annalisa and {Nelson}, Dylan and {Springel}, Volker and {Pakmor}, R{\"u}diger and {Torrey}, Paul and {Weinberger}, Rainer and {Vogelsberger}, Mark and {Marinacci}, Federico and {Genel}, Shy and {van der Wel}, Arjen and {Hernquist}, Lars},
        title = "{First results from the TNG50 simulation: the evolution of stellar and gaseous discs across cosmic time}",
      journal = {\mnras},
         year = 2019,
        month = dec,
       volume = {490},
       number = {3},
        pages = {3196-3233},
          doi = {10.1093/mnras/stz2338},
archivePrefix = {arXiv},
       eprint = {1902.05553},
 primaryClass = {astro-ph.GA},
       adsurl = {https://ui.adsabs.harvard.edu/abs/2019MNRAS.490.3196P}
}

@ARTICLE{vandenBosch2018b,
       author = {{van den Bosch}, Frank C. and {Ogiya}, Go},
        title = "{Dark matter substructure in numerical simulations: a tale of discreteness noise, runaway instabilities, and artificial disruption}",
      journal = {\mnras},
         year = 2018,
        month = apr,
       volume = {475},
       number = {3},
        pages = {4066-4087},
          doi = {10.1093/mnras/sty084},
archivePrefix = {arXiv},
       eprint = {1801.05427},
 primaryClass = {astro-ph.GA},
       adsurl = {https://ui.adsabs.harvard.edu/abs/2018MNRAS.475.4066V}
}

@ARTICLE{McMillan2017,
       author = {{McMillan}, Paul J.},
        title = "{The mass distribution and gravitational potential of the Milky Way}",
      journal = {\mnras},
         year = 2017,
        month = feb,
       volume = {465},
       number = {1},
        pages = {76-94},
          doi = {10.1093/mnras/stw2759},
archivePrefix = {arXiv},
       eprint = {1608.00971},
 primaryClass = {astro-ph.GA},
       adsurl = {https://ui.adsabs.harvard.edu/abs/2017MNRAS.465...76M}
}

@ARTICLE{GaiaDR3,
       author = {{Gaia Collaboration} and {Vallenari}, A. and {Brown}, A.~G.~A. and {Prusti}, T. and {de Bruijne}, J.~H.~J. and {Arenou}, F. and {Babusiaux}, C. and {Biermann}, M. and {Creevey}, O.~L. and {Ducourant}, C. and {Evans}, D.~W. and {Eyer}, L. and {Guerra}, R. and {Hutton}, A. and {Jordi}, C. and {Klioner}, S.~A. and {Lammers}, U.~L. and {Lindegren}, L. and {Luri}, X. and {Mignard}, F. and {Panem}, C. and {Pourbaix}, D. and {Randich}, S. and {Sartoretti}, P. and {Soubiran}, C. and {Tanga}, P. and {Walton}, N.~A. and {Bailer-Jones}, C.~A.~L. and {Bastian}, U. and {Drimmel}, R. and {Jansen}, F. and {Katz}, D. and {Lattanzi}, M.~G. and {van Leeuwen}, F. and {Bakker}, J. and {Cacciari}, C. and {Casta{\~n}eda}, J. and {De Angeli}, F. and {Fabricius}, C. and {Fouesneau}, M. and {Fr{\'e}mat}, Y. and {Galluccio}, L. and {Guerrier}, A. and {Heiter}, U. and {Masana}, E. and {Messineo}, R. and {Mowlavi}, N. and {Nicolas}, C. and {Nienartowicz}, K. and {Pailler}, F. and {Panuzzo}, P. and {Riclet}, F. and {Roux}, W. and {Seabroke}, G.~M. and {Sordo}, R. and {Th{\'e}venin}, F. and {Gracia-Abril}, G. and {Portell}, J. and {Teyssier}, D. and {Altmann}, M. and {Andrae}, R. and {Audard}, M. and {Bellas-Velidis}, I. and {Benson}, K. and {Berthier}, J. and {Blomme}, R. and {Burgess}, P.~W. and {Busonero}, D. and {Busso}, G. and {C{\'a}novas}, H. and {Carry}, B. and {Cellino}, A. and {Cheek}, N. and {Clementini}, G. and {Damerdji}, Y. and {Davidson}, M. and {de Teodoro}, P. and {Nu{\~n}ez Campos}, M. and {Delchambre}, L. and {Dell'Oro}, A. and {Esquej}, P. and {Fern{\'a}ndez-Hern{\'a}ndez}, J. and {Fraile}, E. and {Garabato}, D. and {Garc{\'\i}a-Lario}, P. and {Gosset}, E. and {Haigron}, R. and {Halbwachs}, J.-L. and {Hambly}, N.~C. and {Harrison}, D.~L. and {Hern{\'a}ndez}, J. and {Hestroffer}, D. and {Hodgkin}, S.~T. and {Holl}, B. and {Jan{\ss}en}, K. and {Jevardat de Fombelle}, G. and {Jordan}, S. and {Krone-Martins}, A. and {Lanzafame}, A.~C. and {L{\"o}ffler}, W. and {Marchal}, O. and {Marrese}, P.~M. and {Moitinho}, A. and {Muinonen}, K. and {Osborne}, P. and {Pancino}, E. and {Pauwels}, T. and {Recio-Blanco}, A. and {Reyl{\'e}}, C. and {Riello}, M. and {Rimoldini}, L. and {Roegiers}, T. and {Rybizki}, J. and {Sarro}, L.~M. and {Siopis}, C. and {Smith}, M. and {Sozzetti}, A. and {Utrilla}, E. and {van Leeuwen}, M. and {Abbas}, U. and {{\'A}brah{\'a}m}, P. and {Abreu Aramburu}, A. and {Aerts}, C. and {Aguado}, J.~J. and {Ajaj}, M. and {Aldea-Montero}, F. and {Altavilla}, G. and {{\'A}lvarez}, M.~A. and {Alves}, J. and {Anders}, F. and {Anderson}, R.~I. and {Anglada Varela}, E. and {Antoja}, T. and {Baines}, D. and {Baker}, S.~G. and {Balaguer-N{\'u}{\~n}ez}, L. and {Balbinot}, E. and {Balog}, Z. and {Barache}, C. and {Barbato}, D. and {Barros}, M. and {Barstow}, M.~A. and {Bartolom{\'e}}, S. and {Bassilana}, J.-L. and {Bauchet}, N. and {Becciani}, U. and {Bellazzini}, M. and {Berihuete}, A. and {Bernet}, M. and {Bertone}, S. and {Bianchi}, L. and {Binnenfeld}, A. and {Blanco-Cuaresma}, S. and {Blazere}, A. and {Boch}, T. and {Bombrun}, A. and {Bossini}, D. and {Bouquillon}, S. and {Bragaglia}, A. and {Bramante}, L. and {Breedt}, E. and {Bressan}, A. and {Brouillet}, N. and {Brugaletta}, E. and {Bucciarelli}, B. and {Burlacu}, A. and {Butkevich}, A.~G. and {Buzzi}, R. and {Caffau}, E. and {Cancelliere}, R. and {Cantat-Gaudin}, T. and {Carballo}, R. and {Carlucci}, T. and {Carnerero}, M.~I. and {Carrasco}, J.~M. and {Casamiquela}, L. and {Castellani}, M. and {Castro-Ginard}, A. and {Chaoul}, L. and {Charlot}, P. and {Chemin}, L. and {Chiaramida}, V. and {Chiavassa}, A. and {Chornay}, N. and {Comoretto}, G. and {Contursi}, G. and {Cooper}, W.~J. and {Cornez}, T. and {Cowell}, S. and {Crifo}, F. and {Cropper}, M. and {Crosta}, M. and {Crowley}, C. and {Dafonte}, C. and {Dapergolas}, A. and {David}, M. and {David}, P. and {de Laverny}, P. and {De Luise}, F. and {De March}, R.},
        title = "{Gaia Data Release 3. Summary of the content and survey properties}",
      journal = {\aap},
         year = 2023,
        month = jun,
       volume = {674},
          eid = {A1},
        pages = {A1},
          doi = {10.1051/0004-6361/202243940},
archivePrefix = {arXiv},
       eprint = {2208.00211},
 primaryClass = {astro-ph.GA},
       adsurl = {https://ui.adsabs.harvard.edu/abs/2023A&A...674A...1G}
}

@ARTICLE{McKinnon2026,
       author = {{McKinnon}, Kevin A. and {van der Marel}, Roeland P.},
        title = "{Simulating Roman+Gaia Combined Astrometry, Parallaxes, and Proper Motions}",
      journal = {\pasp},
         year = 2026,
        month = apr,
       volume = {138},
       number = {4},
          eid = {044507},
        pages = {044507},
          doi = {10.1088/1538-3873/ae5a73},
archivePrefix = {arXiv},
       eprint = {2602.00310},
 primaryClass = {astro-ph.IM},
       adsurl = {https://ui.adsabs.harvard.edu/abs/2026PASP..138d4507M}
}

@ARTICLE{Vasiliev2021,
       author = {{Vasiliev}, Eugene and {Belokurov}, Vasily and {Erkal}, Denis},
        title = "{Tango for three: Sagittarius, LMC, and the Milky Way}",
      journal = {\mnras},
         year = 2021,
        month = feb,
       volume = {501},
       number = {2},
        pages = {2279-2304},
          doi = {10.1093/mnras/staa3673},
archivePrefix = {arXiv},
       eprint = {2009.10726},
 primaryClass = {astro-ph.GA},
       adsurl = {https://ui.adsabs.harvard.edu/abs/2021MNRAS.501.2279V}
}

@ARTICLE{Petersen2021,
       author = {{Petersen}, Michael S. and {Pe{\~n}arrubia}, Jorge},
        title = "{Detection of the Milky Way reflex motion due to the Large Magellanic Cloud infall}",
      journal = {Nature Astronomy},
         year = 2021,
        month = jan,
       volume = {5},
        pages = {251-255},
          doi = {10.1038/s41550-020-01254-3},
archivePrefix = {arXiv},
       eprint = {2011.10581},
 primaryClass = {astro-ph.GA},
       adsurl = {https://ui.adsabs.harvard.edu/abs/2021NatAs...5..251P}
}

@ARTICLE{JimenezArranz2023,
       author = {{Jim{\'e}nez-Arranz}, {\'O}. and {Romero-G{\'o}mez}, M. and {Luri}, X. and {McMillan}, P.~J. and {Antoja}, T. and {Chemin}, L. and {Roca-F{\`a}brega}, S. and {Masana}, E. and {Muros}, A.},
        title = "{Kinematic analysis of the Large Magellanic Cloud using Gaia DR3}",
      journal = {\aap},
         year = 2023,
        month = jan,
       volume = {669},
          eid = {A91},
        pages = {A91},
          doi = {10.1051/0004-6361/202244601},
archivePrefix = {arXiv},
       eprint = {2210.01728},
 primaryClass = {astro-ph.GA},
       adsurl = {https://ui.adsabs.harvard.edu/abs/2023A&A...669A..91J}
}

@ARTICLE{Erkal2019,
       author = {{Erkal}, D. and {Belokurov}, V. and {Laporte}, C.~F.~P. and {Koposov}, S.~E. and {Li}, T.~S. and {Grillmair}, C.~J. and {Kallivayalil}, N. and {Price-Whelan}, A.~M. and {Evans}, N.~W. and {Hawkins}, K. and {Hendel}, D. and {Mateu}, C. and {Navarro}, J.~F. and {del Pino}, A. and {Slater}, C.~T. and {Sohn}, S.~T. and {Orphan Aspen Treasury Collaboration}},
        title = "{The total mass of the Large Magellanic Cloud from its perturbation on the Orphan stream}",
      journal = {\mnras},
         year = 2019,
        month = aug,
       volume = {487},
       number = {2},
        pages = {2685-2700},
          doi = {10.1093/mnras/stz1371},
archivePrefix = {arXiv},
       eprint = {1812.08192},
 primaryClass = {astro-ph.GA},
       adsurl = {https://ui.adsabs.harvard.edu/abs/2019MNRAS.487.2685E}
}

@ARTICLE{numpy,
       author = {{Harris}, Charles R. and {Millman}, K. Jarrod and {van der Walt}, St{\'e}fan J. and {Gommers}, Ralf and {Virtanen}, Pauli and {Cournapeau}, David and {Wieser}, Eric and {Taylor}, Julian and {Berg}, Sebastian and {Smith}, Nathaniel J. and {Kern}, Robert and {Picus}, Matti and {Hoyer}, Stephan and {van Kerkwijk}, Marten H. and {Brett}, Matthew and {Haldane}, Allan and {del R{\'\i}o}, Jaime Fern{\'a}ndez and {Wiebe}, Mark and {Peterson}, Pearu and {G{\'e}rard-Marchant}, Pierre and {Sheppard}, Kevin and {Reddy}, Tyler and {Weckesser}, Warren and {Abbasi}, Hameer and {Gohlke}, Christoph and {Oliphant}, Travis E.},
        title = "{Array programming with NumPy}",
      journal = {\nat},
         year = 2020,
        month = sep,
       volume = {585},
       number = {7825},
        pages = {357-362},
          doi = {10.1038/s41586-020-2649-2},
archivePrefix = {arXiv},
       eprint = {2006.10256},
 primaryClass = {cs.MS},
       adsurl = {https://ui.adsabs.harvard.edu/abs/2020Natur.585..357H}
}

@ARTICLE{scipy,
       author = {{Virtanen}, Pauli and {Gommers}, Ralf and {Oliphant}, Travis E. and {Haberland}, Matt and {Reddy}, Tyler and {Cournapeau}, David and {Burovski}, Evgeni and {Peterson}, Pearu and {Weckesser}, Warren and {Bright}, Jonathan and {van der Walt}, St{\'e}fan J. and {Brett}, Matthew and {Wilson}, Joshua and {Millman}, K. Jarrod and {Mayorov}, Nikolay and {Nelson}, Andrew R.~J. and {Jones}, Eric and {Kern}, Robert and {Larson}, Eric and {Carey}, C.~J. and {Polat}, {\.I}lhan and {Feng}, Yu and {Moore}, Eric W. and {VanderPlas}, Jake and {Laxalde}, Denis and {Perktold}, Josef and {Cimrman}, Robert and {Henriksen}, Ian and {Quintero}, E.~A. and {Harris}, Charles R. and {Archibald}, Anne M. and {Ribeiro}, Ant{\^o}nio H. and {Pedregosa}, Fabian and {van Mulbregt}, Paul and {SciPy 1. 0 Contributors}},
        title = "{SciPy 1.0: fundamental algorithms for scientific computing in Python}",
      journal = {Nature Methods},
         year = 2020,
        month = feb,
       volume = {17},
        pages = {261-272},
          doi = {10.1038/s41592-019-0686-2},
archivePrefix = {arXiv},
       eprint = {1907.10121},
 primaryClass = {cs.MS},
       adsurl = {https://ui.adsabs.harvard.edu/abs/2020NatMe..17..261V}
}

@ARTICLE{matplotlib,
       author = {{Hunter}, John D.},
        title = "{Matplotlib: A 2D Graphics Environment}",
      journal = {Computing in Science and Engineering},
         year = 2007,
        month = may,
       volume = {9},
       number = {3},
        pages = {90-95},
          doi = {10.1109/MCSE.2007.55},
       adsurl = {https://ui.adsabs.harvard.edu/abs/2007CSE.....9...90H}
}

@ARTICLE{Stucker2023,
       author = {{St{\"u}cker}, Jens and {Ogiya}, Go and {Angulo}, Raul E. and {Aguirre-Santaella}, Alejandra and {S{\'a}nchez-Conde}, Miguel A.},
        title = "{Tidal stripping in the adiabatic limit}",
      journal = {\mnras},
         year = 2023,
        month = may,
       volume = {521},
       number = {3},
        pages = {4432-4461},
          doi = {10.1093/mnras/stad844},
archivePrefix = {arXiv},
       eprint = {2207.00604},
 primaryClass = {astro-ph.CO},
       adsurl = {https://ui.adsabs.harvard.edu/abs/2023MNRAS.521.4432S}
}

@ARTICLE{Gerhard1985,
       author = {{Gerhard}, O.~E. and {Binney}, J.},
        title = "{Triaxial galaxies containing massive black holes or central density cusps}",
      journal = {\mnras},
         year = 1985,
        month = sep,
       volume = {216},
        pages = {467-502},
          doi = {10.1093/mnras/216.2.467},
       adsurl = {https://ui.adsabs.harvard.edu/abs/1985MNRAS.216..467G}
}

@ARTICLE{Udry1988,
       author = {{Udry}, S. and {Pfenniger}, D.},
        title = "{Stochasticity in elliptical galaxies}",
      journal = {\aap},
         year = 1988,
        month = jun,
       volume = {198},
       number = {1-2},
        pages = {135-149},
       adsurl = {https://ui.adsabs.harvard.edu/abs/1988A&A...198..135U}
}

@ARTICLE{Valluri2010,
       author = {{Valluri}, Monica and {Debattista}, Victor P. and {Quinn}, Thomas and {Moore}, Ben},
        title = "{The orbital evolution induced by baryonic condensation in triaxial haloes}",
      journal = {\mnras},
         year = 2010,
        month = mar,
       volume = {403},
       number = {1},
        pages = {525-544},
          doi = {10.1111/j.1365-2966.2009.16192.x},
archivePrefix = {arXiv},
       eprint = {0906.4784},
 primaryClass = {astro-ph.CO},
       adsurl = {https://ui.adsabs.harvard.edu/abs/2010MNRAS.403..525V}
}

@ARTICLE{Maffione2015,
       author = {{Maffione}, N.~P. and {G{\'o}mez}, F.~A. and {Cincotta}, P.~M. and {Giordano}, C.~M. and {Cooper}, A.~P. and {O'Shea}, B.~W.},
        title = "{On the relevance of chaos for halo stars in the solar neighbourhood}",
      journal = {\mnras},
         year = 2015,
        month = nov,
       volume = {453},
       number = {3},
        pages = {2830-2847},
          doi = {10.1093/mnras/stv1778},
archivePrefix = {arXiv},
       eprint = {1508.00579},
 primaryClass = {astro-ph.GA},
       adsurl = {https://ui.adsabs.harvard.edu/abs/2015MNRAS.453.2830M}
}

@ARTICLE{PriceWhelan2016,
       author = {{Price-Whelan}, Adrian M. and {Johnston}, Kathryn V. and {Valluri}, Monica and {Pearson}, Sarah and {K{\"u}pper}, Andreas H.~W. and {Hogg}, David W.},
        title = "{Chaotic dispersal of tidal debris}",
      journal = {\mnras},
         year = 2016,
        month = jan,
       volume = {455},
       number = {1},
        pages = {1079-1098},
          doi = {10.1093/mnras/stv2383},
archivePrefix = {arXiv},
       eprint = {1507.08662},
 primaryClass = {astro-ph.GA},
       adsurl = {https://ui.adsabs.harvard.edu/abs/2016MNRAS.455.1079P}
}

@ARTICLE{Mestre2020,
       author = {{Mestre}, Mart{\'\i}n and {Llinares}, Claudio and {Carpintero}, Daniel D.},
        title = "{Effects of chaos on the detectability of stellar streams}",
      journal = {\mnras},
         year = 2020,
        month = mar,
       volume = {492},
       number = {3},
        pages = {4398-4408},
          doi = {10.1093/mnras/stz3505},
archivePrefix = {arXiv},
       eprint = {1912.05592},
 primaryClass = {astro-ph.GA},
       adsurl = {https://ui.adsabs.harvard.edu/abs/2020MNRAS.492.4398M}
}

@ARTICLE{GaravitoCamargo2019,
       author = {{Garavito-Camargo}, Nicolas and {Besla}, Gurtina and {Laporte}, Chervin F.~P. and {Johnston}, Kathryn V. and {G{\'o}mez}, Facundo A. and {Watkins}, Laura L.},
        title = "{Hunting for the Dark Matter Wake Induced by the Large Magellanic Cloud}",
      journal = {\apj},
         year = 2019,
        month = oct,
       volume = {884},
       number = {1},
          eid = {51},
        pages = {51},
          doi = {10.3847/1538-4357/ab32eb},
archivePrefix = {arXiv},
       eprint = {1902.05089},
 primaryClass = {astro-ph.GA},
       adsurl = {https://ui.adsabs.harvard.edu/abs/2019ApJ...884...51G}
}

@ARTICLE{Woudenberg2024,
       author = {{Woudenberg}, Hanneke C. and {Helmi}, Amina},
        title = "{First measurement of the triaxiality of the inner dark matter halo of the Milky Way}",
      journal = {\aap},
         year = 2024,
        month = nov,
       volume = {691},
          eid = {A277},
        pages = {A277},
          doi = {10.1051/0004-6361/202451743},
archivePrefix = {arXiv},
       eprint = {2407.21790},
 primaryClass = {astro-ph.GA},
       adsurl = {https://ui.adsabs.harvard.edu/abs/2024A&A...691A.277W}
}

@ARTICLE{DSouza2022,
       author = {{D'Souza}, Richard and {Bell}, Eric F.},
        title = "{Uncertainties associated with the backward integration of dwarf satellites using simple parametric potentials}",
      journal = {\mnras},
         year = 2022,
        month = may,
       volume = {512},
       number = {1},
        pages = {739-760},
          doi = {10.1093/mnras/stac404},
archivePrefix = {arXiv},
       eprint = {2202.05707},
 primaryClass = {astro-ph.GA},
       adsurl = {https://ui.adsabs.harvard.edu/abs/2022MNRAS.512..739D}
}

@ARTICLE{Santistevan2023,
       author = {{Santistevan}, Isaiah B. and {Wetzel}, Andrew and {Tollerud}, Erik and {Sanderson}, Robyn E. and {Samuel}, Jenna},
        title = "{Orbital dynamics and histories of satellite galaxies around Milky Way - mass galaxies in the FIRE simulations}",
      journal = {\mnras},
         year = 2023,
        month = jan,
       volume = {518},
       number = {1},
        pages = {1427-1447},
          doi = {10.1093/mnras/stac3100},
archivePrefix = {arXiv},
       eprint = {2208.05977},
 primaryClass = {astro-ph.GA},
       adsurl = {https://ui.adsabs.harvard.edu/abs/2023MNRAS.518.1427S}
}

@ARTICLE{Nibauer2025,
       author = {{Nibauer}, Jacob and {Bonaca}, Ana},
        title = "{Galactic Accelerations from the GD-1 Stream Suggest a Tilted Dark Matter Halo}",
      journal = {\apjl},
         year = 2025,
        month = may,
       volume = {985},
       number = {1},
          eid = {L22},
        pages = {L22},
          doi = {10.3847/2041-8213/add0a9},
archivePrefix = {arXiv},
       eprint = {2504.07187},
 primaryClass = {astro-ph.GA},
       adsurl = {https://ui.adsabs.harvard.edu/abs/2025ApJ...985L..22N}
}

@ARTICLE{Errani2024,
       author = {{Errani}, Rapha{\"e}l and {Ibata}, Rodrigo and {Navarro}, Julio F. and {Pe{\~n}arrubia}, Jorge and {Walker}, Matthew G.},
        title = "{Microgalaxies in LCDM}",
      journal = {\apj},
         year = 2024,
        month = jun,
       volume = {968},
       number = {2},
          eid = {89},
        pages = {89},
          doi = {10.3847/1538-4357/ad402d},
archivePrefix = {arXiv},
       eprint = {2311.14798},
 primaryClass = {astro-ph.GA},
       adsurl = {https://ui.adsabs.harvard.edu/abs/2024ApJ...968...89E}
}

@ARTICLE{Chemaly2026,
       author = {{Chemaly}, David and {Sola}, Elisabeth and {Koposov}, Sergey and {Zhang}, HanYuan and {Belokurov}, Vasily and {Erkal}, Denis},
        title = "{Constraints on the population level distribution of nearby Dark Matter halo shapes with extragalactic streams}",
      journal = {arXiv e-prints},
         year = 2026,
        month = jul,
          eid = {arXiv:2607.05510},
        pages = {arXiv:2607.05510},
          doi = {10.48550/arXiv.2607.05510},
archivePrefix = {arXiv},
       eprint = {2607.05510},
 primaryClass = {astro-ph.GA},
       adsurl = {https://ui.adsabs.harvard.edu/abs/2026arXiv260705510C}
}

@ARTICLE{Vasiliev2023,
       author = {{Vasiliev}, Eugene},
        title = "{The Effect of the LMC on the Milky Way System}",
      journal = {Galaxies},
         year = 2023,
        month = apr,
       volume = {11},
       number = {2},
          eid = {59},
        pages = {59},
          doi = {10.3390/galaxies11020059},
archivePrefix = {arXiv},
       eprint = {2304.09136},
 primaryClass = {astro-ph.GA},
       adsurl = {https://ui.adsabs.harvard.edu/abs/2023Galax..11...59V}
}

@ARTICLE{Vasiliev2024,
       author = {{Vasiliev}, Eugene},
        title = "{Dear Magellanic Clouds, welcome back!}",
      journal = {\mnras},
         year = 2024,
        month = jan,
       volume = {527},
       number = {1},
        pages = {437-456},
          doi = {10.1093/mnras/stad2612},
archivePrefix = {arXiv},
       eprint = {2306.04837},
 primaryClass = {astro-ph.GA},
       adsurl = {https://ui.adsabs.harvard.edu/abs/2024MNRAS.527..437V}
}

@ARTICLE{Lucchini2025,
       author = {{Lucchini}, Scott and {Han}, Jiwon Jesse and {Mishra}, Sapna and {Fox}, Andrew J.},
        title = "{The LMC Corona Favors a First Passage}",
      journal = {arXiv e-prints},
         year = 2025,
        month = oct,
          eid = {arXiv:2510.03395},
        pages = {arXiv:2510.03395},
          doi = {10.48550/arXiv.2510.03395},
archivePrefix = {arXiv},
       eprint = {2510.03395},
 primaryClass = {astro-ph.GA}
}

@ARTICLE{MartinezGarcia2026,
       author = {{Mart{\'\i}nez-Garc{\'\i}a}, Alberto Manuel and {del Pino}, Andr{\'e}s and {van der Marel}, Roeland P. and {Battaglia}, Giuseppina and {{\L}okas}, Ewa L. and {Vitral}, Eduardo and {McKinnon}, Kevin A. and {Watkins}, Laura L. and {Kallivayalil}, Nitya and {Sohn}, Sangmo Tony and {Thomas}, Guillaume F. and {Cardona-Barrero}, Salvador and {Anguiano}, Borja and {Alzate-Trujillo}, Jairo A. and {Nogueras-Lara}, Francisco and {Bennet}, Paul and {Hidalgo-Pinilla}, Adri{\'a}n},
        title = "{Reconstructing the orbits of Milky Way dwarf galaxies: An LMC perspective}",
      journal = {arXiv e-prints},
         year = 2026,
        month = jun,
          eid = {arXiv:2606.13787},
        pages = {arXiv:2606.13787},
          doi = {10.48550/arXiv.2606.13787},
archivePrefix = {arXiv},
       eprint = {2606.13787},
 primaryClass = {astro-ph.GA},
       adsurl = {https://ui.adsabs.harvard.edu/abs/2026arXiv260613787M}
}

@ARTICLE{Dillamore2026,
       author = {{Dillamore}, Adam M. and {Sanders}, Jason L. and {Brooks}, Richard A.~N.},
        title = "{GSE versus LMC: reshaping of radially biased stellar haloes by satellites}",
      journal = {\mnras},
         year = 2026,
        month = jul,
       volume = {549},
       number = {4},
          eid = {stag1110},
        pages = {stag1110},
          doi = {10.1093/mnras/stag1110},
archivePrefix = {arXiv},
       eprint = {2603.11159},
 primaryClass = {astro-ph.GA},
       adsurl = {https://ui.adsabs.harvard.edu/abs/2026MNRAS.549g1110D}
}

@ARTICLE{Cavieres2025,
       author = {{Cavieres}, Manuel and {Chanam{\'e}}, Julio and {Navarrete}, Camila and {Ordenes-Brice{\~n}o}, Yasna and {Garavito-Camargo}, Nicol{\'a}s and {Besla}, Gurtina and {Hempel}, Maren and {Vivas}, A. Katherina and {G{\'o}mez}, Facundo},
        title = "{The Distant Milky Way Halo from the Southern Hemisphere: Characterization of the LMC-induced Dynamical Friction Wake}",
      journal = {\apj},
         year = 2025,
        month = apr,
       volume = {983},
       number = {1},
          eid = {83},
        pages = {83},
          doi = {10.3847/1538-4357/adbf08},
archivePrefix = {arXiv},
       eprint = {2410.00114},
 primaryClass = {astro-ph.GA},
       adsurl = {https://ui.adsabs.harvard.edu/abs/2025ApJ...983...83C}
}

@ARTICLE{Kallivayalil2013,
       author = {{Kallivayalil}, Nitya and {van der Marel}, Roeland P. and {Besla}, Gurtina and {Anderson}, Jay and {Alcock}, Charles},
        title = "{Third-epoch Magellanic Cloud Proper Motions. I. Hubble Space Telescope/WFC3 Data and Orbit Implications}",
      journal = {\apj},
         year = 2013,
        month = feb,
       volume = {764},
       number = {2},
          eid = {161},
        pages = {161},
          doi = {10.1088/0004-637X/764/2/161},
archivePrefix = {arXiv},
       eprint = {1301.0832},
 primaryClass = {astro-ph.CO},
       adsurl = {https://ui.adsabs.harvard.edu/abs/2013ApJ...764..161K}
}

@ARTICLE{Patel2020,
       author = {{Patel}, Ekta and {Kallivayalil}, Nitya and {Garavito-Camargo}, Nicolas and {Besla}, Gurtina and {Weisz}, Daniel R. and {van der Marel}, Roeland P. and {Boylan-Kolchin}, Michael and {Pawlowski}, Marcel S. and {G{\'o}mez}, Facundo A.},
        title = "{The Orbital Histories of Magellanic Satellites Using Gaia DR2 Proper Motions}",
      journal = {\apj},
         year = 2020,
        month = apr,
       volume = {893},
       number = {2},
          eid = {121},
        pages = {121},
          doi = {10.3847/1538-4357/ab7b75},
archivePrefix = {arXiv},
       eprint = {2001.01746},
 primaryClass = {astro-ph.GA},
       adsurl = {https://ui.adsabs.harvard.edu/abs/2020ApJ...893..121P}
}

@ARTICLE{Sheng2024,
       author = {{Sheng}, Yanjun and {Ting}, Yuan-Sen and {Xue}, Xiang-Xiang and {Chang}, Jiang and {Tian}, Hao},
        title = "{Uncovering the first-infall history of the LMC through its dynamical impact in the Milky Way halo}",
      journal = {\mnras},
         year = 2024,
        month = nov,
       volume = {534},
       number = {3},
        pages = {2694-2714},
          doi = {10.1093/mnras/stae2259},
archivePrefix = {arXiv},
       eprint = {2404.08975},
 primaryClass = {astro-ph.GA},
       adsurl = {https://ui.adsabs.harvard.edu/abs/2024MNRAS.534.2694S}
}

@ARTICLE{Simon2018,
       author = {{Simon}, Joshua D.},
        title = "{Gaia Proper Motions and Orbits of the Ultra-faint Milky Way Satellites}",
      journal = {\apj},
         year = 2018,
        month = aug,
       volume = {863},
       number = {1},
          eid = {89},
        pages = {89},
          doi = {10.3847/1538-4357/aacdfb},
archivePrefix = {arXiv},
       eprint = {1804.10230},
 primaryClass = {astro-ph.GA},
       adsurl = {https://ui.adsabs.harvard.edu/abs/2018ApJ...863...89S}
}

@ARTICLE{Conroy2021,
       author = {{Conroy}, Charlie and {Naidu}, Rohan P. and {Garavito-Camargo}, Nicol{\'a}s and {Besla}, Gurtina and {Zaritsky}, Dennis and {Bonaca}, Ana and {Johnson}, Benjamin D.},
        title = "{All-sky dynamical response of the Galactic halo to the Large Magellanic Cloud}",
      journal = {\nat},
         year = 2021,
        month = apr,
       volume = {592},
       number = {7855},
        pages = {534-536},
          doi = {10.1038/s41586-021-03385-7},
archivePrefix = {arXiv},
       eprint = {2104.09515},
 primaryClass = {astro-ph.GA},
       adsurl = {https://ui.adsabs.harvard.edu/abs/2021Natur.592..534C}
}

@ARTICLE{GaiaHelmi2018,
       author = {{Gaia Collaboration} and {Helmi}, A. and {van Leeuwen}, F. and {McMillan}, P.~J. and {Massari}, D. and {Antoja}, T. and {Robin}, A.~C. and {Lindegren}, L. and {Bastian}, U. and {Arenou}, F. and {Babusiaux}, C. and {Biermann}, M. and {Breddels}, M.~A. and {Hobbs}, D. and {Jordi}, C. and {Pancino}, E. and {Reyl{\'e}}, C. and {Veljanoski}, J. and {Brown}, A.~G.~A. and {Vallenari}, A. and {Prusti}, T.},
        title = "{Gaia Data Release 2. Kinematics of globular clusters and dwarf galaxies around the Milky Way}",
      journal = {\aap},
         year = 2018,
        month = aug,
       volume = {616},
          eid = {A12},
        pages = {A12},
          doi = {10.1051/0004-6361/201832698},
archivePrefix = {arXiv},
       eprint = {1804.09381},
 primaryClass = {astro-ph.GA},
       adsurl = {https://ui.adsabs.harvard.edu/abs/2018A&A...616A..12G}
}

@ARTICLE{Li2021,
       author = {{Li}, Hefan and {Hammer}, Francois and {Babusiaux}, Carine and {Pawlowski}, Marcel S. and {Yang}, Yanbin and {Arenou}, Frederic and {Du}, Cuihua and {Wang}, Jianling},
        title = "{Gaia EDR3 Proper Motions of Milky Way Dwarfs. I. 3D Motions and Orbits}",
      journal = {\apj},
         year = 2021,
        month = jul,
       volume = {916},
       number = {1},
          eid = {8},
        pages = {8},
          doi = {10.3847/1538-4357/ac0436},
archivePrefix = {arXiv},
       eprint = {2104.03974},
 primaryClass = {astro-ph.GA},
       adsurl = {https://ui.adsabs.harvard.edu/abs/2021ApJ...916....8L}
}

@ARTICLE{Patel2017,
       author = {{Patel}, Ekta and {Besla}, Gurtina and {Sohn}, Sangmo Tony},
        title = "{Orbits of massive satellite galaxies - I. A close look at the Large Magellanic Cloud and a new orbital history for M33}",
      journal = {\mnras},
         year = 2017,
        month = feb,
       volume = {464},
       number = {4},
        pages = {3825-3849},
          doi = {10.1093/mnras/stw2616},
archivePrefix = {arXiv},
       eprint = {1609.04823},
 primaryClass = {astro-ph.GA},
       adsurl = {https://ui.adsabs.harvard.edu/abs/2017MNRAS.464.3825P}
}

@ARTICLE{Sohn2020,
       author = {{Sohn}, Sangmo Tony and {Patel}, Ekta and {Fardal}, Mark A. and {Besla}, Gurtina and {van der Marel}, Roeland P. and {Geha}, Marla and {Guhathakurta}, Puragra},
        title = "{HST Proper Motions of NGC 147 and NGC 185: Orbital Histories and Tests of a Dynamically Coherent Andromeda Satellite Plane}",
      journal = {\apj},
         year = 2020,
        month = sep,
       volume = {901},
       number = {1},
          eid = {43},
        pages = {43},
          doi = {10.3847/1538-4357/abaf49},
archivePrefix = {arXiv},
       eprint = {2008.06055},
 primaryClass = {astro-ph.GA},
       adsurl = {https://ui.adsabs.harvard.edu/abs/2020ApJ...901...43S}
}

@ARTICLE{vanderMarel2019,
       author = {{van der Marel}, Roeland P. and {Fardal}, Mark A. and {Sohn}, Sangmo Tony and {Patel}, Ekta and {Besla}, Gurtina and {del Pino}, Andr{\'e}s and {Sahlmann}, Johannes and {Watkins}, Laura L.},
        title = "{First Gaia Dynamics of the Andromeda System: DR2 Proper Motions, Orbits, and Rotation of M31 and M33}",
      journal = {\apj},
         year = 2019,
        month = feb,
       volume = {872},
       number = {1},
          eid = {24},
        pages = {24},
          doi = {10.3847/1538-4357/ab001b},
archivePrefix = {arXiv},
       eprint = {1805.04079},
 primaryClass = {astro-ph.GA},
       adsurl = {https://ui.adsabs.harvard.edu/abs/2019ApJ...872...24V}
}

@ARTICLE{Samuel2022,
       author = {{Samuel}, Jenna and {Wetzel}, Andrew and {Santistevan}, Isaiah and {Tollerud}, Erik and {Moreno}, Jorge and {Boylan-Kolchin}, Michael and {Bailin}, Jeremy and {Pardasani}, Bhavya},
        title = "{Extinguishing the FIRE: environmental quenching of satellite galaxies around Milky Way-mass hosts in simulations}",
      journal = {\mnras},
         year = 2022,
        month = aug,
       volume = {514},
       number = {4},
        pages = {5276-5295},
          doi = {10.1093/mnras/stac1706},
archivePrefix = {arXiv},
       eprint = {2203.07385},
 primaryClass = {astro-ph.GA},
       adsurl = {https://ui.adsabs.harvard.edu/abs/2022MNRAS.514.5276S}
}

@ARTICLE{Fillingham2019,
       author = {{Fillingham}, Sean P. and {Cooper}, Michael C. and {Kelley}, Tyler and {Rodriguez Wimberly}, M.~K. and {Boylan-Kolchin}, Michael and {Bullock}, James S. and {Garrison-Kimmel}, Shea and {Pawlowski}, Marcel S. and {Wheeler}, Coral},
        title = "{Characterizing the Infall Times and Quenching Timescales of Milky Way Satellites with $Gaia$ Proper Motions}",
      journal = {arXiv e-prints},
         year = 2019,
        month = jun,
          eid = {arXiv:1906.04180},
        pages = {arXiv:1906.04180},
          doi = {10.48550/arXiv.1906.04180},
archivePrefix = {arXiv},
       eprint = {1906.04180},
 primaryClass = {astro-ph.GA},
       adsurl = {https://ui.adsabs.harvard.edu/abs/2019arXiv190604180F}
}

@ARTICLE{Geha2024,
       author = {{Geha}, Marla and {Mao}, Yao-Yuan and {Wechsler}, Risa H. and {Asali}, Yasmeen and {Kado-Fong}, Erin and {Kallivayalil}, Nitya and {Nadler}, Ethan O. and {Tollerud}, Erik J. and {Weiner}, Benjamin and {de los Reyes}, Mithi A.~C. and {Wang}, Yunchong and {Wu}, John F.},
        title = "{The SAGA Survey. IV. The Star Formation Properties of 101 Satellite Systems around Milky Way─mass Galaxies}",
      journal = {\apj},
         year = 2024,
        month = nov,
       volume = {976},
       number = {1},
          eid = {118},
        pages = {118},
          doi = {10.3847/1538-4357/ad61e7},
archivePrefix = {arXiv},
       eprint = {2404.14499},
 primaryClass = {astro-ph.GA},
       adsurl = {https://ui.adsabs.harvard.edu/abs/2024ApJ...976..118G}
}

@ARTICLE{Via2026,
       author = {{The Via Collaboration}},
        title = "{The Via Project: Overview of the Science, Instrument, and Survey}",
      journal = {arXiv e-prints},
         year = 2026,
        month = jun,
          eid = {arXiv:2606.18332},
        pages = {arXiv:2606.18332},
          doi = {10.48550/arXiv.2606.18332},
archivePrefix = {arXiv},
       eprint = {2606.18332},
 primaryClass = {astro-ph.IM},
       adsurl = {https://ui.adsabs.harvard.edu/abs/2026arXiv260618332T}
}

@ARTICLE{Taylor2001,
       author = {{Taylor}, James E. and {Babul}, Arif},
        title = "{The Dynamics of Sinking Satellites around Disk Galaxies: A Poor Man's Alternative to High-Resolution Numerical Simulations}",
      journal = {\apj},
         year = 2001,
        month = oct,
       volume = {559},
       number = {2},
        pages = {716-735},
          doi = {10.1086/322276},
archivePrefix = {arXiv},
       eprint = {astro-ph/0012305},
 primaryClass = {astro-ph},
       adsurl = {https://ui.adsabs.harvard.edu/abs/2001ApJ...559..716T}
}

@ARTICLE{Erkal2016,
       author = {{Erkal}, Denis and {Sanders}, Jason L. and {Belokurov}, Vasily},
        title = "{Stray, swing and scatter: angular momentum evolution of orbits and streams in aspherical potentials}",
      journal = {\mnras},
         year = 2016,
        month = sep,
       volume = {461},
       number = {2},
        pages = {1590-1604},
          doi = {10.1093/mnras/stw1400},
archivePrefix = {arXiv},
       eprint = {1603.08922},
 primaryClass = {astro-ph.GA},
       adsurl = {https://ui.adsabs.harvard.edu/abs/2016MNRAS.461.1590E}
}

@ARTICLE{Chen2017,
       author = {{Chen}, Shu-Rong and {Schive}, Hsi-Yu and {Chiueh}, Tzihong},
        title = "{Jeans analysis for dwarf spheroidal galaxies in wave dark matter}",
      journal = {\mnras},
         year = 2017,
        month = jun,
       volume = {468},
       number = {2},
        pages = {1338-1348},
          doi = {10.1093/mnras/stx449},
archivePrefix = {arXiv},
       eprint = {1606.09030},
 primaryClass = {astro-ph.GA},
       adsurl = {https://ui.adsabs.harvard.edu/abs/2017MNRAS.468.1338C}
}

@ARTICLE{Dalal2022,
       author = {{Dalal}, Neal and {Kravtsov}, Andrey},
        title = "{Excluding fuzzy dark matter with sizes and stellar kinematics of ultrafaint dwarf galaxies}",
      journal = {\prd},
         year = 2022,
        month = sep,
       volume = {106},
       number = {6},
          eid = {063517},
        pages = {063517},
          doi = {10.1103/PhysRevD.106.063517},
archivePrefix = {arXiv},
       eprint = {2203.05750},
 primaryClass = {astro-ph.CO},
       adsurl = {https://ui.adsabs.harvard.edu/abs/2022PhRvD.106f3517D}
}

@ARTICLE{Mitra2024,
       author = {{Mitra}, Kaustav and {van den Bosch}, Frank C. and {Lange}, Johannes U.},
        title = "{BASILISK II. Improved constraints on the galaxy-halo connection from satellite kinematics in SDSS}",
      journal = {\mnras},
         year = 2024,
        month = sep,
       volume = {533},
       number = {3},
        pages = {3647-3675},
          doi = {10.1093/mnras/stae2030},
archivePrefix = {arXiv},
       eprint = {2409.03105},
 primaryClass = {astro-ph.CO},
       adsurl = {https://ui.adsabs.harvard.edu/abs/2024MNRAS.533.3647M}
}

@ARTICLE{Mitra2025,
       author = {{Mitra}, Kaustav and {van den Bosch}, Frank C. and {Baggen}, Josephine and {Lange}, Johannes U.},
        title = "{BASILISK IV. No $S_8$ Tension with Satellite Kinematics}",
      journal = {arXiv e-prints},
         year = 2025,
        month = dec,
          eid = {arXiv:2512.14889},
        pages = {arXiv:2512.14889},
          doi = {10.48550/arXiv.2512.14889},
archivePrefix = {arXiv},
       eprint = {2512.14889},
 primaryClass = {astro-ph.CO},
       adsurl = {https://ui.adsabs.harvard.edu/abs/2025arXiv251214889M}
}

@ARTICLE{Marsh2019,
       author = {{Marsh}, David J.~E. and {Niemeyer}, Jens C.},
        title = "{Strong Constraints on Fuzzy Dark Matter from Ultrafaint Dwarf Galaxy Eridanus II}",
      journal = {\prl},
         year = 2019,
        month = aug,
       volume = {123},
       number = {5},
          eid = {051103},
        pages = {051103},
          doi = {10.1103/PhysRevLett.123.051103},
archivePrefix = {arXiv},
       eprint = {1810.08543},
 primaryClass = {astro-ph.CO},
       adsurl = {https://ui.adsabs.harvard.edu/abs/2019PhRvL.123e1103M}
}

@ARTICLE{Chiang2026b,
       author = {{Chiang}, Barry T. and {Dutra}, Isaque and {Natarajan}, Priyamvada},
        title = "{Constraining the Nature of Dark Matter from Tidal Radii of Cluster Galaxy Subhalos}",
      journal = {\apj},
         year = 2026,
        month = jan,
       volume = {997},
       number = {1},
          eid = {106},
        pages = {106},
          doi = {10.3847/1538-4357/ae23c1},
archivePrefix = {arXiv},
       eprint = {2511.14726},
 primaryClass = {astro-ph.CO},
       adsurl = {https://ui.adsabs.harvard.edu/abs/2026ApJ...997..106C}
}

@ARTICLE{Natarajan2026,
       author = {{Natarajan}, Priyamvada and {Chiang}, Barry T. and {Dutra}, Isaque},
        title = "{New Cold Dark Matter Crisis Revealed by Multiscale Cluster Lensing}",
      journal = {\apjl},
         year = 2026,
        month = apr,
       volume = {1001},
       number = {1},
          eid = {L12},
        pages = {L12},
          doi = {10.3847/2041-8213/ae53ea},
archivePrefix = {arXiv},
       eprint = {2601.07909},
 primaryClass = {astro-ph.CO},
       adsurl = {https://ui.adsabs.harvard.edu/abs/2026ApJ..1001L..12N}
}

@ARTICLE{Calabrese2016,
       author = {{Calabrese}, Erminia and {Spergel}, David N.},
        title = "{Ultra-light dark matter in ultra-faint dwarf galaxies}",
      journal = {\mnras},
         year = 2016,
        month = aug,
       volume = {460},
       number = {4},
        pages = {4397-4402},
          doi = {10.1093/mnras/stw1256},
archivePrefix = {arXiv},
       eprint = {1603.07321},
 primaryClass = {astro-ph.CO},
       adsurl = {https://ui.adsabs.harvard.edu/abs/2016MNRAS.460.4397C}
}

@ARTICLE{Nadler2019,
       author = {{Nadler}, Ethan O. and {Gluscevic}, Vera and {Boddy}, Kimberly K. and {Wechsler}, Risa H.},
        title = "{Constraints on Dark Matter Microphysics from the Milky Way Satellite Population}",
      journal = {\apjl},
         year = 2019,
        month = jun,
       volume = {878},
       number = {2},
          eid = {L32},
        pages = {L32},
          doi = {10.3847/2041-8213/ab1eb2},
archivePrefix = {arXiv},
       eprint = {1904.10000},
 primaryClass = {astro-ph.CO},
       adsurl = {https://ui.adsabs.harvard.edu/abs/2019ApJ...878L..32N}
}

@ARTICLE{Nadler2021,
       author = {{Nadler}, E.~O. and {Drlica-Wagner}, A. and {Bechtol}, K. and {Mau}, S. and {Wechsler}, R.~H. and {Gluscevic}, V. and {Boddy}, K. and {Pace}, A.~B. and {Li}, T.~S. and {McNanna}, M. and {Riley}, A.~H. and {Garc{\'\i}a-Bellido}, J. and {Mao}, Y.-Y. and {Green}, G. and {Burke}, D.~L. and {Peter}, A. and {Jain}, B. and {Abbott}, T.~M.~C. and {Aguena}, M. and {Allam}, S. and {Annis}, J. and {Avila}, S. and {Brooks}, D. and {Carrasco Kind}, M. and {Carretero}, J. and {Costanzi}, M. and {da Costa}, L.~N. and {De Vicente}, J. and {Desai}, S. and {Diehl}, H.~T. and {Doel}, P. and {Everett}, S. and {Evrard}, A.~E. and {Flaugher}, B. and {Frieman}, J. and {Gerdes}, D.~W. and {Gruen}, D. and {Gruendl}, R.~A. and {Gschwend}, J. and {Gutierrez}, G. and {Hinton}, S.~R. and {Honscheid}, K. and {Huterer}, D. and {James}, D.~J. and {Krause}, E. and {Kuehn}, K. and {Kuropatkin}, N. and {Lahav}, O. and {Maia}, M.~A.~G. and {Marshall}, J.~L. and {Menanteau}, F. and {Miquel}, R. and {Palmese}, A. and {Paz-Chinch{\'o}n}, F. and {Plazas}, A.~A. and {Romer}, A.~K. and {Sanchez}, E. and {Scarpine}, V. and {Serrano}, S. and {Sevilla-Noarbe}, I. and {Smith}, M. and {Soares-Santos}, M. and {Suchyta}, E. and {Swanson}, M.~E.~C. and {Tarle}, G. and {Tucker}, D.~L. and {Walker}, A.~R. and {Wester}, W. and {DES Collaboration}},
        title = "{Constraints on Dark Matter Properties from Observations of Milky Way Satellite Galaxies}",
      journal = {\prl},
         year = 2021,
        month = mar,
       volume = {126},
       number = {9},
          eid = {091101},
        pages = {091101},
          doi = {10.1103/PhysRevLett.126.091101},
archivePrefix = {arXiv},
       eprint = {2008.00022},
 primaryClass = {astro-ph.CO},
       adsurl = {https://ui.adsabs.harvard.edu/abs/2021PhRvL.126i1101N}
}

@ARTICLE{Liu2026,
       author = {{Liu}, Jianxiang and {Gong}, Yan and {Liao}, Kai},
        title = "{Joint Constraints on Fuzzy and Warm Dark Matter from Satellite Populations of the Milky Way and Andromeda}",
      journal = {\apj},
         year = 2026,
        month = mar,
       volume = {1000},
       number = {1},
          eid = {88},
        pages = {88},
          doi = {10.3847/1538-4357/ae48e9},
archivePrefix = {arXiv},
       eprint = {2512.01361},
 primaryClass = {astro-ph.CO},
       adsurl = {https://ui.adsabs.harvard.edu/abs/2026ApJ..1000...88L}
}

@ARTICLE{Dekker2022,
       author = {{Dekker}, Ariane and {Ando}, Shin'ichiro and {Correa}, Camila A. and {Ng}, Kenny C.~Y.},
        title = "{Warm dark matter constraints using Milky Way satellite observations and subhalo evolution modeling}",
      journal = {\prd},
         year = 2022,
        month = dec,
       volume = {106},
       number = {12},
          eid = {123026},
        pages = {123026},
          doi = {10.1103/PhysRevD.106.123026},
archivePrefix = {arXiv},
       eprint = {2111.13137},
 primaryClass = {astro-ph.CO},
       adsurl = {https://ui.adsabs.harvard.edu/abs/2022PhRvD.106l3026D}
}

@ARTICLE{Zavala2013,
       author = {{Zavala}, J. and {Vogelsberger}, M. and {Walker}, M.~G.},
        title = "{Constraining self-interacting dark matter with the Milky way's dwarf spheroidals.}",
      journal = {\mnras},
         year = 2013,
        month = apr,
       volume = {431},
        pages = {L20-L24},
          doi = {10.1093/mnrasl/sls053},
archivePrefix = {arXiv},
       eprint = {1211.6426},
 primaryClass = {astro-ph.CO},
       adsurl = {https://ui.adsabs.harvard.edu/abs/2013MNRAS.431L..20Z}
}

@ARTICLE{Correa2021,
       author = {{Correa}, Camila A.},
        title = "{Constraining velocity-dependent self-interacting dark matter with the Milky Way's dwarf spheroidal galaxies}",
      journal = {\mnras},
         year = 2021,
        month = may,
       volume = {503},
       number = {1},
        pages = {920-937},
          doi = {10.1093/mnras/stab506},
archivePrefix = {arXiv},
       eprint = {2007.02958},
 primaryClass = {astro-ph.GA},
       adsurl = {https://ui.adsabs.harvard.edu/abs/2021MNRAS.503..920C}
}

@ARTICLE{Chiang2021,
       author = {{Chiang}, Barry T. and {Schive}, Hsi-Yu and {Chiueh}, Tzihong},
        title = "{Soliton oscillations and revised constraints from Eridanus II of fuzzy dark matter}",
      journal = {\prd},
         year = 2021,
        month = may,
       volume = {103},
       number = {10},
          eid = {103019},
        pages = {103019},
          doi = {10.1103/PhysRevD.103.103019},
archivePrefix = {arXiv},
       eprint = {2104.13359},
 primaryClass = {astro-ph.CO},
       adsurl = {https://ui.adsabs.harvard.edu/abs/2021PhRvD.103j3019C}
}

@ARTICLE{vandenBosch2004,
       author = {{van den Bosch}, Frank C. and {Norberg}, Peder and {Mo}, H.~J. and {Yang}, Xiaohu},
        title = "{Probing dark matter haloes with satellite kinematics}",
      journal = {\mnras},
         year = 2004,
        month = aug,
       volume = {352},
       number = {4},
        pages = {1302-1314},
          doi = {10.1111/j.1365-2966.2004.08021.x},
archivePrefix = {arXiv},
       eprint = {astro-ph/0404033},
 primaryClass = {astro-ph},
       adsurl = {https://ui.adsabs.harvard.edu/abs/2004MNRAS.352.1302V}
}

@ARTICLE{More2011,
       author = {{More}, Surhud and {van den Bosch}, Frank C. and {Cacciato}, Marcello and {Skibba}, Ramin and {Mo}, H.~J. and {Yang}, Xiaohu},
        title = "{Satellite kinematics - III. Halo masses of central galaxies in SDSS}",
      journal = {\mnras},
         year = 2011,
        month = jan,
       volume = {410},
       number = {1},
        pages = {210-226},
          doi = {10.1111/j.1365-2966.2010.17436.x},
archivePrefix = {arXiv},
       eprint = {1003.3203},
 primaryClass = {astro-ph.CO},
       adsurl = {https://ui.adsabs.harvard.edu/abs/2011MNRAS.410..210M}
}

@ARTICLE{Lange2019,
       author = {{Lange}, Johannes U. and {van den Bosch}, Frank C. and {Zentner}, Andrew R. and {Wang}, Kuan and {Villarreal}, Antonia Sierra},
        title = "{Updated results on the galaxy-halo connection from satellite kinematics in SDSS}",
      journal = {\mnras},
         year = 2019,
        month = aug,
       volume = {487},
       number = {3},
        pages = {3112-3129},
          doi = {10.1093/mnras/stz1466},
archivePrefix = {arXiv},
       eprint = {1811.03596},
 primaryClass = {astro-ph.CO},
       adsurl = {https://ui.adsabs.harvard.edu/abs/2019MNRAS.487.3112L}
}

@ARTICLE{Bose2016,
       author = {{Bose}, Sownak and {Hellwing}, Wojciech A. and {Frenk}, Carlos S. and {Jenkins}, Adrian and {Lovell}, Mark R. and {Helly}, John C. and {Li}, Baojiu},
        title = "{The Copernicus Complexio: statistical properties of warm dark matter haloes}",
      journal = {\mnras},
         year = 2016,
        month = jan,
       volume = {455},
       number = {1},
        pages = {318-333},
          doi = {10.1093/mnras/stv2294},
archivePrefix = {arXiv},
       eprint = {1507.01998},
 primaryClass = {astro-ph.CO},
       adsurl = {https://ui.adsabs.harvard.edu/abs/2016MNRAS.455..318B}
}

@ARTICLE{Giocoli2026,
       author = {{Giocoli}, C. and {Despali}, G. and {Moscardini}, L. and {Meneghetti}, M. and {Sheth}, R.~K. and {Pillepich}, A. and {Vogelsberger}, M.},
        title = "{The AIDA-TNG project: 3D halo shapes}",
      journal = {\aap},
         year = 2026,
        month = feb,
       volume = {706},
          eid = {A340},
        pages = {A340},
          doi = {10.1051/0004-6361/202558400},
archivePrefix = {arXiv},
       eprint = {2512.15856},
 primaryClass = {astro-ph.CO},
       adsurl = {https://ui.adsabs.harvard.edu/abs/2026A&A...706A.340G}
}

@ARTICLE{Peter2013,
       author = {{Peter}, Annika H.~G. and {Rocha}, Miguel and {Bullock}, James S. and {Kaplinghat}, Manoj},
        title = "{Cosmological simulations with self-interacting dark matter - II. Halo shapes versus observations}",
      journal = {\mnras},
         year = 2013,
        month = mar,
       volume = {430},
       number = {1},
        pages = {105-120},
          doi = {10.1093/mnras/sts535},
archivePrefix = {arXiv},
       eprint = {1208.3026},
 primaryClass = {astro-ph.CO},
       adsurl = {https://ui.adsabs.harvard.edu/abs/2013MNRAS.430..105P}
}

@ARTICLE{Schive2014,
       author = {{Schive}, Hsi-Yu and {Chiueh}, Tzihong and {Broadhurst}, Tom},
        title = "{Cosmic structure as the quantum interference of a coherent dark wave}",
      journal = {Nature Physics},
         year = 2014,
        month = jul,
       volume = {10},
       number = {7},
        pages = {496-499},
          doi = {10.1038/nphys2996},
archivePrefix = {arXiv},
       eprint = {1406.6586},
 primaryClass = {astro-ph.GA},
       adsurl = {https://ui.adsabs.harvard.edu/abs/2014NatPh..10..496S}
}

@ARTICLE{Vargya2022,
       author = {{Vargya}, Drona and {Sanderson}, Robyn and {Sameie}, Omid and {Boylan-Kolchin}, Michael and {Hopkins}, Philip F. and {Wetzel}, Andrew and {Graus}, Andrew},
        title = "{Shapes of Milky-Way-mass galaxies with self-interacting dark matter}",
      journal = {\mnras},
         year = 2022,
        month = oct,
       volume = {516},
       number = {2},
        pages = {2389-2405},
          doi = {10.1093/mnras/stac2069},
archivePrefix = {arXiv},
       eprint = {2104.14069},
 primaryClass = {astro-ph.GA},
       adsurl = {https://ui.adsabs.harvard.edu/abs/2022MNRAS.516.2389V}
}

@ARTICLE{deZeeuw1985,
       author = {{de Zeeuw}, T.},
        title = "{Elliptical galaxies with separable potentials}",
      journal = {\mnras},
         year = 1985,
        month = sep,
       volume = {216},
        pages = {273-334},
          doi = {10.1093/mnras/216.2.273},
       adsurl = {https://ui.adsabs.harvard.edu/abs/1985MNRAS.216..273D}
}

@ARTICLE{Statler1987,
       author = {{Statler}, Thomas S.},
        title = "{Self-consistent Models of Perfect Triaxial Galaxies}",
      journal = {\apj},
         year = 1987,
        month = oct,
       volume = {321},
        pages = {113},
          doi = {10.1086/165619},
       adsurl = {https://ui.adsabs.harvard.edu/abs/1987ApJ...321..113S}
}

@ARTICLE{BeraldoeSilva2023,
       author = {{Beraldo e Silva}, Leandro and {Debattista}, Victor P. and {Anderson}, Stuart Robert and {Valluri}, Monica and {Erwin}, Peter and {Daniel}, Kathryne J. and {Deg}, Nathan},
        title = "{Orbital Support and Evolution of Flat Profiles of Bars (Shoulders)}",
      journal = {\apj},
         year = 2023,
        month = sep,
       volume = {955},
       number = {1},
          eid = {38},
        pages = {38},
          doi = {10.3847/1538-4357/ace976},
archivePrefix = {arXiv},
       eprint = {2303.04828},
 primaryClass = {astro-ph.GA},
       adsurl = {https://ui.adsabs.harvard.edu/abs/2023ApJ...955...38B}
}

%%%%%%%%%%%%%%%%%%%%%%%%%%%%%%%%%%%%%%%%%%
\appendix

\section{Drift-corrected Lyapunov exponents}
\label{app:lyapunov}

We measure the maximal Lyapunov exponent $\lamc$ of $10^5$ orbits per host with the shadow-trajectory method of \citet{Benettin1976} \citep[see also][]{PriceWhelan2016}. For each orbit we integrate a companion offset in position by $\delta_0 = 10^{-4}\kpc$, a scale seven orders of magnitude below the orbital scale yet far above the positional error of the integrator (relative and absolute tolerances of $10^{-12}$). Every $0.1\Gyr$ we record the phase-space separation of the pair, $\Delta = \left(|\delta\mathbf{x}|^2 + T_U^2\,|\delta\mathbf{v}|^2\right)^{1/2}$, with the velocity-rescaling constant $T_U = 1\kpc\,{\rm km^{-1}\,s}$ (the results are insensitive to the choice of $T_U > 0$, formally and as verified numerically). Both time-scales are short compared with the orbital periods, whose medians are $\Trad \simeq 8$--$14\Gyr$ across the three energy bins (\fref{fig:periods}); each $0.1\Gyr$ segment thus corresponds to $\simeq\!10^{-2}\,\Trad$, and $T_U \simeq 0.98\Gyr$ to $0.07$--$0.12\,\Trad$. After each step, the companion orbit is then re-seeded with its phase-space offset reset back to $\delta_0$ along the same direction, and a fresh $0.1\Gyr$ segment begins from there, while the fiducial orbit itself is integrated continuously throughout. The running sum $\lamft(t) = t^{-1} \sum_k \ln(\Delta_k/\delta_0)$ over the segments defines the finite-time estimate, which converges to $\lamc$ as $t \to \infty$. This repeated re-seeding keeps the pair within the linear regime where the Lyapunov exponent is defined; a single shadow orbit integrated indefinitely would instead saturate at the orbital scale within a few e-folds in a gravitationally bound system\footnote{Independently, frequency-mapping techniques diagnose chaos from the drift of the fundamental orbital frequencies, over far shorter integration baselines than Lyapunov exponents require \citep{Valluri1998}. We nevertheless adopt the Lyapunov approach as we are interested not only in \textit{whether} an orbit is chaotic but also in \textit{how fast} it diverges. Specifically, the reconstruction-error budget of \S\ref{ssec:divergence} requires the per-orbit e-folding time $\tchaos = 1/\lamc$ in physical units, to be compared directly with the Hubble time; frequency drift provides a chaos indicator and a diffusion rate in frequency space, but not directly the configuration-space divergence time-scale.}.

\begin{figure}
    \includegraphics[width=\linewidth]{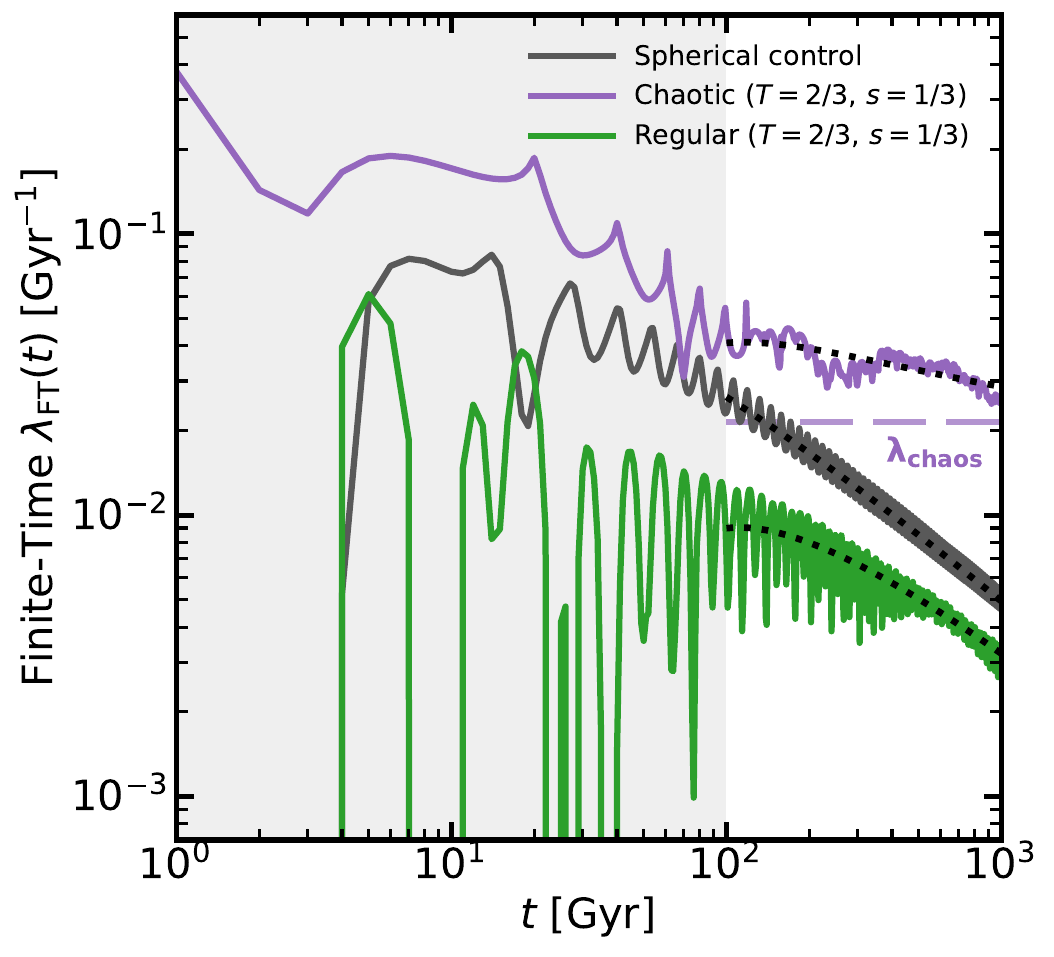}
    \caption{Finite-time Lyapunov estimate $\lamft(t)$ for three representative orbits: one in the spherical control (dark grey), and a regular (green) and a chaotic (purple) orbit in the $T = 2/3$, $s = 1/3$ host. Dotted curves show the fitted model of \eref{eq:drift}; the dashed horizontal line marks the fitted $\lamc$ of the chaotic orbit. The grey band marks the early transient excluded from the fit.}
    \label{fig:driftfit}
\end{figure}

For orbits in a static potential, the finite-time estimate carries a well-known bias as demonstrated in \fref{fig:driftfit}. A regular orbit has $\lamc = 0$ by definition, but nearby regular orbits still dephase, separating linearly in time through their frequency difference, so $\lamft$ decays only as $\ln t/t$ and never reaches zero at any finite time. In \fref{fig:driftfit}, the regular triaxial orbit (green) is indistinguishable from the integrable spherical control (dark grey) at every epoch. A chaotic orbit (purple) instead flattens onto a positive plateau, but even after $1000\Gyr$ its raw endpoint still sits visibly above that plateau. We therefore fit every orbit with
\begin{equation}\label{eq:drift}
\lamft(t) = \lamc + \frac{A \ln t + B}{t}\,,
\end{equation}
which is linear in its three coefficients, and adopt the fitted constant $\lamc$ as the chaos measure. The spherical control both validates the estimator and calibrates the detection threshold. Because the spherical host is integrable, every orbit there has $\lamc = 0$ formally, so the fitted $\lamc$ values of its $10^5$ orbits, processed through the identical pipeline, sample the pure noise distribution of the estimator, with a median of $2 \times 10^{-5}\Gyrmin$. Adopting the $99$th percentile of this noise distribution ($6.6 \times 10^{-4}\Gyrmin$) as the chaotic orbit classification threshold fixes the false-positive rate at $1\%$ by construction. The population statistics of \S\ref{ssec:census} change by $\lesssim\!1$ percentage point when the fit is truncated at $900\Gyr$ instead of $1000\Gyr$.

The threshold is an \textit{operational definition}, not a physical boundary. The $\lamc$ distribution is continuous, with weak chaos extending down towards the noise floor, so any chaos census must cut such a statistic at a chosen sensitivity, and the quoted chaotic fractions depend on the adopted convention. Our construction parallels that of \citet{Valluri2010}, who likewise set their frequency-drift threshold at the $99.5$th percentile of a spherical control. Importantly, the results are qualitatively insensitive to the exact choice adopted. Raising the threshold tenfold lowers $\fchaos$ in \fref{fig:census} by only $\sim\!10$ percentage points in the flattened hosts ($44$--$53\%$ to $34$--$41\%$), while the median $\tchaos$ of the surviving chaotic orbits remains $\geq\!28\Gyr$. As an independent check, we also performed the NAFF frequency drift analysis using the public \textsc{naif} implementation \citep{BeraldoeSilva2023}. These two methodologically distinct approaches agree on the ranking of both the orbits and the hosts: in every host, the Lyapunov-chaotic orbits drift systematically more than the regular ones, and both identify chaos as most prevalent in the $s = 1/3$, $T \geq 1/3$ hosts, a level of qualitative consistency also noted by \citet{PriceWhelan2016}.

\label{lastpage}
\end{document}